\documentclass[11pt,oneside,fleqn,reqno]{article}
\usepackage{amsmath}
\usepackage{amsthm}
\usepackage{amssymb}
\usepackage[pdftex]{graphicx,color}
\usepackage{mathtools}
\usepackage{wrapfig}
\usepackage{natbib} \citeindextrue
\usepackage{subfigure}
\usepackage{dsfont}
\usepackage{multirow}
\usepackage{multicol}
\usepackage{algorithm}
\usepackage{algorithmic}
\usepackage{bm}
\usepackage{comment}

\usepackage{epstopdf}

\allowdisplaybreaks

\def \qed{\hfill $\Box$}

\def \hbar{\bar h}

\def \v1 {\mathbb{F}}

\newcommand{\doublespace}{\addtolength{\baselineskip}{.5\baselineskip}}
\newcounter{stepno}

\newtheorem{thm}{Theorem}[section]

\newtheorem{prop}{Proposition}[section]

\newcommand{\bnn}{\\ \\$\begin{array}{rcll}}
\newcommand{\enn}{\end{array}$\\ \\}
\newcommand{\ba}{\begin{eqnarray}}
\newcommand{\ea}{\end{eqnarray}}
\newcommand{\bas}{\begin{eqnarray*}}
\newcommand{\eas}{\end{eqnarray*}}
\newcommand{\bdm}{\begin{displaymath}}
\newcommand{\edm}{\end{displaymath}}
\newcommand{\be}{\begin{equation}}
\newcommand{\ee}{\end{equation}}
\newcommand{\bn}{\begin{eqnarray}}
\newcommand{\en}{\end{eqnarray}}
\newcommand{\bns}{\begin{eqnarray*}}
\newcommand{\ens}{\end{eqnarray*}}

\newcommand{\defvarbegin}{\begin{quotation}\vspace{-15pt}\begin{tabbing}}
\newcommand{\defvarend}  {\end{tabbing}\vspace{-10pt}\end{quotation}}

\newcommand{\bnarray}{\begin{equation}\begin{array}{rcll}}
\newcommand{\enarray}{\end{array}\end{equation}}

\newcommand{\barr}{\begin{array}}
\newcommand{\earr}{\end{array}}

\newcounter{cnum}

\newcommand{\beginalg}{\setcounter{stepno}{1}
                \begin{list}{\bf Step~\arabic{stepno}}
                         {\usecounter{stepno}\settowidth{\labelwidth}{\bf Step~9m}
                \addtolength{\leftmargin}{2\parindent}}
                }
\newcommand{\eg}{\end{list}}

\def \define{\begin{quote}\begin{itemize}}
\def \enddefine{\end{itemize}\end{quote}}

\newlength{\boxedparwidth} 
  {\begin{center} \begin{tabular}{|@{\hspace{.15in}}c@{\hspace{.15in}}|}
                 \hline \\ \begin{minipage}[t]{\boxedparwidth}
                 \setlength{\parindent}{0.0in}}%
   {\end{minipage} \\ \\ \hline \end{tabular} \end{center}}
\newcounter{chapterctr}
\newcounter{example}[chapterctr]
\newcounter{ctr}
\begin{document}

\title{Individual and group fairness in geographical partitioning}
\author{Ilya O. Ryzhov \and John Gunnar Carlsson \and Yinchu Zhu}

\date{\today}

\maketitle

\begin{abstract}
Socioeconomic segregation often arises in school districting and other contexts, causing some groups to be over- or under-represented within a particular district. This phenomenon is closely linked with disparities in opportunities and outcomes. We formulate a new class of geographical partitioning problems in which the population is heterogeneous, and it is necessary to ensure fair representation for each group at each facility. We prove that the optimal solution is a novel generalization of the additively weighted Voronoi diagram, and we propose a simple and efficient algorithm to compute it, thus resolving an open question dating back to \cite{DvWaWo51}. The efficacy and potential for practical insight of the approach are demonstrated in a realistic case study involving seven demographic groups and $78$ district offices.
\end{abstract}

\doublespace

\section{Introduction}

Consider a service system in which individuals are served by facilities at different locations within a geographical region. For example, the facilities could represent schools, polling places, or commercial fulfillment centers. The geographical partitioning problem \citep{CaDe13} divides the region into non-overlapping districts, such that all individuals residing in the same district are served by the same facility. The goal is to choose a partition that optimizes some measure of social welfare, most commonly the average travel cost per individual \citep{CaCaDe16}.

We formulate and study a novel variant of this problem where the population is heterogeneous, consisting of multiple demographic groups, each with a different spatial distribution throughout the region. Again we optimize the expected cost, but now we also impose a new \textit{group fairness} condition: each subpopulation can be neither over- nor under-represented at any facility. In other words, the districts are designed in such a way that the proportion of the population belonging to a particular group in any district must match that group's incidence in the entire population. This condition is also known as ``demographic parity'' in the literature \citep{Dwork12}.

We are motivated by the well-known phenomenon of socioeconomic segregation in school districts \citep{ReBi11}. As income inequality increased over the past several decades, both high- and low-income households exhibited a greater tendency to cluster together geographically, resulting in segregation not only between school districts but also between individual schools in the same district \citep{OwReJe16}. Much of this behavior has been driven by families with children \citep{Ow16}, for whom school boundaries are a major factor in residential decisions. Income segregation between school districts has been linked to gaps in student achievement \citep{RyVa07,Ow18}, as many high-income districts receive more resources due to higher revenues from local taxes \citep{Baker17}. Unequal representation of different socioeconomic groups across districts is thus closely linked with disparities in opportunities and outcomes.

The causes of income segregation are complex and, generally, beyond the scope of what can be ``fixed'' by a single optimization model. Nonetheless, incorporating demographic parity into districting decisions could be seen as one possible way to remove a perceived incentive for economic segregation. In this connection, it is interesting that \cite{Sa17a,Sa17} found empirical evidence that segregation is less pronounced in districts that are ``irregularly'' shaped, i.e., not contained within a small area. Drawing on these observations, a policy paper by \cite{Br19} proposed, ``Another option to disrupt the segregating influence of catchment areas is to gerrymander them in reverse -- that is, to draw catchment area lines to purposely include a diverse population.'' At the same time, such drastic measures can come at a substantial cost \citep{CaSiBrWi24}, even for those groups that are meant to benefit from them: for example, some disadvantaged households may experience significant increases in travel time as a result of rezoning. Formulating and solving a fair districting problem, such as the one in this paper, can help quantify the cost of fairness and contribute to a more informed policy.

In addition to group fairness, we also impose the condition of \textit{individual fairness}, meaning that the assignment of an individual to a facility depends only on the geographical location of the individual, and not on the demographic group to which the individual belongs. It is well-known that these attributes are correlated \citep{BaMcRu04}, but explicit dependence of the decision on the demographic group is perceived as unfair and has been successfully challenged in court \citep{Parents07,Students23}. In our setting, individual fairness means that we have a single partition that is universally applied to the entire population.

Individual and group fairness have been extensively studied in very general contexts, and the distinction between them can be made more precise as follows. Suppose that we randomly select an individual from the population. The individual is described by attributes $\left(X,Z\right)$. In our setting, $X$ represents the geographical location of the individual (a point on the map), but in the more general literature it may model other information; for example, \cite{ZhRy24} considers a hiring problem where $X$ is a vector describing the qualifications of a job applicant. The \textit{protected attribute} $Z$ is a categorical variable indicating the demographic or socioeconomic group to which the individual belongs. We act upon the individual with some decision $Y$ (in our setting, an assignment to a facility). Because $Y$ depends on the observed random attributes, it is also a random variable. We then have the following definitions:

\begin{itemize}
\item Individual fairness \citep{BaSe16}: $Y$ is conditionally independent of $Z$ given $X$.
\item Group fairness \citep{Dwork12}: $Y$ is independent of $Z$.
\end{itemize}


One does not at all imply the other, and it is quite uncommon to achieve both individual and group fairness simultaneously in the same problem. In fact, an influential paper by \cite{KlMuRa17} argues that this is impossible in general, with unavoidable tradeoffs between different definitions of fairness. In the specific context of partitioning, however, a striking result by \cite{DvWaWo51} proved the existence of partitions satisfying both criteria for any number of facilities and subpopulations. Unfortunately, the proof was non-constructive, and the question of how such partitions can be computed remained open. This question is fully resolved in our paper: we provide a complete characterization of partitions that not only satisfy both types of fairness, but are \textit{optimal} with respect to expected cost.

Our approach is inspired by the theory of semidiscrete optimal transport \citep{HaSc20,Vi21}, though our problem does not fully fit that framework due to the presence of the group fairness condition. We relax the partitioning problem to allow probabilistic assignments, so that a fixed location can be assigned to any facility randomly according to a distribution to be determined by the decision-maker. This version of the problem can be formulated as an infinite-dimensional linear program. Crucially, however, the dual of this problem can be written as a nonlinear, but finite-dimensional concave program, which can be tractably solved. The dual variables characterize the optimal assignment policy, which turns out to be deterministic; in other words, even with probabilistic assignments allowed, it is still optimal to assign each location to a single facility. The assignment rule has a geometric interpretation as a novel generalization of the additively weighted Voronoi diagram \citep{Au91}, which is well-known to arise in classical districting problems. In our context, each facility has multiple weights (one for each group) rather than just one, and these weights are mixed differently at each location depending on the relative incidence of each group there.

We then show how the optimal partition can be efficiently computed. Although the objective function of the finite-dimensional dual involves a difficult integral, it turns out that it can be tractably optimized using the method of stochastic approximation \citep{PaKi11}. We only require the ability to simulate individuals from the spatial distribution of any demographic group; in the case of population densities, these data are publicly available. With simulation, the problem can be solved using a very simple algorithm, which can be coded in several lines, and which provably converges at a near-optimal rate. Because the procedure only requires realizations of costs, we can actually solve the problem without knowing the closed form of the cost function. In fact, in our case study, we obtain costs from a database that uses real network data and accounts for traffic patterns in urban environments.

We demonstrate the efficacy and utility of our method on a subregion of Los Angeles with 78 school district offices and seven demographic groups for which census data were available. The case study starkly illustrates the costs and tradeoffs that arise when imposing demographic parity. We find that the imposition of fairness constraints significantly increases travel costs across the board for all groups, including disadvantaged ones. In addition, the presence of segregation can lead to undesirable geometry in the partition, e.g., discontinuities in the service zones. At the same time, our results also provide insight into how these negative impacts can be mitigated. For example, extreme outliers of the travel cost distribution can be reduced by optimizing expected \textit{squared} travel time, which is trivial to do under our modeling and algorithmic framework. Furthermore, reducing the number of distinct groups in the model will improve costs across the board for \textit{all} groups. Potentially, such an analysis could help policymakers decide the set of groups for which fair representation is to be enforced; in some cases, the cost of fairness may outweigh the benefits for certain subpopulations, while also increasing the burden on everyone else. The presence of such cases can also be interpreted as a measure of how much segregation is present in the region under consideration.

In summary, this paper makes the following contributions. 1) We formulate a new class of geographical partitioning problems with a heterogeneous population and demographic parity constraints, neither of which has previously been considered in computational geometry. 2) We prove that the optimal solution is completely characterized by a finite set of weights, which comprise a novel generalization of the additively weighted Voronoi diagram. 3) We show how the optimal weights can be found using a stochastic approximation algorithm, which is trivial to implement, converges at a near-optimal rate, and requires only realizations from the cost function. In this way, we answer a long-standing open question about the computation of fair partitions. 4) We empirically evaluate our approach on a realistic setting, providing insight into the cost of fairness for each subgroup. Overall, our work combines concepts from algorithmic fairness and computational geometry in a novel way, and is of both theoretical and practical interest.

\section{Literature review}\label{sec:review}

Our work bridges three distinct streams of literature, including elements of algorithmic fairness, optimal transport, and geographical partitioning. Below, we position our work relative to each stream.

\textit{Algorithmic fairness}. This literature, surveyed by \cite{BaGoLo20}, has mainly grown out of the computer science community. Both individual \citep{JoKeMoRo16,GuKa21} and group \citep{Dwork12,KaZh21} fairness, in the sense of our paper, have attracted considerable attention. Only one of these definitions is generally adopted in any given study, due to the widespread belief that the differences between them are irreconcilable. In this respect, our work stands out as an example of a setting where two distinct definitions can be satisfied simultaneously.

Computer scientists and statisticians have also proposed other definitions specialized for prediction problems \citep{HaPrPrSr16,Ch17}, which are the main focus of machine learning. For example, given a limited sample, one may wish to balance prediction error \citep{CaVe10}, or reduce prediction bias \citep{ZiRo20}, across multiple demographic groups. These studies do not explicitly model any decision problem, and thus do not apply to situations where the goal is to design a fair \textit{decision} rather than a fair prediction. See also \cite{ZhRy24} for an in-depth example of a problem where fair decisions are very different from fair predictions.

\textit{Optimal transport}. Recent work \citep{HaSc20} has shown that optimal transport theory \citep{PeCu19} offers a useful lens through which to approach geographical partitioning problems. Semidiscrete optimal transport is an infinite-dimensional linear program with a finite-dimensional dual \citep{CaCaDe16}, which can be solved efficiently using the method of stochastic approximation \citep{GeCuPeBa16}, a classical technique from the simulation literature \citep{PaKi11}.

Recently, \cite{JoLu19} and \cite{Chzhen20} leveraged optimal transport theory to find policies satisfying group fairness conditions, by connecting the latter to the Wasserstein barycenter problem \citep{AgCa11}. This approach is limited to quadratic cost functions, and also allows the decision to depend directly on the demographic group, thus violating individual fairness. In fact, when both types of fairness are considered, the problem is no longer reducible to classical optimal transport. However, we show that infinite-dimensional duality still applies and provides a roadmap for efficient computation.

\textit{Geographical partitioning}. The field of computational geometry \citep{CaDe13} has extensively studied additively weighted Voronoi diagrams \citep{AuKl00}, a class of geometric structures where the assignment of an individual to a facility is made by minimizing the sum of travel distance and a facility-specific ``weight.'' Applications include the design of delivery \citep{HaHoLa07} and political \citep{RiScSi08} districts, as well as various problems in facility logistics \citep{YuJaUk12}. Most of these papers, like ours, view partitioning as a central planning problem, with \cite{CaPeRy24} being one exception where both centralized and decentralized behavior is considered. While this literature has examined some notions of fairness, mainly in the form of load-balancing constraints between facilities \citep{ArCaKa09}, our paper is the first to explicitly model multiple demographic groups with different population densities. As a result, we discover a novel generalization of the Voronoi diagram in which each facility uses a location-dependent mixture of weights.

There is also a mature literature on districting that uses mixed-integer programming models; examples include \cite{SwKiJa23}, \cite{ShBu25}, \cite{OzSmGo25} and other references cited therein. Various measures of fair representation can be incorporated into the constraints of these models \citep{ArMaPeRi21}. This work is primarily application-driven. While our approach is more stylized, it enables us to derive a clean analytical characterization of partitions satisfying both individual and group fairness, providing an understanding of how group fairness impacts the geometric structure of the optimal partition.

\section{Problem statement}

Let $\mathcal{X}$ be a compact subset of $\mathbb{R}^d$ representing the geographical region under consideration (in most practical applications, we will have $d=2$). Let $X$ be a random variable following a probability density $f$ supported on $\mathcal{X}$. We may think of $f$ as the population density and $X$ as the geographical location of a randomly selected individual from the population.

Let $Z$ be a discrete random variable taking values in $\left\{1,...,M\right\}$ with $q_z = P\left(Z=z\right)$. Each $z$ represents a different subgroup of the population, with $f_z\left(x\right) = P\left(X \in dx\mid Z = z\right)$ denoting the conditional density for that group. In this paper, $q_z$ and $f_z$ are all known; in practice, one would extract them from census data.

Suppose that there are $K$ facilities located at points $x_1,...,x_K \in \mathcal{X}$. Let $c\left(x,k\right)$ be the cost of assigning an individual at location $x$ to the $k$th facility. A natural and widely used cost function is the Euclidean distance $c\left(x,k\right) = \|x-x_k\|_2$, but our analysis is not limited to this choice. We only require the condition that the pairwise difference $c\left(X,j\right) - c\left(X,k\right)$, for any $j \neq k$, is absolutely continuous with respect to Lebesgue measure.

We formulate the problem
\begin{equation}\label{eq:partition}
\min_{\mathcal{A}_1,...,\mathcal{A}_K} \sum_k \mathbb{E}\left(c\left(X,k\right)1_{\left\{X\in\mathcal{A}_k\right\}}\right)
\end{equation}
subject to
\begin{eqnarray}
P\left(X \in \mathcal{A}_j\cap\mathcal{A}_k\right) &=& 0, \quad j\neq k,\label{eq:disjoint}\\
\bigcup_k \mathcal{A}_k &=& \mathcal{X},\label{eq:union}\\
P\left(X \in \mathcal{A}_k\mid Z=z\right) &=& P\left(X \in \mathcal{A}_k\mid Z=z'\right), \quad z,z'\in\left\{1,...,M\right\}, \quad k=1,...,K.\label{eq:fair}
\end{eqnarray}
This problem partitions $\mathcal{X}$ into sets $\mathcal{A}_k$. Constraints (\ref{eq:disjoint}) ensure that the sets are disjoint except possibly on a boundary that has measure zero. Constraint (\ref{eq:union}) ensures that the sets comprise a partition. An individual appearing at a location $x \in \mathcal{A}_k$ is presumed to be assigned to facility $k$, incurring cost $c\left(x,k\right)$. The objective (\ref{eq:partition}) thus chooses the partition to minimize the expected cost incurred by a randomly selected individual. Thus far, these elements of the problem are typical of geographical partitioning problems, e.g., those in \cite{CaCaDe16}.

In marked contrast with those problems, however, we also include the \textit{group fairness} constraints (\ref{eq:fair}), which ensure that the region served by facility $k$ is probabilistically independent of $Z$. In other words, each facility must serve the same proportion of each subgroup, or, to put it yet another way, the demographics of the individuals served by facility $k$ should match those of the population as a whole. As discussed in Section \ref{sec:review}, group fairness has been studied in a variety of settings, such as prediction \citep{Chzhen20} and hiring \citep{ZhRy24}. The critical distinction here is that (\ref{eq:partition})-(\ref{eq:fair}) also imposes \textit{individual fairness}, that is, we design only a single partition that is applied to all demographic groups. Another way to say this is that the assignment of an individual is based only on the individual's location $X$ and not their demographic $Z$, whereas virtually all existing papers on group fairness would allow the decision to explicitly depend on both.

In the context of geographical partitioning, it is possible to simultaneously achieve both types of fairness. \cite{DvWaWo51} proved that the system (\ref{eq:disjoint})-(\ref{eq:fair}) is feasible, though this paper did not show how the partition could be computed, and indeed it did not explicitly model any objective function. In our framework, the objective (\ref{eq:partition}) introduces the notion of an \textit{optimal} fair partition. In Sections \ref{sec:ot}-\ref{sec:sa}, we will show how such a partition can be characterized and computed. First, however, we present a simple example to illustrate the concept.

\begin{figure}[t]
        \centering
        \subfigure[Red density.]{
            \includegraphics[width=0.31\textwidth]{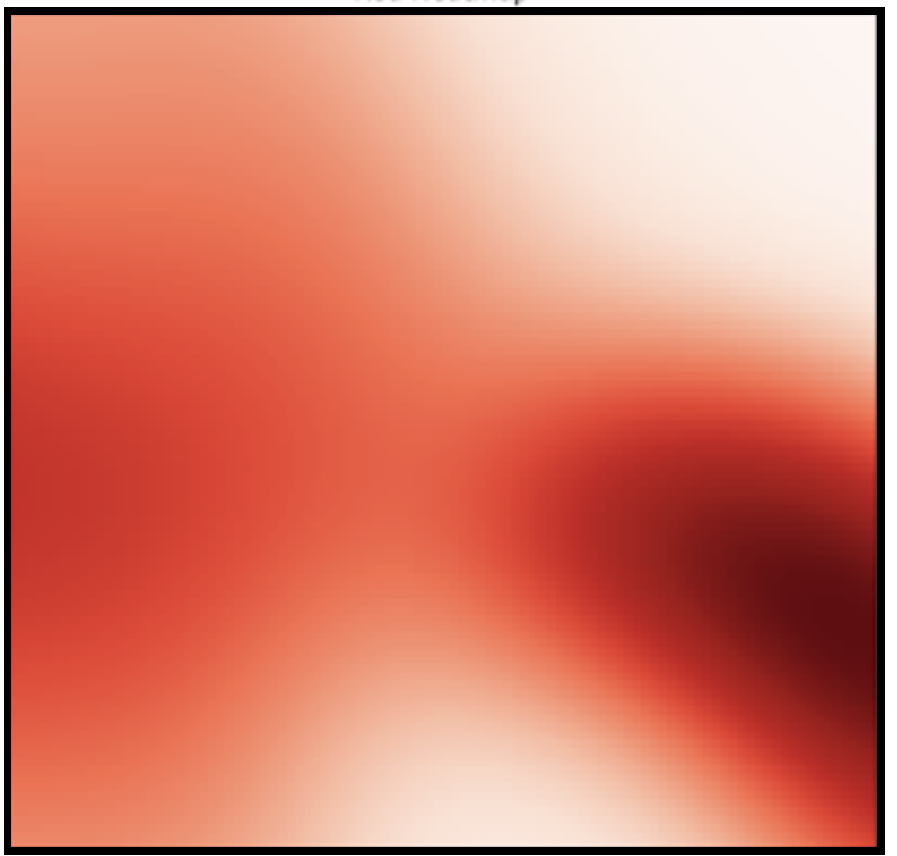}
        }
        \subfigure[Green density.]{
            \includegraphics[width=0.31\textwidth]{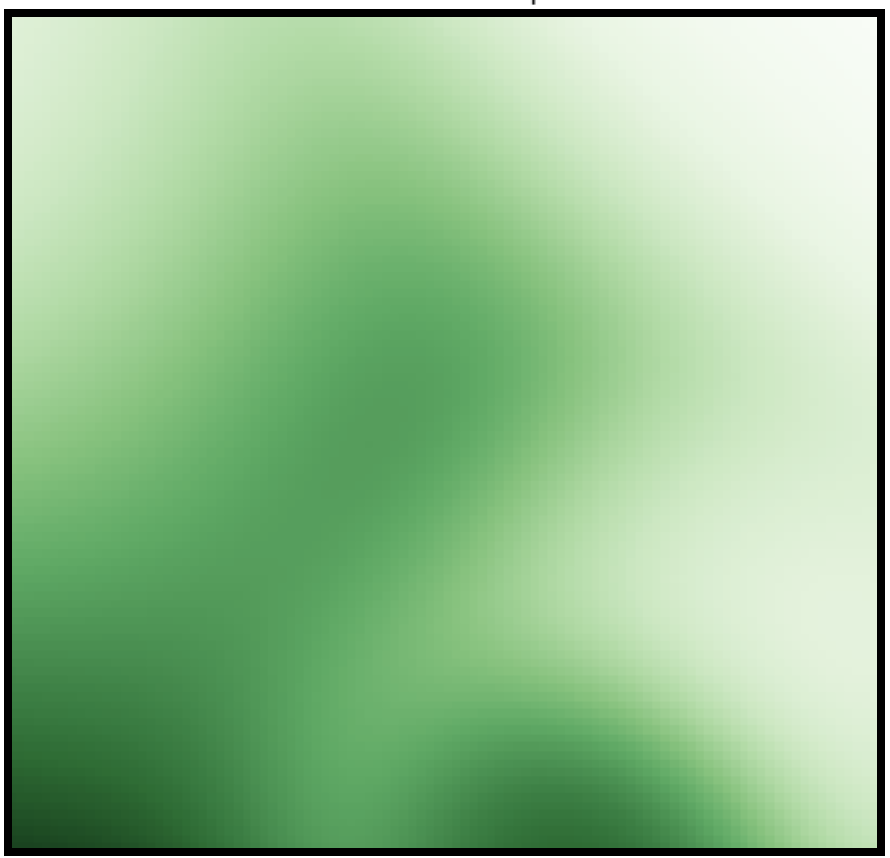}
        }
        \subfigure[Blue density.]{
            \includegraphics[width=0.31\textwidth]{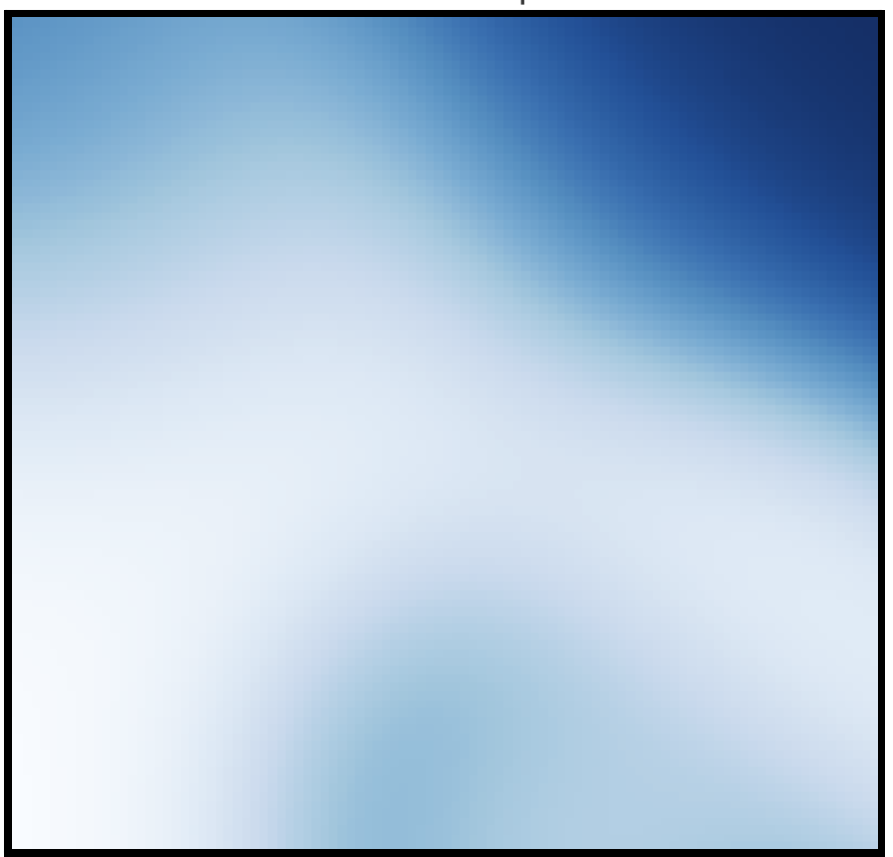}
        }
        \caption{Illustrative example with three subpopulations.}\label{fig:example1}
\end{figure}

Figure \ref{fig:example1} shows an instance where the geographical region $\mathcal{X} = \left[0,1\right]^2$ contains three subpopulations. The population densities $f_z$, for $z \in\left\{1,2,3\right\}$, are shown separately. We see that the ``red'' group is mainly concentrated in the bottom-right portion of the unit square, with some presence near the left edge; the ``green'' group mainly resides along the bottom edge, with some presence in the middle; and the ``blue'' group is heavily concentrated in the top-right. Each subpopulation has some presence in every part of the map, but overall the region is quite segregated.

\begin{figure}[b]
        \centering
        \subfigure[Voronoi partition.]{
            \includegraphics[width=0.47\textwidth]{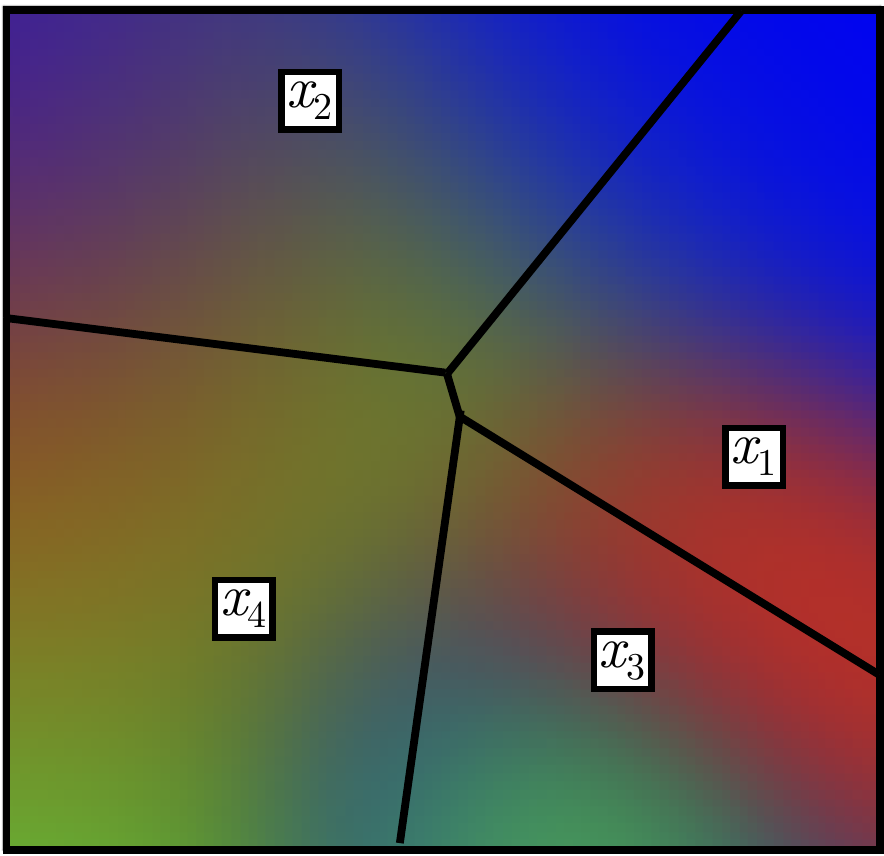}\label{fig:exampleunfair}
        }
        \subfigure[Fair partition.]{
            \includegraphics[width=0.47\textwidth]{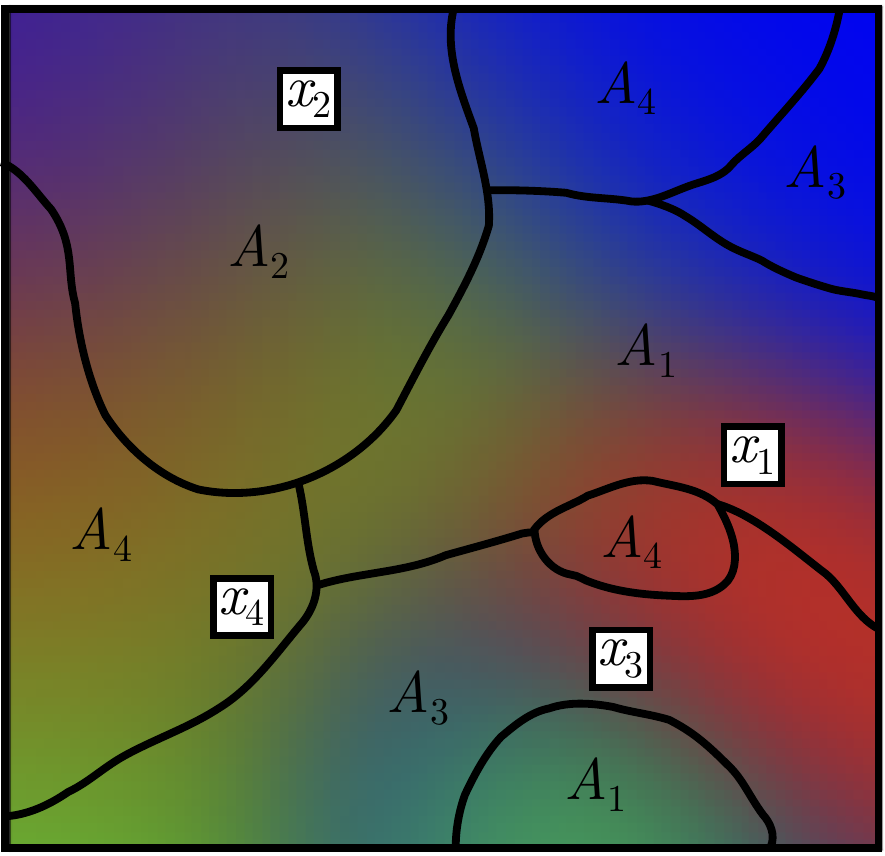}\label{fig:examplefair}
        }
        \caption{Illustrative example with two partitions.}\label{fig:example2}
\end{figure}

In Figure \ref{fig:example2}, we superimpose the three densities onto the same plots and consider the partitioning problem with four facilities represented by white squares. Figure \ref{fig:exampleunfair} shows a partition in which every individual is assigned to the closest facility, i.e., the facility with the smallest Euclidean distance from their location. This type of partition is very well-known in the literature under the name ``Voronoi diagram.'' Individual fairness is clearly satisfied, because assignment is a deterministic function of location and has no additional dependence on demographic. However, group fairness is not satisfied: for example, the blue group is over-represented at facility 1 and under-represented at facility 4.

On the other hand, Figure \ref{fig:examplefair} shows the \textit{optimal fair partition} for this instance. The objective is still to minimize expected travel distance (measured using the Euclidean norm), but we now impose the group fairness constraints (\ref{eq:fair}). The region served by facility $z$ is now labeled. We see that the regions have become non-contiguous: for example, the top-right corner is now divided between facilities $1$, $3$, and $4$, as the latter two have no other way to achieve representation for the blue group. Such geometry is likely undesirable in practice, and, in our case study in Section \ref{sec:la}, we discuss some ways to reduce these effects by changing the cost function $c$. However, if the populations are already very heavily segregated, even an optimal partition will unavoidably have disconnected regions. The most that can be guaranteed is that the enclaves and exclaves will be chosen in a way that minimizes the overall cost at the population level.

If we were to adopt the fair partition in this example, some individuals from the blue group would experience a significant increase in travel cost. At the same time, if the blue group were disadvantaged to begin with, the more distant facilities might be preferable in other ways (e.g., because those schools have more resources). Furthermore, whatever the resulting travel cost might be, the partition in Figure \ref{fig:examplefair} is optimal, and any attempt to make the borders ``nicer'' would inevitably make the objective value even worse. One could also make the argument (though this issue is outside the scope of this paper) that group fairness will eventually change the population densities themselves in a manner that reduces segregation. Lastly, a comparison of the two partitions in Figure \ref{fig:example2} could allow policymakers to quantify the cost of eliminating segregation, regardless of whether the fair districting plan is implemented. In other words, the study of (\ref{eq:partition})-(\ref{eq:fair}) can be informative and insightful even when not interpreted as a direct ``call to action.''

\section{Optimal transport formulation and solution}\label{sec:ot}

To solve (\ref{eq:partition})-(\ref{eq:fair}), we relax the problem to allow probabilistic assignments, analogously to the well-known Kantorovich formulation of semidiscrete optimal transport \citep{Vi21}. Let $Y$ be a discrete random variable taking values in $\left\{1,...,K\right\}$. The quantity $g\left(x,k\right) = P\left(Y=k\mid X=x\right)$ represents the conditional probability of assigning an individual to facility $k$, given that they appear at location $x$. In the original formulation, this probability would be $1$ if $x \in \mathcal{A}_k$ and $0$ otherwise, but this may no longer be the case in the relaxation. We will eventually show, however, that the \textit{optimal} solution of the relaxed problem is also feasible for (\ref{eq:partition})-(\ref{eq:fair}). This result is similar to the well-known equivalence of the Monge and Kantorovich formulations of optimal transport \citep{Pr07}, with the difference that neither of those formulations contains fairness constraints.

In Section \ref{sec:semidiscrete}, we formulate and solve the relaxed problem under the additional assumption that $P\left(Y = k\right) = p_k$ for some pre-specified vector $p$ of probabilities. This is like modifying (\ref{eq:fair}) to require $P\left(X \in \mathcal{A}_k\mid Z\right) = p_k$. Then, in Section \ref{sec:optimize}, we show how to optimize over $p$. We break the problem down into these two parts for ease of understanding, and also because the case of fixed $p$ may be of stand-alone interest in itself (for example, if the $k$th facility only has enough capacity to handle a proportion $p_k$ of the population).

\subsection{Solution with fixed region sizes}\label{sec:semidiscrete}

We formulate the problem
\begin{equation}\label{eq:kantorovich}
\min_g \sum_k \iint_{\mathcal{X}} \sum_z q_z c\left(x,k\right)g\left(x,k\right)f_z\left(x\right)dx
\end{equation}
subject to $g \geq 0$ and
\begin{eqnarray}
\sum_k g\left(x,k\right) &=& 1, \quad x\in\mathcal{X},\label{eq:sumto1}\\
\iint_{\mathcal{X}} g\left(x,k\right) f_z\left(x\right) dx &=& p_k, \quad k = 1,...,K, \quad z = 1,...,M.\label{eq:marginalk}
\end{eqnarray}
The objective function in (\ref{eq:kantorovich}) is simply $\mathbb{E}\left(c\left(X,Y\right)\right)$ written out by conditioning on $Z$. This is the same objective as (\ref{eq:partition}), with the sole difference being that the assignment $Y$ is no longer required to be a deterministic function of $X$. Constraints (\ref{eq:sumto1}) require every location to be assigned to a facility, analogously to (\ref{eq:union}). Constraints (\ref{eq:marginalk}) force $Y$ to be probabilistically independent of $Z$, similarly to (\ref{eq:fair}), but also impose the fixed pmf $p$ on the marginal distribution of $Y$. Individual fairness is ensured by the fact that, for fixed $x$ and $k$, the decision $g\left(x,k\right)$ has no dependence on $Z$.

Problem (\ref{eq:kantorovich})-(\ref{eq:marginalk}) resembles semidiscrete optimal transport \citep{HaSc20,ZhRy24b}. The difference is that, in the classical semidiscrete formulation, the index $z$ is not present at all, and we can then simply optimize over the joint density of $\left(X,Y\right)$ rather than the conditional pmf of $Y$ given $X$. The constraints in semidiscrete optimal transport fix the marginal distributions of both $X$ and $Y$, without the additional complication of fairness.

These differences notwithstanding, problem (\ref{eq:kantorovich})-(\ref{eq:marginalk}) has one aspect in common with semidiscrete OT, namely, that it is an infinite-dimensional linear program. We may therefore derive its Kantorovich dual using the following result.


\begin{thm}\label{thm:duality}
The Kantorovich dual of problem (\ref{eq:kantorovich})-(\ref{eq:marginalk}) is given by
\begin{equation}\label{eq:dualobj}
\sup_{\phi,\psi_z} \iint_{\mathcal{X}} \phi\left(x\right)dx + \sum_z \sum_k p_k \psi_z\left(k\right)
\end{equation}
subject to the constraints
\begin{equation}\label{eq:dualcons}
\phi\left(x\right) + \sum_z f_z\left(x\right) \psi_z\left(k\right) \leq c\left(x,k\right)\left(\sum_z q_z f_z\left(x\right)\right), \quad x\in\mathcal{X}, \quad k = 1,...,K.
\end{equation}
\end{thm}

The proof of Theorem \ref{thm:duality} is given in the Appendix. The dual variables $\phi\left(x\right)$ and $\psi_z\left(k\right)$ correspond to the primal constraints (\ref{eq:sumto1}) and (\ref{eq:marginalk}), respectively. Note that (\ref{eq:dualobj})-(\ref{eq:dualcons}) is also an infinite-dimensional linear program, with $\phi$ being a functional variable. However, as we will now show, it is possible to reformulate the problem as a nonlinear, but finite-dimensional concave program.

We may optimize $\phi$ by taking
\begin{eqnarray*}
\phi\left(x\right) &=& \min_k\left[ c\left(x,k\right)\left(\sum_z q_z f_z\left(x\right)\right) - \sum_z f_z\left(x\right) \psi_z\left(k\right)\right]\\
&=& \left(\sum_z q_z f_z\left(x\right)\right)\min_k\left[ c\left(x,k\right) - \sum_z \frac{f_z\left(x\right)}{\left(\sum_{z'} q_{z'} f_{z'}\left(x\right)\right)} \psi_z\left(k\right)\right].
\end{eqnarray*}
Letting $w_{k,z} = \frac{\psi_z\left(k\right)}{q_z}$, we may rewrite (\ref{eq:dualobj})-(\ref{eq:dualcons}) as the unconstrained (and finite-dimensional) optimization problem
\begin{equation}\label{eq:reformulation1}
\max_{w_{k,z}} \iint_{\mathcal{X}} \left(\sum_z q_z f_z\left(x\right)\right)\min_k\left[ c\left(x,k\right) - \sum_z \frac{q_z f_z\left(x\right)}{\left(\sum_{z'} q_{z'} f_{z'}\left(x\right)\right)} w_{k,z}\right]dx + \sum_z\sum_k p_k q_z w_{k,z}.
\end{equation}
Observe, furthermore, that
\begin{equation*}
\sum_z q_z f_z\left(x\right) = \sum_z P\left(X\in dx\mid Z=z\right)P\left(Z=z\right) = P\left(X\in dx\right),
\end{equation*}
whereas
\begin{equation*}
\frac{q_z f_z\left(x\right)}{\sum_{z'} q_{z'} f_{z'}\left(x\right)} = \frac{P\left(X\in dx\mid Z=z\right)P\left(Z=z\right)}{P\left(X\in dx\right)} = P\left(Z=z\mid X=x\right).
\end{equation*}
Consequently, (\ref{eq:reformulation1}) may be further rewritten as
\begin{equation}\label{eq:reformulation2}
\max_{w_{k,z}} \mathbb{E}\left\{\min_k\left[c\left(X,k\right) - \mathbb{E}\left(w_{k,Z}\mid X\right)\right]\right\} + \sum_z \sum_k p_k q_z w_{k,z}.
\end{equation}
Before proceeding further, let us contrast (\ref{eq:reformulation2}) with the well-known dual reformulation of semidiscrete OT \citep{GeCuPeBa16}. In that problem, $Z$ is not present (the population is homogeneous), so the primal LP is essentially a special case of (\ref{eq:kantorovich})-(\ref{eq:marginalk}) with $M = 1$. In that case, (\ref{eq:reformulation2}) reduces to
\begin{equation}\label{eq:nofairness}
\max_{w_k} \mathbb{E}\left(\min_k c\left(X,k\right) - w_k\right) + p^\top w,
\end{equation}
that is, for each facility $k$ there is a single, fixed weight $w_k$. In the presence of group fairness constraints, however, the $k$th facility now has multiple weights $w_{k,z}$ corresponding to different subgroups, and the objective function calculates a mixture of these weights according to the conditional distribution of $Z$ given $X$. However, since the conditional expectation is a linear function of the weights, the function
\begin{equation*}
F\left(w,x\right) = \min_k\left[c\left(x,k\right) - \mathbb{E}\left(w_{k,Z}\mid X =x\right)\right],
\end{equation*}
viewed as a function of $w$, is a minimum of a finite number of linear functions and therefore concave. Since expectations preserve concavity, (\ref{eq:reformulation2}) is an unconstrained concave maximization problem, and therefore tractable in principle.

We can further characterize the optimal solution $w^*$ of (\ref{eq:reformulation2}). Since $\mathcal{X}$ is closed and bounded, we may differentiate under the integral sign and obtain
\begin{equation*}
\nabla_w \mathbb{E}\left[F\left(w,X\right)\right] = \mathbb{E}\left[\nabla_w F\left(w,X\right)\right].
\end{equation*}
For fixed $x$ and $k,z$,
\begin{equation*}
\frac{\partial}{\partial w_{k,z}} F\left(w,x\right) = -P\left(Z = z\mid X = x\right)\cdot 1_{\left\{k = \arg\min_j c\left(x,j\right) - \mathbb{E}\left(w_{j,Z}\mid X =x\right)\right\}}.
\end{equation*}
Therefore,
\begin{eqnarray}
\frac{\partial}{\partial w_{k,z}} \mathbb{E}\left[F\left(w,X\right)\right] &=& -\mathbb{E}\left[P\left(Z = z\mid X\right)\cdot 1_{\left\{k = \arg\min_j c\left(X,j\right) - \mathbb{E}\left(w_{j,Z}\mid X\right)\right\}}\right]\nonumber\\
&=& -\mathbb{E}\left[1_{\left\{Z=z\right\}}\cdot 1_{\left\{k = \arg\min_j c\left(X,j\right) - \mathbb{E}\left(w_{j,Z}\mid X\right)\right\}}\right],\label{eq:usecinlar}
\end{eqnarray}
where (\ref{eq:usecinlar}) follows by the projection property of conditional expectations (\citealp{Ci11}, ch. IV, sec. 1, eq. (1.4)). Therefore, $w^*$ solves the system
\begin{equation}\label{eq:root}
\mathbb{E}\left[1_{\left\{Z=z\right\}}\cdot 1_{\left\{k = \arg\min_j c\left(X,j\right) - \mathbb{E}\left(w_{j,Z}\mid X\right)\right\}}\right] = p_k q_z, \quad k = 1,...,K, \quad z = 1,...,M.
\end{equation}

From (\ref{eq:reformulation2}), we can also characterize the optimal primal solution as a function of the optimal dual variables. The following result shows how to construct $Y$ to optimize the relaxation (\ref{eq:kantorovich})-(\ref{eq:marginalk}).

\begin{prop}\label{prop:optsolution}
Let $w^*$ denote the optimal solution of (\ref{eq:reformulation2}), and define
\begin{equation*}
Y^* = \arg\min_j c\left(X,j\right) - \mathbb{E}\left(w^*_{j,Z}\mid X\right),
\end{equation*}
with ties broken arbitrarily. Then, the joint distribution of $\left(X,Y^*\right)$ optimally solves (\ref{eq:kantorovich})-(\ref{eq:marginalk}).
\end{prop}

\noindent\textbf{Proof:} By definition of $Y^*$, (\ref{eq:root}) is exactly $P\left(Z=z,Y^*=k\right) = p_k P\left(Z=z\right)$, whence $P\left(Y^*=k\mid Z=z\right) = p_k$ for all $z$. This choice of $Y$ is therefore feasible for (\ref{eq:sumto1})-(\ref{eq:marginalk}).

By weak duality,
\begin{equation*}
\mathbb{E}\left\{\min_k\left[c\left(X,k\right) - \mathbb{E}\left(w^*_{k,Z}\mid X\right)\right]\right\} + \sum_z \sum_k q_z p_k w^*_{k,z} \leq \min_g \sum_k \mathbb{E}\left( c\left(X,k\right)1_{\left\{Y=k\right\}}\right) dx,
\end{equation*}
where $g$ is the conditional pmf of $Y$ given $X$, satisfying (\ref{eq:sumto1})-(\ref{eq:marginalk}) as before. We then write
\begin{eqnarray}
\mathbb{E} \sum_k c\left(X,k\right)1_{\left\{Y^*=k\right\}} &=& \mathbb{E} \sum_k \left(c\left(X,k\right)-\mathbb{E}\left(w^*_{k,Z}\mid X\right)\right)1_{\left\{Y^*=k\right\}} + \mathbb{E}\sum_k \mathbb{E}\left(w^*_{k,Z}\mid X\right)1_{\left\{Y^*=k\right\}}\nonumber\\
&=& \mathbb{E}\left\{\min_k\left[c\left(X,k\right) - \mathbb{E}\left(w^*_{k,Z}\mid X\right)\right] \right\} + \sum_k \mathbb{E}\left(w^*_{k,Z} 1_{\left\{Y^*=k\right\}}\right)\label{eq:projagain}\\
&=& \mathbb{E}\left\{\min_k\left[c\left(X,k\right) - \mathbb{E}\left(w^*_{k,Z}\mid X\right)\right] \right\} + \sum_k \mathbb{E}\left(w^*_{k,Z}\right) P\left(Y^*=k\right)\label{eq:useindep}\\
&=& \mathbb{E}\left\{\min_k\left[c\left(X,k\right) - \mathbb{E}\left(w^*_{k,Z}\mid X\right)\right] \right\} + \sum_k \sum_z p_k q_z w^*_{k,z}.\nonumber
\end{eqnarray}
where (\ref{eq:projagain}) again follows by the projection property of conditional expectations (since $Y^*$ is a deterministic function of $X$), whereas (\ref{eq:useindep}) uses the independence of $Z$ and $Y^*$. The desired result follows.\qed

The optimal $Y^*$ from Proposition \ref{prop:optsolution} then induces the partition
\begin{equation}\label{eq:induced}
\mathcal{A}_k = \left\{x \in \mathcal{X}\,:\, k = \arg\min_j c\left(x,j\right) - \mathbb{E}\left(w^*_{j,Z}\mid X = x\right)\right\},
\end{equation}
which is feasible for the original problem (\ref{eq:partition})-(\ref{eq:fair}) under some mild conditions on the distribution of $X$. Specifically, it suffices for the pairwise differences $c\left(X,j\right) - c\left(X,k\right)$ to have a density for any $j \neq k$.

Once again, it is interesting to compare (\ref{eq:induced}) with the partition obtained in the standard semidiscrete OT problem, where group fairness constraints are not present. Recall that this standard problem has the dual objective (\ref{eq:nofairness}), where there is no $Z$ variable and only one weight for each $k$ value. Letting $w^*_k$ be the optimal weights for that problem, (\ref{eq:nofairness}) induces a partition where an individual appearing at location $x$ is assigned to the facility corresponding to $\arg\min_j c\left(x,j\right) - w^*_j$. This geometric structure, where assignments are made by minimizing the sum of travel distance and a facility-specific weight, is precisely the well-known additively weighted Voronoi diagram \citep{Au91}. In our setting, (\ref{eq:reformulation2}) induces the partition (\ref{eq:induced}), which can be viewed as a novel generalization of this concept. The imposition of group fairness constraints leads to a diagram in which each facility has multiple weights, rather than just one, and the assignment uses a mixture of these weights that depends on the location of the individual. In other words, if a subgroup $z$ is more prevalent at location $x$, then the corresponding weights $w^*_{j,z}$ play a greater role in determining the facility serving that location.

\subsection{Solution with optimal region sizes}\label{sec:optimize}

We will now show how the vector $p$ may be eliminated entirely, yielding an optimal solution to (\ref{eq:partition})-(\ref{eq:fair}) under the best possible marginal distribution of $Y$. Let $w^*\left(p\right)$ denote the optimal weights for problem (\ref{eq:reformulation2}) with a fixed $p$, and let
\begin{equation}\label{eq:vstar}
V^*\left(p\right) = \mathbb{E}\left\{\min_k\left[c\left(X,k\right) - \mathbb{E}\left(w^*_{k,Z}\left(p\right)\mid X\right)\right]\right\} + \sum_z \sum_k p_k q_z w^*_{k,z}\left(p\right).
\end{equation}
By the envelope theorem,
\begin{equation*}
\frac{\partial}{\partial p_k} V^*\left(p\right) = \sum_z q_z w^*_{k,z}\left(p\right),
\end{equation*}
because the only direct dependence of (\ref{eq:vstar}) on $p$ occurs in the last term.

Consider the outer optimization problem
\begin{equation*}
\min_p V^*\left(p\right)
\end{equation*}
subject to $p_k \geq 0$ and $\sum_k p_k = 1$. Let $\nu_k$ and $\lambda$ be the Lagrange multipliers of the inequality and equality constraints, respectively. Then, the optimal $p^*,\lambda^*,\nu^*$ satisfy the first-order optimality conditions
\begin{eqnarray}
\sum_z q_z w^*_{k,z}\left(p^*\right) + \lambda^* - \nu^*_k &=& 0, \qquad k = 1,...,K,\label{eq:kkt1}\\
\nu^*_k p^*_k &=& 0, \qquad k = 1,...,K,\label{eq:kkt2}\\
\nu^*_k &\geq& 0, \qquad k = 1,...,K.\label{eq:kkt3}
\end{eqnarray}
Adding a constant $\lambda^*$ to all the weights will not affect the partition, so we may assume $\lambda^* = 0$ without loss of generality. Then, the complementary slackness conditions imply that
\begin{eqnarray*}
\sum_z q_z w^*_{k,z}\left(p^*\right) = 0 \quad &\Leftrightarrow& \quad p^*_k > 0,\\
\sum_z q_z w^*_{k,z}\left(p^*\right) = \nu^*_k \quad &\Leftrightarrow& \quad p^*_k = 0.
\end{eqnarray*}
For $k$ satisfying $p^*_k = 0$, define the modified weights $\bar{w}_{k,z} = w^*_{k,z}\left(p^*\right) - \nu^*_k$. Then, clearly $\sum_z q_z \bar{w}_{k,z} = 0$. To show that the modified weights do not change the partition, it is sufficient to show that the set $\mathcal{A}_k$ induced by $w^*_{k,z}\left(p^*\right)$ according to (\ref{eq:induced}) remains unchanged by the modification. Since $p^*_k = 0$, we have
\begin{equation}\label{eq:contradiction}
c\left(x,k\right) - \mathbb{E}\left(w^*_{k,Z}\left(p^*\right)\mid X=x\right) \geq \min_{j\neq k:p^*_j>0} c\left(x,j\right) - \mathbb{E}\left(w^*_{j,Z}\left(p^*\right)\mid X=x\right), \qquad x\in\mathcal{X},
\end{equation}
that is, for any $x$ there is always one facility that is preferable to $k$. Suppose now that, under the modified weights, (\ref{eq:contradiction}) is violated: that is, there exists $x_0$ such that
\begin{equation*}
c\left(x_0,k\right) - \mathbb{E}\left(\bar{w}_{k,Z}\mid X=x_0\right) < \min_{j\neq k:p^*_j>0} c\left(x_0,j\right) - \mathbb{E}\left(w^*_{j,Z}\left(p^*\right)\mid X=x_0\right).
\end{equation*}
However,
\begin{eqnarray*}
c\left(x_0,k\right) - \mathbb{E}\left(\bar{w}_{k,Z}\mid X=x_0\right) &=& c\left(x_0,k\right) - \mathbb{E}\left(w^*_{k,Z}\left(p^*\right)\mid X=x_0\right) + \nu^*_k\\
&\geq & c\left(x_0,k\right) - \mathbb{E}\left(w^*_{k,Z}\left(p^*\right)\mid X=x_0\right),
\end{eqnarray*}
which leads to a contradiction with (\ref{eq:contradiction}). Thus, the modified weights do not change the original partition; the objective value is also unchanged since it is determined only by the non-empty regions. Therefore, the optimality conditions (\ref{eq:kkt1})-(\ref{eq:kkt3}) of the outer problem can be rewritten more succinctly as
\begin{equation*}
\sum_z q_z w^*_{k,z}\left(p^*\right) = 0, \quad k = 1,...,K,
\end{equation*}
that is, $\mathbb{E}\left(w^*_{k,Z}\right) = 0$ for all $k$.

\section{Characterization and computation of optimal partition}\label{sec:sa}

We may now combine the results of Sections \ref{sec:semidiscrete}-\ref{sec:optimize} and give a complete characterization of the optimal partition. Our goal is to find a solution $\left(w^*,p^*\right)$ of the system of nonlinear equations
\begin{eqnarray}
\mathbb{E}\left[1_{\left\{Z=z\right\}}\cdot 1_{\left\{k = \arg\min_j c\left(X,j\right) - \mathbb{E}\left(w_{j,Z}\mid X\right)\right\}}\right] &=& p_k q_z, \quad k =1,...,K, \quad z = 1,...,M,\label{eq:lagrangian}\\
\sum_z q_z w_{k,z} &=& 0, \quad k = 1,...,K.\label{eq:constraint}
\end{eqnarray}
Note that the left-hand side of (\ref{eq:lagrangian}) is $P\left(Z=z,Y=k\right)$, so this constraint implies $p_k \geq 0$, $\sum_k p_k = 1$ and we may omit those conditions.

It can be readily seen that (\ref{eq:lagrangian})-(\ref{eq:constraint}) are precisely the KKT conditions of the constrained concave program
\begin{equation}\label{eq:nop}
\max_{w_{k,z}} \mathbb{E}\left\{\min_k\left[c\left(X,k\right) - \mathbb{E}\left(w_{k,Z}\mid X\right)\right]\right\}
\end{equation}
subject to the linear equalities (\ref{eq:constraint}). Here, the optimal region sizes $p^*_k$ become the Lagrange multipliers of the equality constraints. This problem trivially satisfies Slater's condition, so (\ref{eq:lagrangian})-(\ref{eq:constraint}) are necessary and sufficient for optimality. Even if there are multiple $\left(w^*,p^*\right)$ satisfying (\ref{eq:lagrangian})-(\ref{eq:constraint}), each solution achieves the same value of (\ref{eq:nop}) as well as $V^*\left(p^*\right)$.


We may eliminate the constraints entirely by using the transformation
\begin{equation}\label{eq:proj}
w_{k,z} = v_{k,z} - \frac{\sum_{z'} q_{z'}v_{k,z'}}{q^\top q}q_z = v_{k,z} - \frac{\mathbb{E}\left(v_{k,Z}\right)}{q^\top q}q_z,
\end{equation}
which projects any arbitrary $\left(v_{k,1},...,v_{k,M}\right)$ onto the orthogonal complement of $q$. This yields the unconstrained optimization problem
\begin{equation}\label{eq:unconstrained}
\max_{v_{k,z}} \mathbb{E}\left\{\min_k\left[c\left(X,k\right) - \mathbb{E}\left(v_{k,Z}\mid X\right)+ \frac{\mathbb{E}\left(q_Z\mid X\right)}{q^\top q}\mathbb{E}\left(v_{k,Z}\right)\right]\right\}.
\end{equation}
Since (\ref{eq:proj}) is a linear function of $v$, the objective function in (\ref{eq:unconstrained}) is still concave. Note that $\mathbb{E}\left(q_Z\mid X\right)$ does not depend on $v$. Therefore, for fixed $k$, $z$, and $x$,
\begin{equation*}
\frac{\partial}{\partial v_{k,z}} \mathbb{E}\left(w_{k,Z}\mid X=x\right) = P\left(Z=z\mid X=x\right) - \frac{\mathbb{E}\left(q_Z\mid X=x\right)}{q^\top q}q_z.
\end{equation*}
Applying the projection property of conditional expectation once more, the optimality conditions of (\ref{eq:unconstrained}) are
\begin{equation}\label{eq:finalopt}
\mathbb{E}\left\{ 1_{\left\{k = \arg\min_j c\left(X,j\right) - \mathbb{E}\left(v_{j,Z}\mid X\right) + \frac{\mathbb{E}\left(q_Z\mid X\right)}{q^\top q}\mathbb{E}\left(v_{k,Z}\right) \right\}}\left(1_{\left\{Z=z\right\}}-\frac{q_Z}{q^\top q}q_z\right) \right\} = 0.
\end{equation}
for $k = 1,...,K$ and $z = 1,...,M$.

The system (\ref{eq:unconstrained}) may be approached as a stochastic root-finding problem \citep{PaKi11}. Because the expectation in (\ref{eq:finalopt}) is difficult to compute, we may instead approximate it using simulation, in a procedure known as stochastic approximation \citep{KuYi03}. Given an approximation $v^n$ of the solution, we observe a sample $\left(X^{n+1},Z^{n+1}\right)$ from the underlying joint distribution, and evaluate the random variable inside the expectation in (\ref{eq:finalopt}) using these values. On average, this random variable will point us toward the root. We then compute $v^{n+1}$ by adjusting $v^n$ in the indicated (noisy) direction, using a stepsize to smooth out noise. Algorithm \ref{fig:sa} formally states the procedure.

\begin{algorithm}[t]
 \mbox{}\hrulefill\mbox{}
 \begin{description}
    \item[Step 0:] Initialize $n =0$ and $v^0_{k,z}$ for $k=1,...,K$, $z = 1,...,M$. Choose a suitable deterministic stepsize sequence $\left\{\alpha_n\right\}^{\infty}_{n=0}$.
    \item[Step 1:] Observe $\left(X^{n+1},Z^{n+1}\right)$ and iterate
    \begin{equation}\label{eq:sa}
    v^{n+1}_{k,z} = v^n_{k,z} + \alpha_n \cdot 1_{\left\{k = \arg\min_j c\left(X^{n+1},j\right) - \mathbb{E}\left(w^n_{j,Z}\mid X = X^{n+1} \right)\right\}}\left(1_{\left\{Z^{n+1}=z\right\}} - \frac{q_{Z^{n+1}}}{q^\top q}q_z\right),
    \end{equation}
    where $w^n$ is computed by plugging $v^n$ into (\ref{eq:proj}).
    \item[Step 2:] Increment $n$ by $1$ and return to Step 1.
 \end{description}
 \mbox{}\hrulefill\mbox{}
 \vspace{0.1in}
 \caption{Stochastic approximation procedure for solving (\ref{eq:unconstrained}).}\label{fig:sa}
\end{algorithm}

Since the random variable in (\ref{eq:sa}) is bounded, the convergence of $\left\{v^n\right\}$ to an optimum of (\ref{eq:unconstrained}) easily follows by classical SA theory. It is possible to characterize the convergence rate in greater depth. The following theorem (proved in the Appendix) derives a bound that matches the near-optimal rates in \cite{BaMo11}, but uses weaker conditions. More precisely, Thms. 4-7 in \cite{BaMo11} require the stochastic gradient to be Lipschitz continuous, which does not hold in our setting because (\ref{eq:sa}) uses indicator functions. We drop this requirement, but obtain the same guarantee. It is worth noting that the rate bound is obtained, not on the sequence $\left\{v^n\right\}$ directly, but on the \textit{average} of these iterates. This technique, known as Polyak averaging, has been known to speed up convergence since \cite{PoJu92}. From an implementation perspective, it is quite simple to incorporate averaging into Algorithm \ref{fig:sa}.

\begin{thm}\label{thm:sa}
Let $v^0 \in \mathbb{R}^\ell$ for arbitrary dimension $\ell$. Let $h:\mathbb{R}^\ell \rightarrow \mathbb{R}$ be a concave, differentiable function with $v^* \in \arg\max_v h\left(v\right)$. Let $\left\{\zeta^n\right\}^{\infty}_{n=1}$ be a sequence of random functions mapping $\mathbb{R}^\ell$ into $\mathbb{R}^\ell$ satisfying $\mathbb{E}\left(\zeta^n\left(v\right)\right) = \nabla_v h\left(v\right)$ and $P\left(\left|\zeta^n\left(v\right)\right| \leq C\right) = 1$ for some constant $C > 0$. Define
\begin{equation*}
v^{n+1} = v_n + \alpha_n \zeta^n\left(v^n\right), \qquad n \geq 0,
\end{equation*}
with $\alpha_n = \frac{\alpha}{\sqrt{n+1}}$, and let $\bar{v}^n = \frac{1}{n}\sum^n_{m=1} v^m$ be the average of the iterates.

Then, there exists a constant $D>0$, depending only on $\alpha$, $C$, and $\|v_0\|_2$, such that
\begin{equation*}
h\left(v^*\right) - \mathbb{E}\left(h\left(\bar{v}^n\right)\right) \leq D\cdot \frac{\log n}{\sqrt{n}}.
\end{equation*}
\end{thm}

\section{Case study: Districting in LA County}\label{sec:la}

In this section, we demonstrate the practicability of our approach on real population and demographic data. Our setting is the San Fernando Valley (SFV) region of Los Angeles County, a diverse area with significant demographic variation across neighborhoods. Some areas show high concentrations of specific demographic groups, a pattern typical of urban segregation documented in the literature \citep{ReBi11}.

The census distinguishes between seven\footnote[1]{Unfortunately, ``Hispanic'' is not among the available options in the census, and therefore we were not able to include it as a distinct category in this study.} racial/ethnic categories: Multiracial ($13.43\%$ of the total population in the region), White ($45.39\%$), Black ($3.86\%$), American Indian ($1.30\%$), Asian ($11.41\%$), Pacific Islander ($0.11\%$), and Other ($24.50\%$). In Section \ref{sec:laseven}, we apply our modeling and algorithmic framework to construct a partition that achieves fair representation for all seven categories, to demonstrate the capabilities of our approach. In Section \ref{sec:lathree}, we repeat the same analysis with three groups: Black, Asian, and all the others combined into one. Our intention here is simply to demonstrate how the number of groups influences the costs for the population as a whole. For example, a comparison between the two versions of the case study will show that, when we require fair representation for all seven groups, costs increase for everyone across the board, including the disadvantaged populations whom the fair partition aims to benefit. This should be an important consideration for policymakers when deliberating the issue of fair representation.

\subsection{Analysis for seven subgroups}\label{sec:laseven}

We obtained latitude and longitude coordinates for school district administrative offices from the National Center for Education Statistics Education Demographic and Geographic Estimates Program geocoded data files \citep{nces}. From this dataset, we selected $78$ administrative offices located within the SFV boundary using a San Fernando Valley shapefile obtained from ArcGIS Online \cite{esri}. We also incorporated detailed census tract population data from the American Community Survey for all seven demographic groups.

The cost $c\left(x,k\right)$ of traveling to district office $k$ for a household located at coordinates $x$ was computed using the Open Source Routing Machine (OSRM) with real road network data, providing realistic driving times and distances rather than simply Euclidean norms. This approach captures the true cost of travel in an urban environment with complex street networks and traffic patterns. Using real costs does not create any additional challenges for our solution technique since Algorithm \ref{fig:sa} only requires the ability to evaluate $c\left(X,k\right)$ for a realization $X$ from the population. We ran our algorithm for two variants of the optimization problem: one where $c\left(x,k\right)$ was set equal to the real travel distance (which we also refer to as ``regular distance''), and another one where it was set equal to the \textit{squared} travel distance. In computational geometry, partitions based on squared distance are known as ``power diagrams'' and tend to create more compact and balanced regions, because they heavily penalize extreme deviations from district centers \citep{CoKlYo18}.

Both variants include group fairness constraints. As a baseline, we also compute a standard Voronoi diagram where facility $k$ serves the set $\left\{x : k = \arg\min_j c\left(x,j\right)\right\}$, that is, each household is assigned to the closest district office. This ``unconstrained'' partition will clearly not achieve demographic parity. Of course, in practice, the determination of school district boundaries involves other factors aside from simply travel distance. Nonetheless, Voronoi diagrams have been previously studied for this purpose \citep{Pe00,GoAr20}, and because our focus here is on evaluating the price of fairness, they provide a natural basis for comparison. In these experiments, we do not fix the capacity of each facility (that is, we optimize over $p_k$), but if one wished to fix them in advance, this could be done using the techniques of Section \ref{sec:semidiscrete}.


\begin{table}[t]
\begin{centering}
\begin{tabular}{|c||c|c|c||c|c|c|}
\hline
Race & Median (u) & Median (c) & Median (c-sq) & 90th \% (u) & 90th \% (c) & 90th \% (c-sq)\tabularnewline
\hline
\hline
Multiracial & 2034 & 3137 & 5394 & 3739 & 13536 & 10696\tabularnewline
\hline
White & 2217 & 3439 & 5955 & 4709 & 14578 & 12327\tabularnewline
\hline
Black & 2154 & 3439 & 5958 & 3739 & 18353 & 12510\tabularnewline
\hline
American Indian & 1853 & 2860 & 5363 & 3487 & 13536 & 10045\tabularnewline
\hline
Asian & 2213 & 3562 & 6058 & 4104 & 14319 & 12116\tabularnewline
\hline
Pacific Islander & 2068 & 4185 & 6104 & 3901 & 14590 & 11863\tabularnewline
\hline
Other & 1843 & 2991 & 5540 & 3455 & 14037 & 10045\tabularnewline
\hline
\end{tabular}
\par\end{centering}
\caption{\label{tab:Travel-distances-and}Travel distances (in meters) for unconstrained (u), equity-constrained (c), and equity-constrained with squared distance (c-sq) models, categorized by race.}
\end{table}

\begin{figure}[t]
\centering
\includegraphics[width=0.99\textwidth]{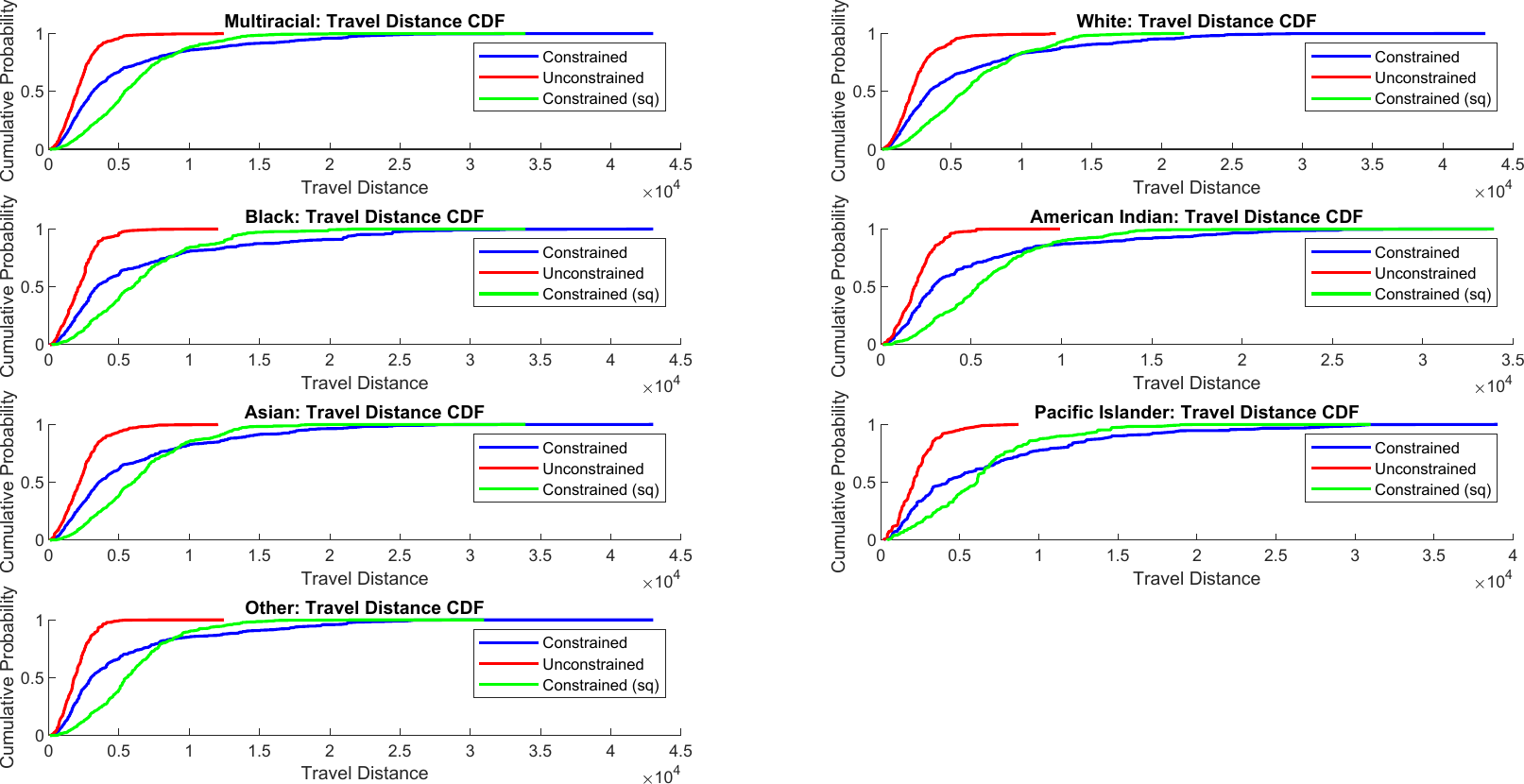}\label{fig:cdfs1}
\caption{CDFs for travel distances for each demographic.}
\end{figure}

Table \ref{tab:Travel-distances-and} shows the mean and 90th percentile of the travel distance for each demographic group under each of the three models (for the model that optimizes squared cost, we still report statistics for the unsquared distance). Figure \ref{fig:cdfs1} further shows the empirical cumulative distribution functions of the travel distance for each demographic group/model combination. We see that the price of fairness is quite substantial, with all seven demographic groups experiencing significant increases in travel time relative to the unconstrained model. Optimizing the squared distance reduces outliers on both the low and high ends: the median travel time increases for each group relative to the standard equity-constrained model, with the benefit of reducing extremely high times.

\begin{figure}[H]
	\centering
	\subfigure[Facility 12 (travel distance).]{
		\includegraphics[width=0.33\textwidth]{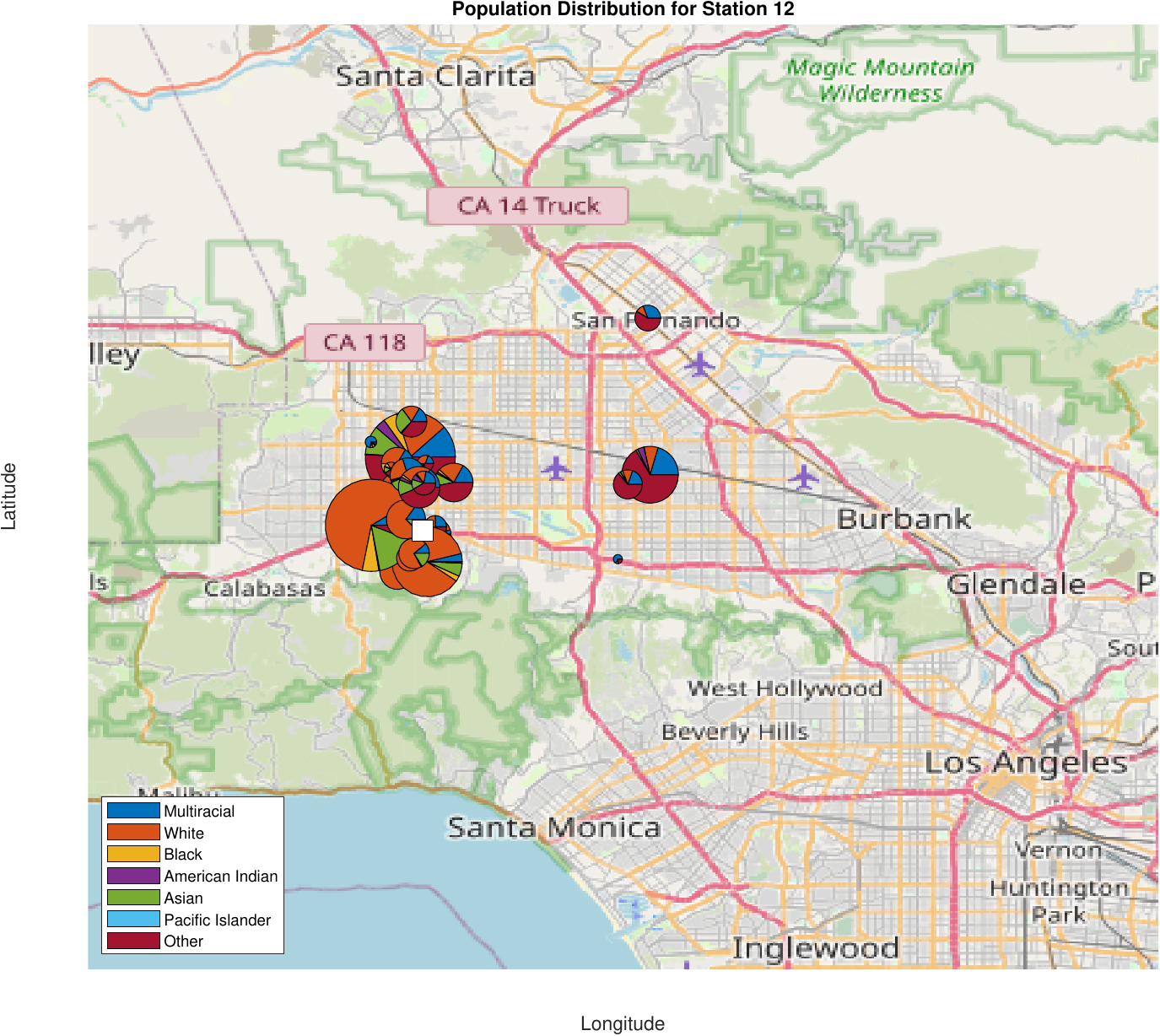}
		\label{fig:3r}
	}
	\subfigure[Facility 12 (squared distance).]{
		\includegraphics[width=0.33\textwidth]{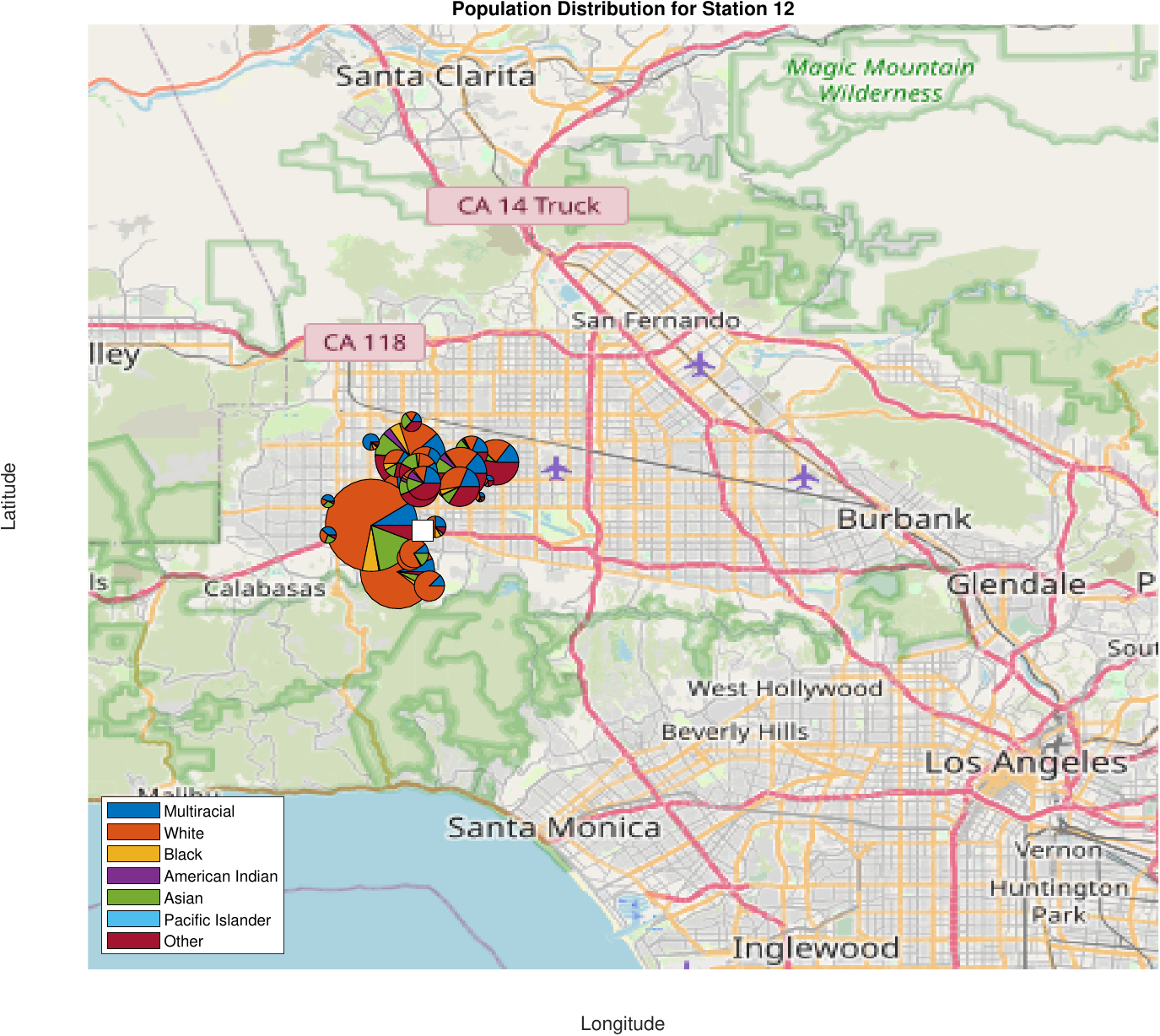}
		\label{fig:3s}
	}
	\subfigure[Facility 34 (travel distance).]{
		\includegraphics[width=0.33\textwidth]{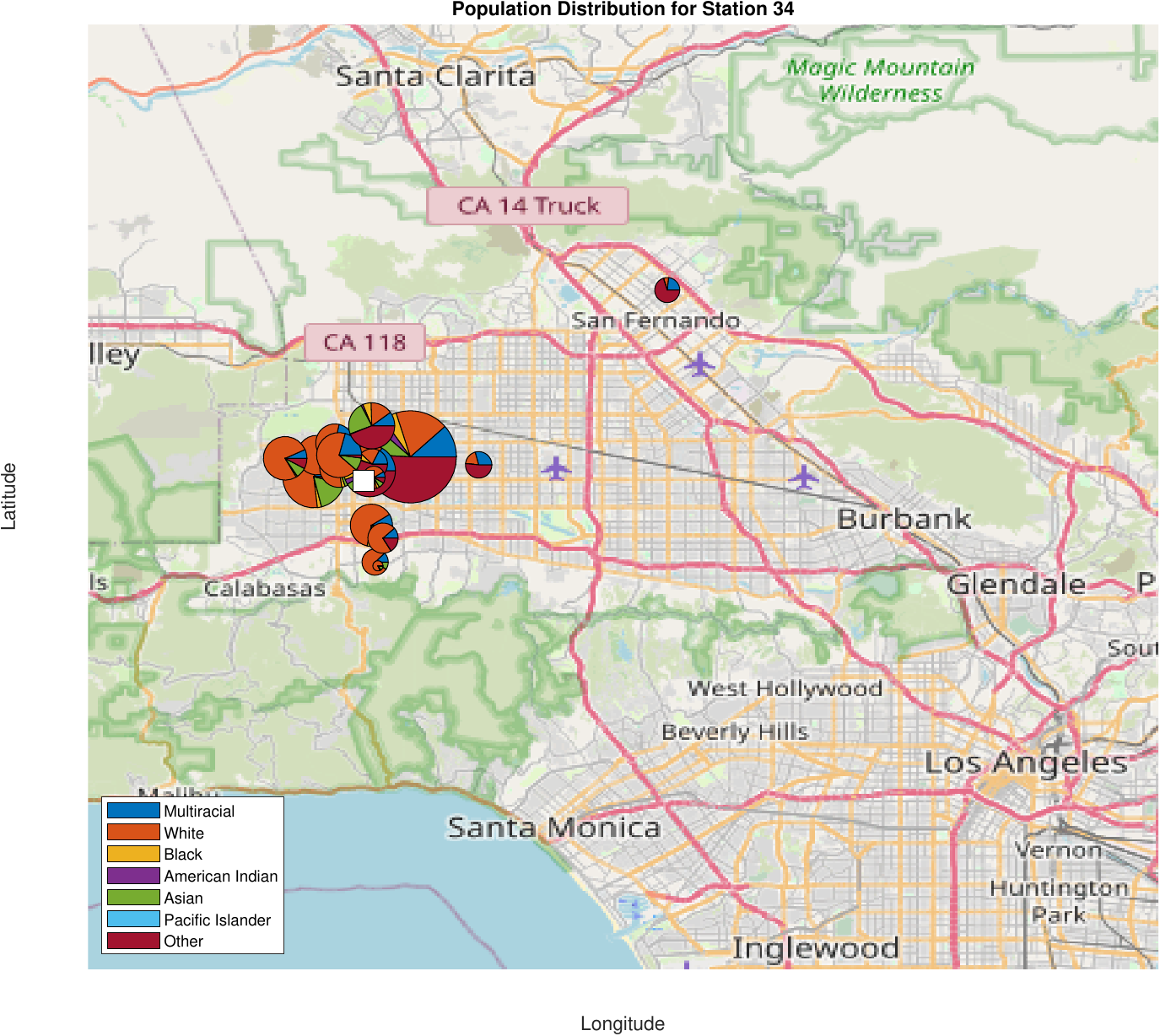}
		\label{fig:12r}
	}
	\subfigure[Facility 34 (squared distance).]{
		\includegraphics[width=0.33\textwidth]{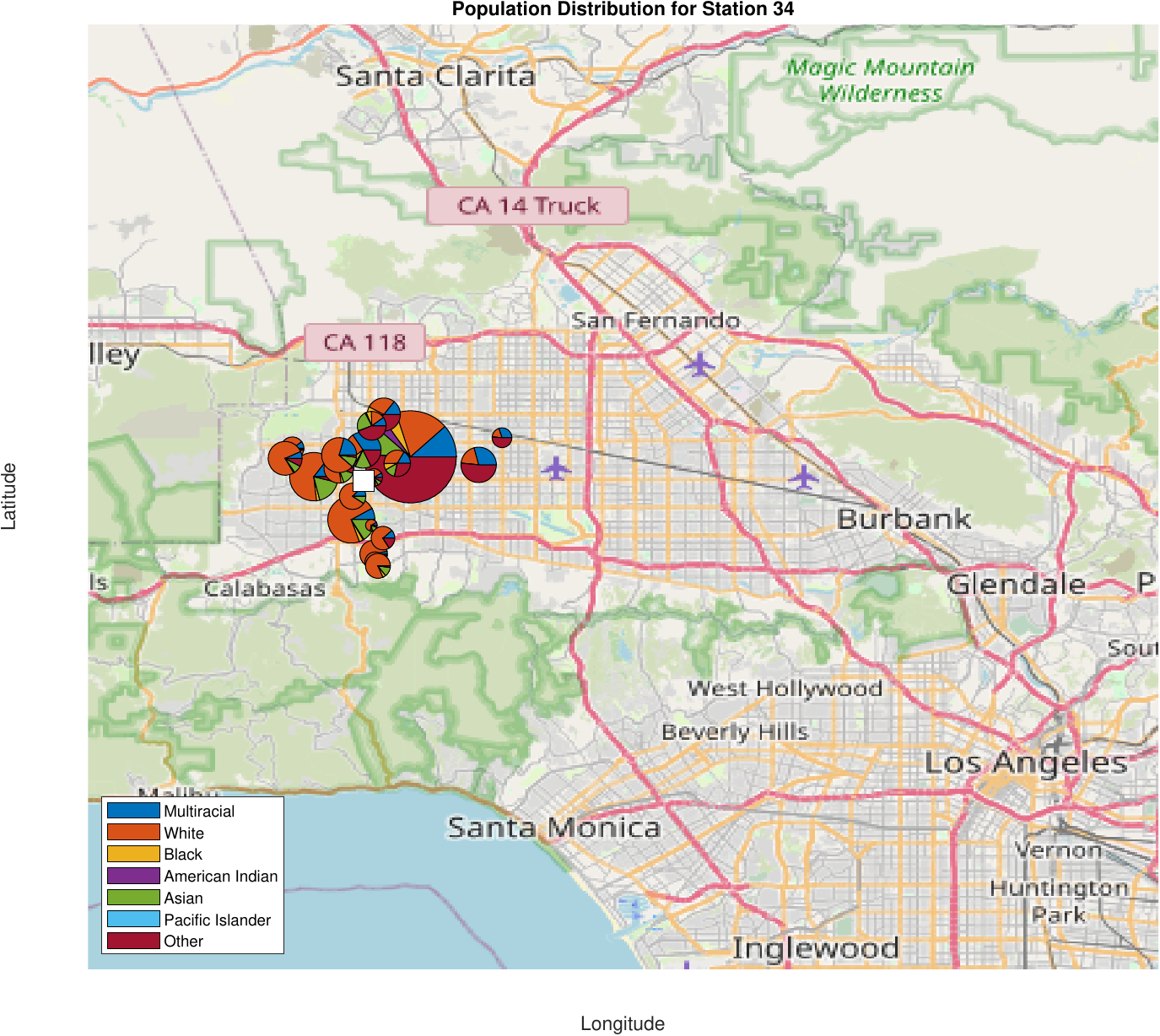}
		\label{fig:12s}
	}
	\subfigure[Facility 63 (travel distance).]{
		\includegraphics[width=0.33\textwidth]{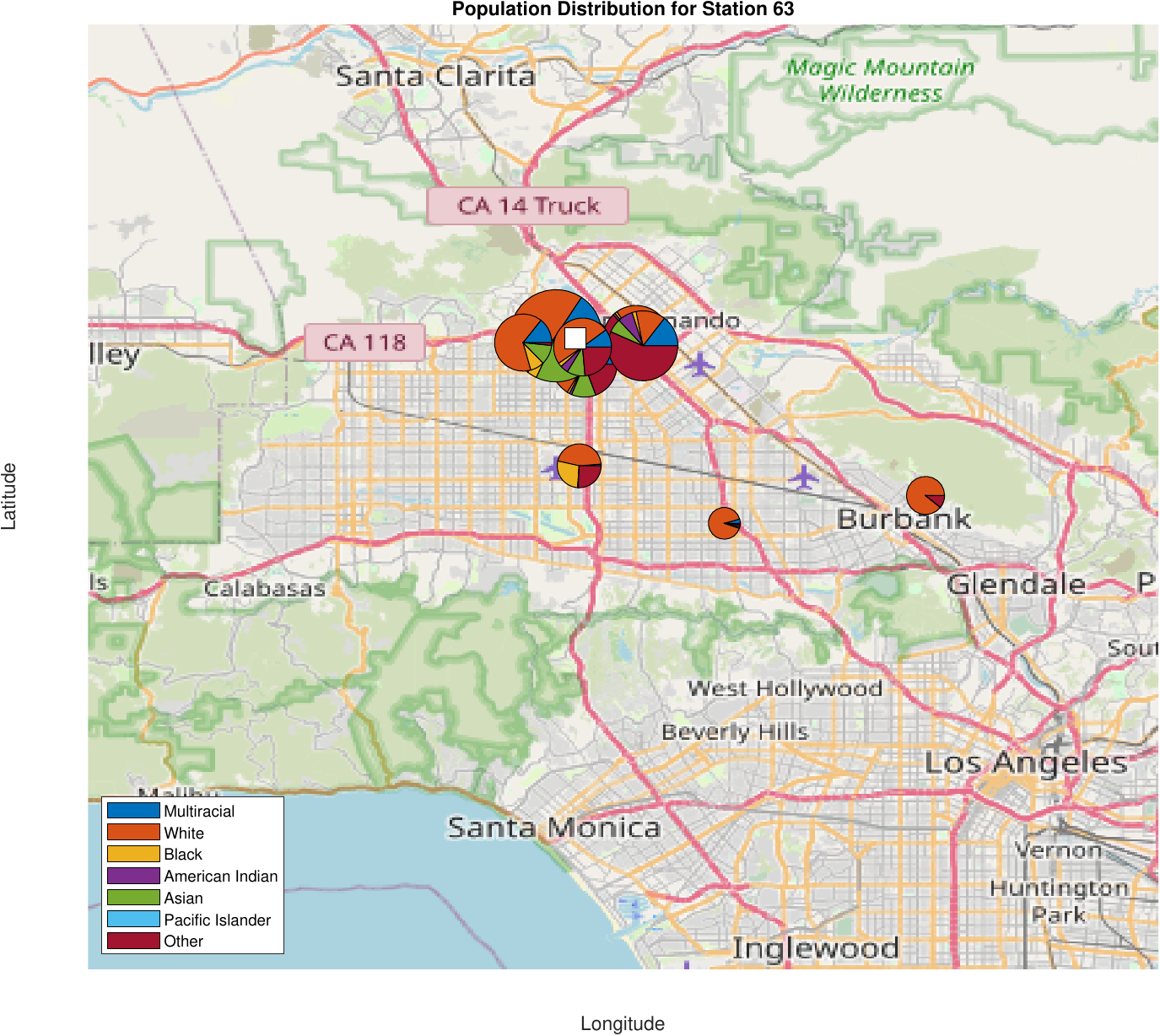}
		\label{fig:27r}
	}
	\subfigure[Facility 63 (squared distance).]{
		\includegraphics[width=0.33\textwidth]{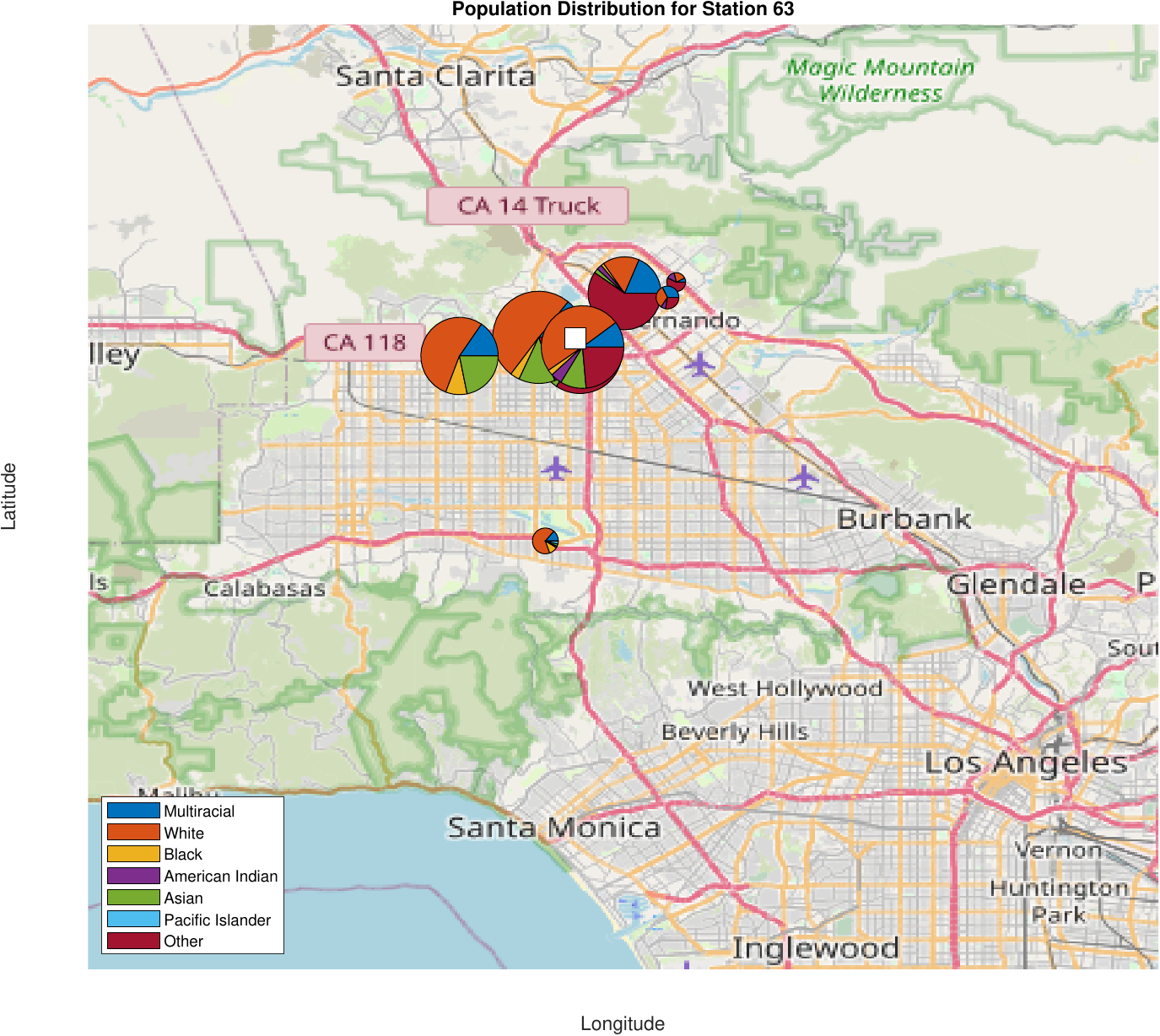}
		\label{fig:27s}
	}
	\subfigure[Facility 76 (travel distance).]{
		\includegraphics[width=0.33\textwidth]{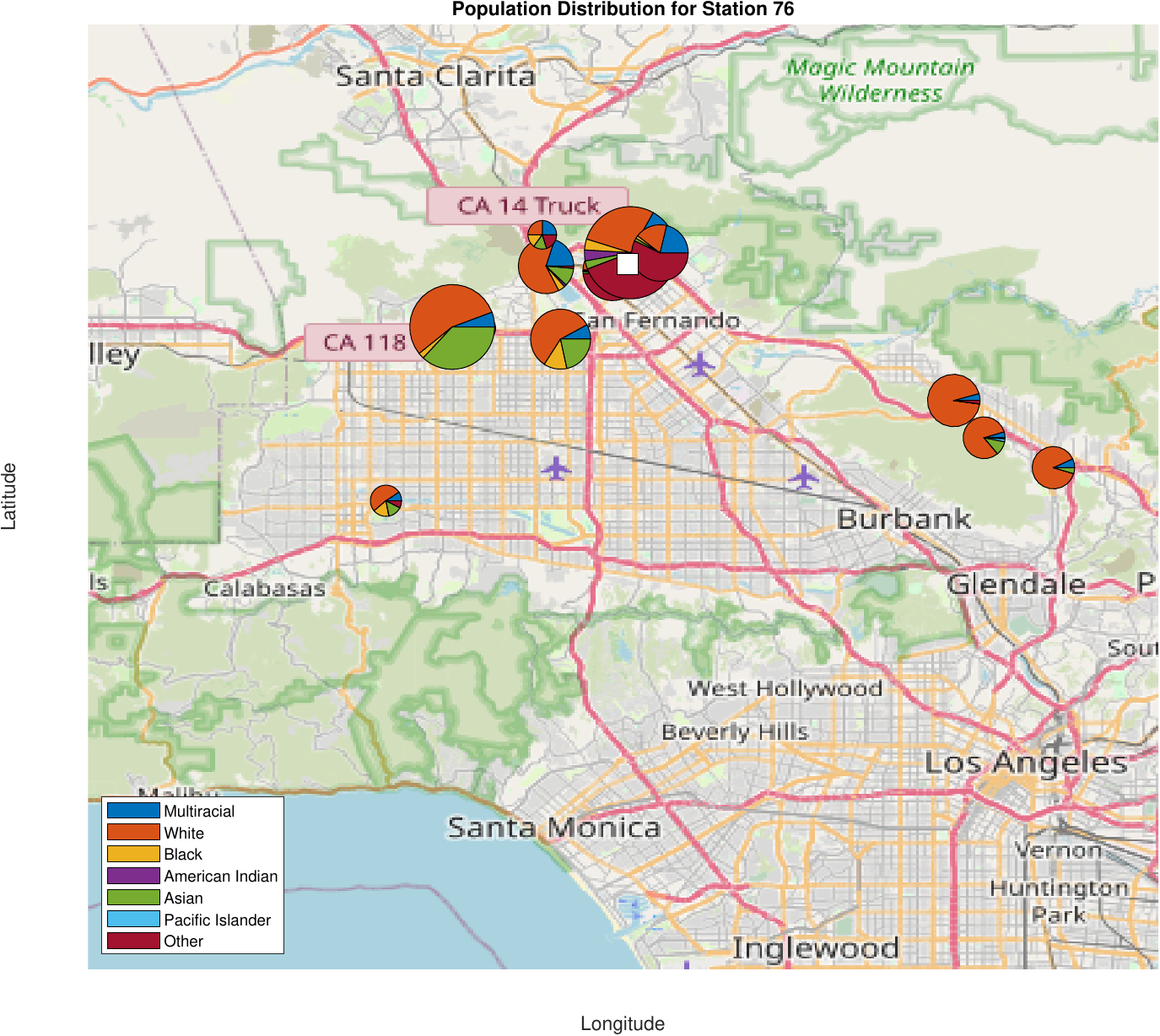}
		\label{fig:32r}
	}
	\subfigure[Facility 76 (squared distance).]{
		\includegraphics[width=0.33\textwidth]{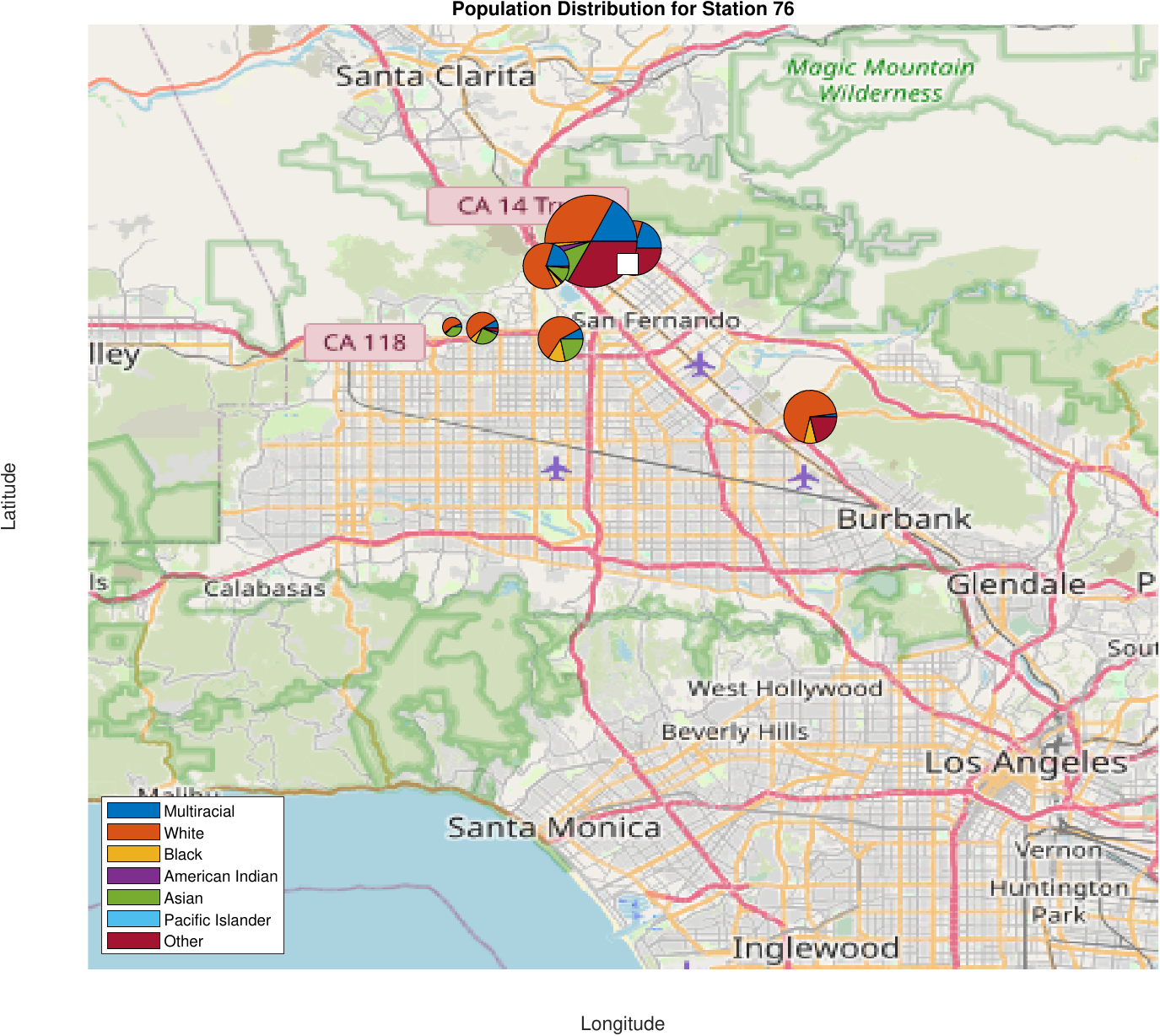}
		\label{fig:32s}
	}
\caption{Four representative districts, drawn using two different cost functions (regular and squared distance).}\label{fig:facilities}
\end{figure}

Additional insights can be obtained by examining individual district boundaries. Figure \ref{fig:facilities} shows four representative districts under both cost functions (a complete set of plots for all $78$ facilities is included in the Appendix). The district offices are represented by white squares, and the census tracts assigned to the office in question are shown as pie charts whose size reflects the local population. Thus, for example, Facility 12 is located in the center-west of the map (along Route 101, east of Calabasas). Under regular distance, several census tracts from the other side of I-405 are assigned to this office, leading to high travel times for those households; under squared distance, these assignments are no longer made. Similar behavior can be observed for Facilities 34, 63, and 76: in all three cases, switching to squared distance eliminates the worst outliers, though it may not eliminate discontinuities completely. The optimal district sizes (that is, the optimal choices of $p_k$) need not the same for both cost functions. We will revisit these four examples in Section \ref{sec:lathree} when the entire case study is rerun for a reduced set of demographic groups.

It is also noteworthy that fair partitioning may result in the closure of school administrative offices. This does not happen for the unconstrained model, since for any facility, there is always some subset of the population that is closest to it. However, under equity constraints and regular cost, $7$ out of $78$ district offices receive no student assignments, and this number increases to $12$ when we optimize the squared distance. Five offices are closed under both models. These situations can arise when a particular demographic group is hyper-concentrated in a small geographical area: if that area is very far away from a certain office, it may be more cost-effective to close that office entirely rather than shuttle individuals across the map to it.

While the results of this analysis may not be directly implementable, they can nonetheless be useful to inform discussion of equity in districting. First, it bears repeating that an equitable partition is shown to be \textit{feasible}, which is not obvious without our analysis. Second, we are able to quantify the cost paid by each demographic group for fair representation. As Table \ref{tab:Travel-distances-and} shows, the cost may even be higher for disadvantaged groups: for example, median travel times (under regular distance) increase by $54\%$ for the white population, but by $59\%$ for the Black population. This is a serious concern, and potentially a governing body might elect not to change district boundaries at all on such grounds. Third, if the authorities wished to adjust boundaries without fully adopting our approach, visuals such as those in Figure \ref{fig:facilities} may help guide these modifications by suggesting where representation of certain groups might be drawn from geographically. Finally, it is important to note that certain undesirable aspects of the boundaries (such as non-contiguity) are a consequence of a high level of segregation in the population. In a sense, the difficulty of creating an equitable partition is itself an illustration of how much segregation is already present.

\subsection{Analysis for three subgroups}\label{sec:lathree}

As mentioned previously, we also ran a version of the same case study in which only three demographic groups were considered: Black, Asian, and all the others combined. Thus, every facility must have $84.73\%$ of its assigned population coming from this last group, but the precise breakdown within that group (White, Multiracial, Other and so on) is no longer constrained. The main purpose of these additional experiments is to show how the number of groups included in the model influences travel costs for all groups, whether explicitly included or not.

\begin{table}[t]
\begin{centering}
\begin{tabular}{|c||c|c|c||c|c|c|}
\hline
Race & Median (u) & Median (c) & Median (c-sq) & 90th \% (u) & 90th \% (c) & 90th \% (c-sq)\tabularnewline
\hline
\hline
Multiracial & 2034 & 2473 & 3702 & 3739 & 6870 & 7783\tabularnewline
\hline
White & 2217 & 2547 & 3535 & 4709 & 7500 & 7958\tabularnewline
\hline
Black & 2154 & 2984 & 4261 & 3739 & 16592 & 11750\tabularnewline
\hline
American Indian & 1853 & 2241 & 3434 & 3487 & 5463 & 7490\tabularnewline
\hline
Asian & 2213 & 2943 & 4148 & 4104 & 14823 & 9963\tabularnewline
\hline
Pacific Islander & 2068 & 2683 & 3480 & 3901 & 8701 & 8042\tabularnewline
\hline
Other & 1843 & 2233 & 3606 & 3455 & 5390 & 7160\tabularnewline
\hline
\end{tabular}
\par\end{centering}
\caption{\label{tab:Travel-distances-three-races}Travel distances (in meters) for unconstrained (u), equity-constrained (c), and equity-constrained with squared distance (c-sq) models, categorized by race (three-group formulation).}
\end{table}

Table \ref{tab:Travel-distances-three-races} presents the median and 90th percentile of the travel time for each group. Note that, while the \textit{model} only distinguishes between three groups, we report results for \textit{all seven} of the groups from the original analysis. Once more, the baseline is a model with no fairness constraints (Voronoi diagram), so those statistics are the same as in Table \ref{tab:Travel-distances-and}.

When fairness constraints are included, the results are very different between Tables \ref{tab:Travel-distances-and} and \ref{tab:Travel-distances-three-races}. Let us first consider the two groups (Black and Asian) that achieve fair representation under both versions of the case study. In Table \ref{tab:Travel-distances-and} (seven groups), the median travel cost for the Black subpopulation increases by $59.6\%$ relative to the baseline (under the regular cost function), but in Table \ref{tab:Travel-distances-three-races}, this increase is only $38.5\%$. Likewise, for the Asian subpopulation, the increase in the median is $60.9\%$ with seven groups in the model, but only $32.9\%$ with three groups.

\begin{figure}[b]
\centering
\includegraphics[width=0.99\textwidth]{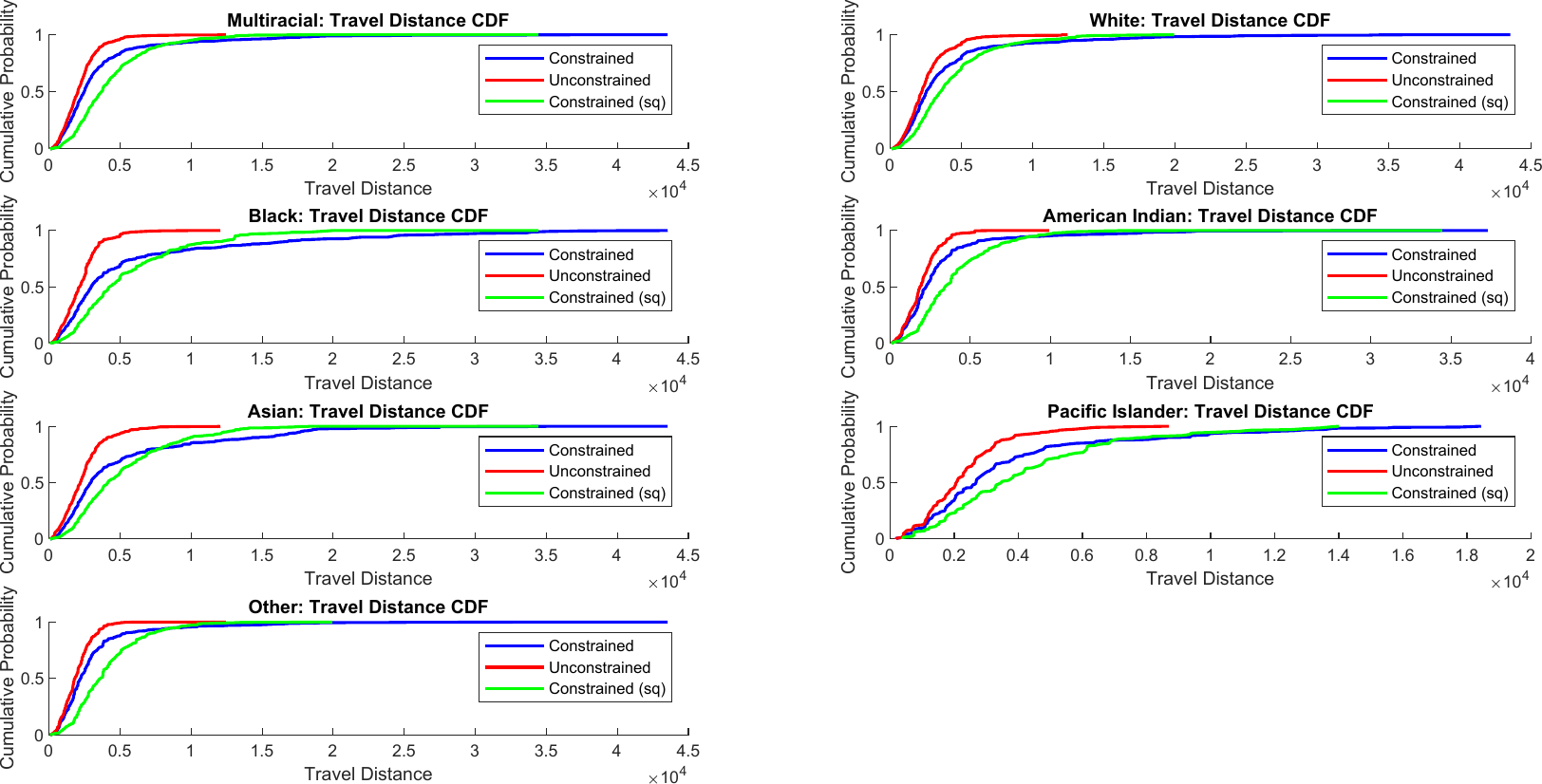}\label{fig:cdfs2}
\caption{CDFs for travel distances for each demographic (three-group formulation).}
\end{figure}

For the remaining five groups, the difference is even greater. In Table \ref{tab:Travel-distances-and}, the median travel cost increases by over $54\%$ for each of the five relative to the baseline model (again, under the regular cost function). In Table \ref{tab:Travel-distances-three-races}, the increase in the median is below $30\%$ for all five groups. A more detailed comparison of the cdfs for various groups and models can be seen in Figure \ref{fig:cdfs2}; in general, the distributions for the constrained models are closer to the baseline than they were in Figure \ref{fig:cdfs1}.

\begin{figure}[H]
	\centering
	\subfigure[Facility 12 (travel distance).]{
		\includegraphics[width=0.33\textwidth]{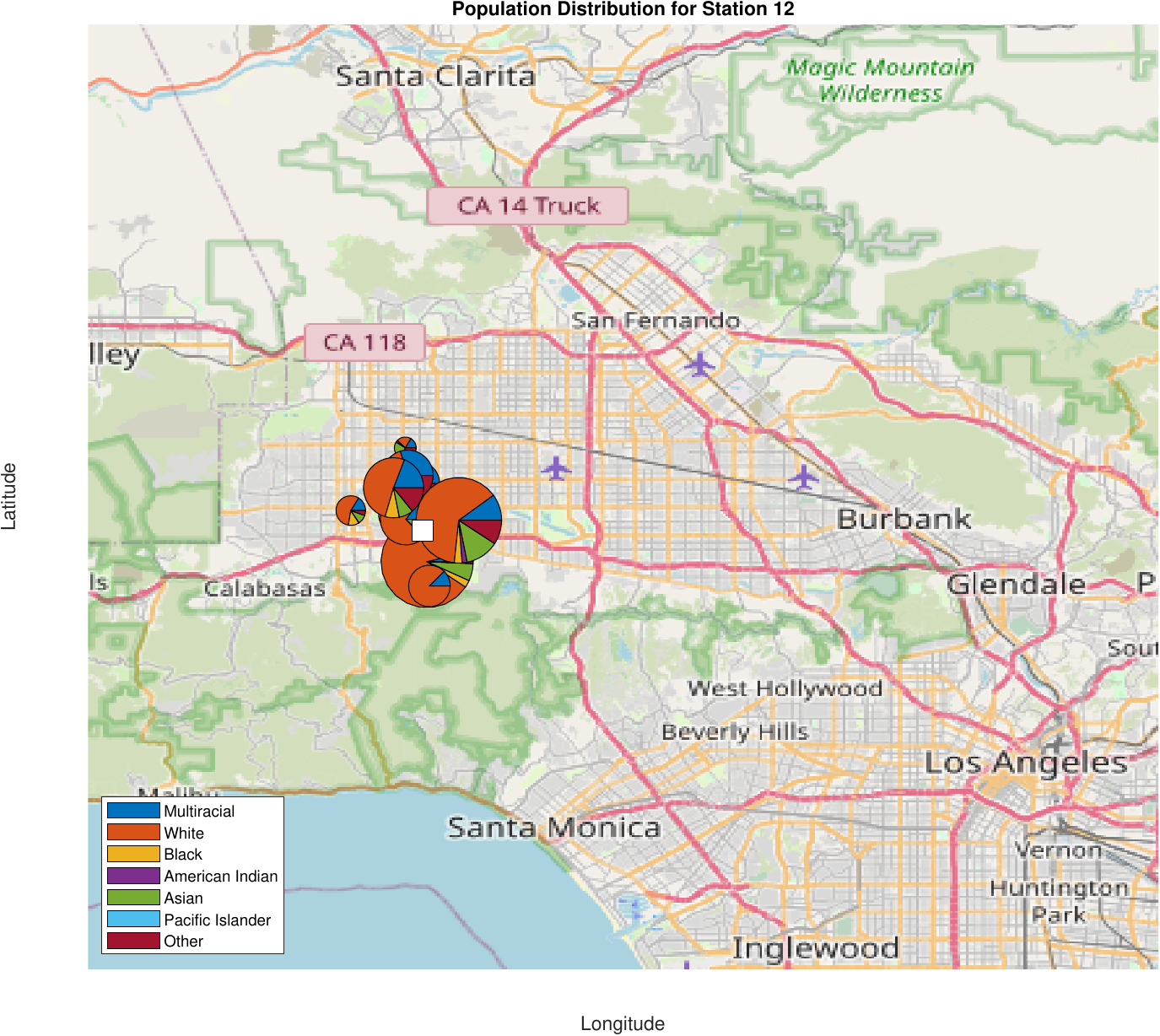}
		\label{fig:3r}
	}
	\subfigure[Facility 12 (squared distance).]{
		\includegraphics[width=0.33\textwidth]{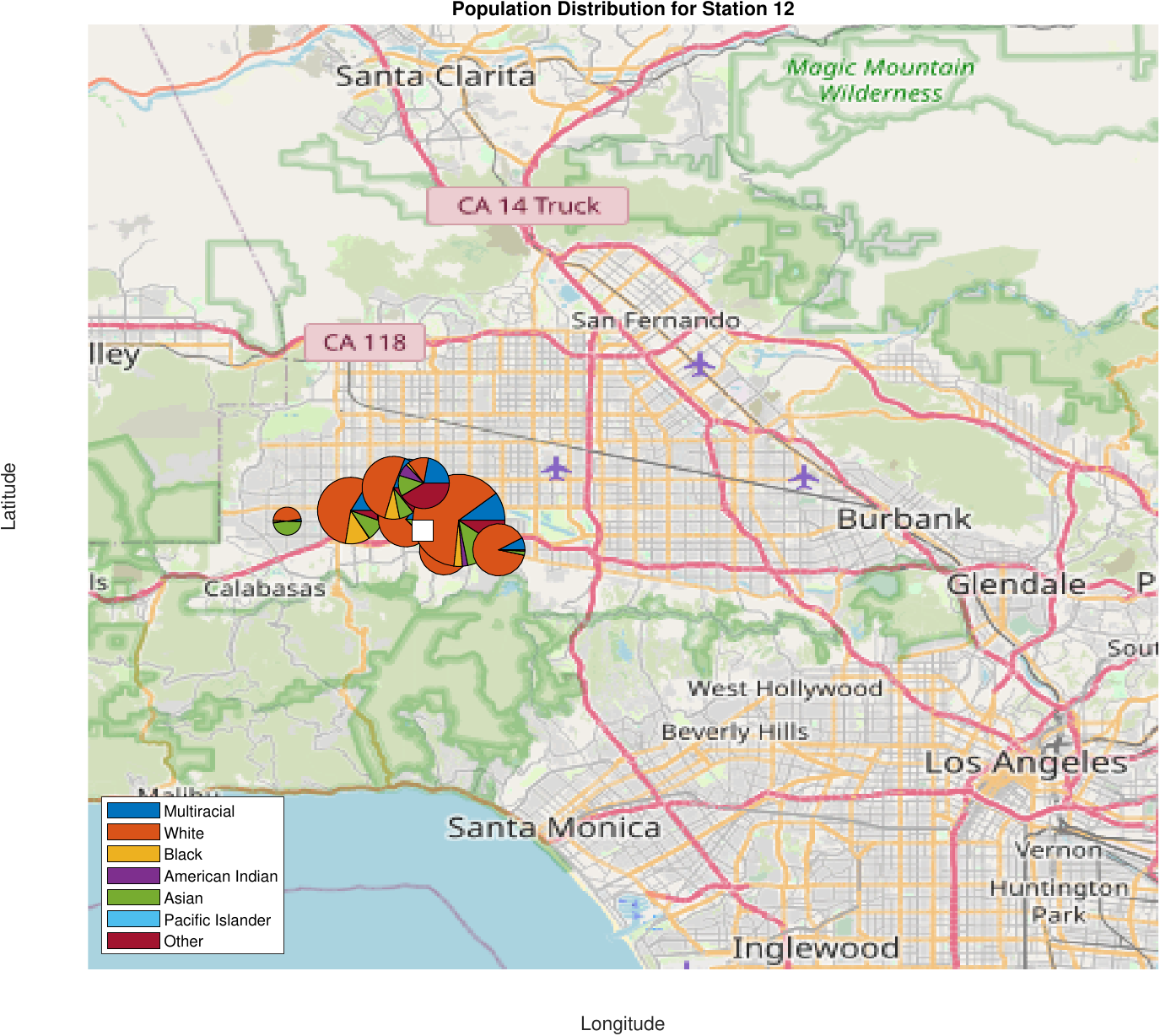}
		\label{fig:3s}
	}
	\subfigure[Facility 34 (travel distance).]{
		\includegraphics[width=0.33\textwidth]{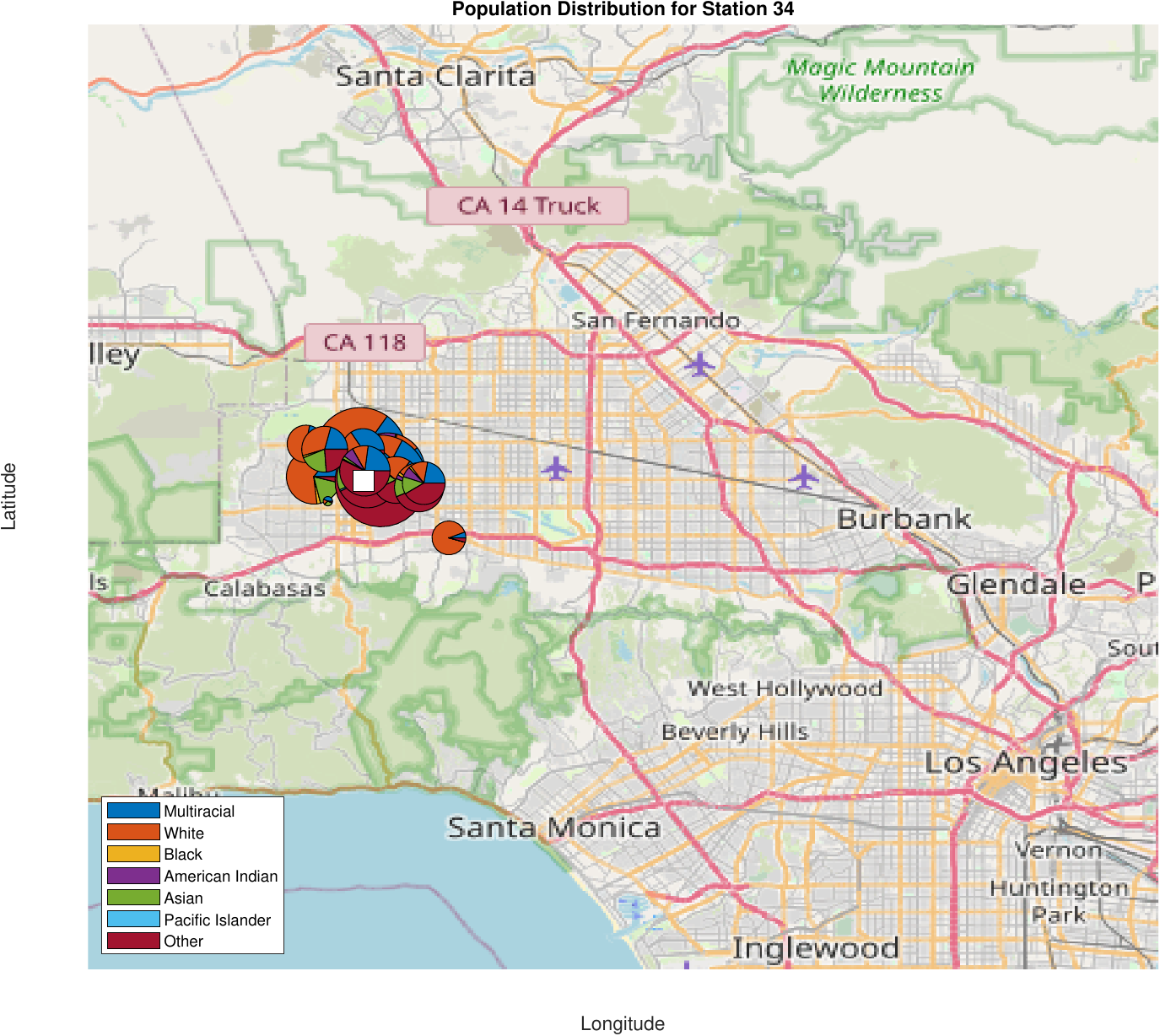}
		\label{fig:12r}
	}
	\subfigure[Facility 34 (squared distance).]{
		\includegraphics[width=0.33\textwidth]{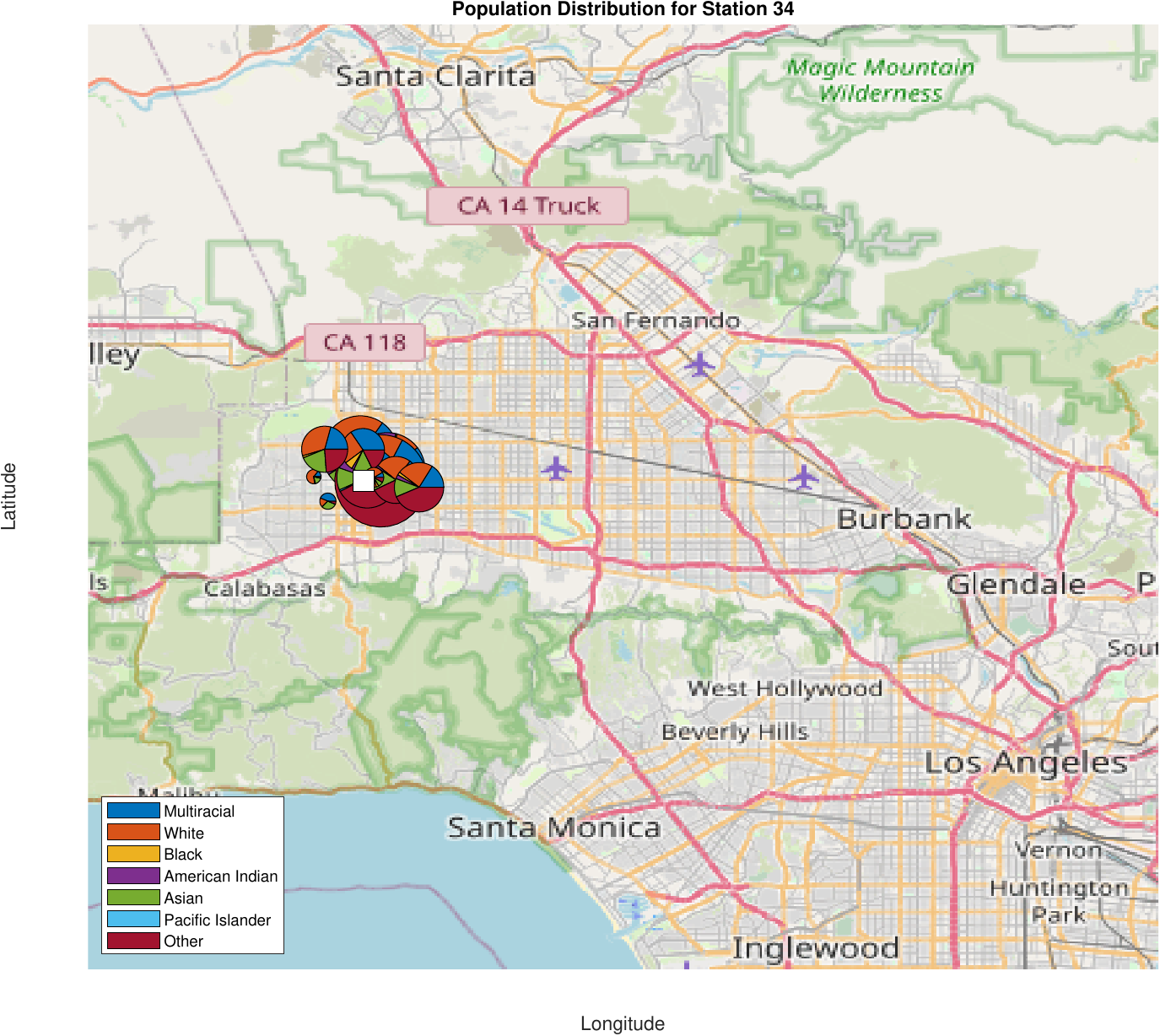}
		\label{fig:12s}
	}
	\subfigure[Facility 63 (travel distance).]{
		\includegraphics[width=0.33\textwidth]{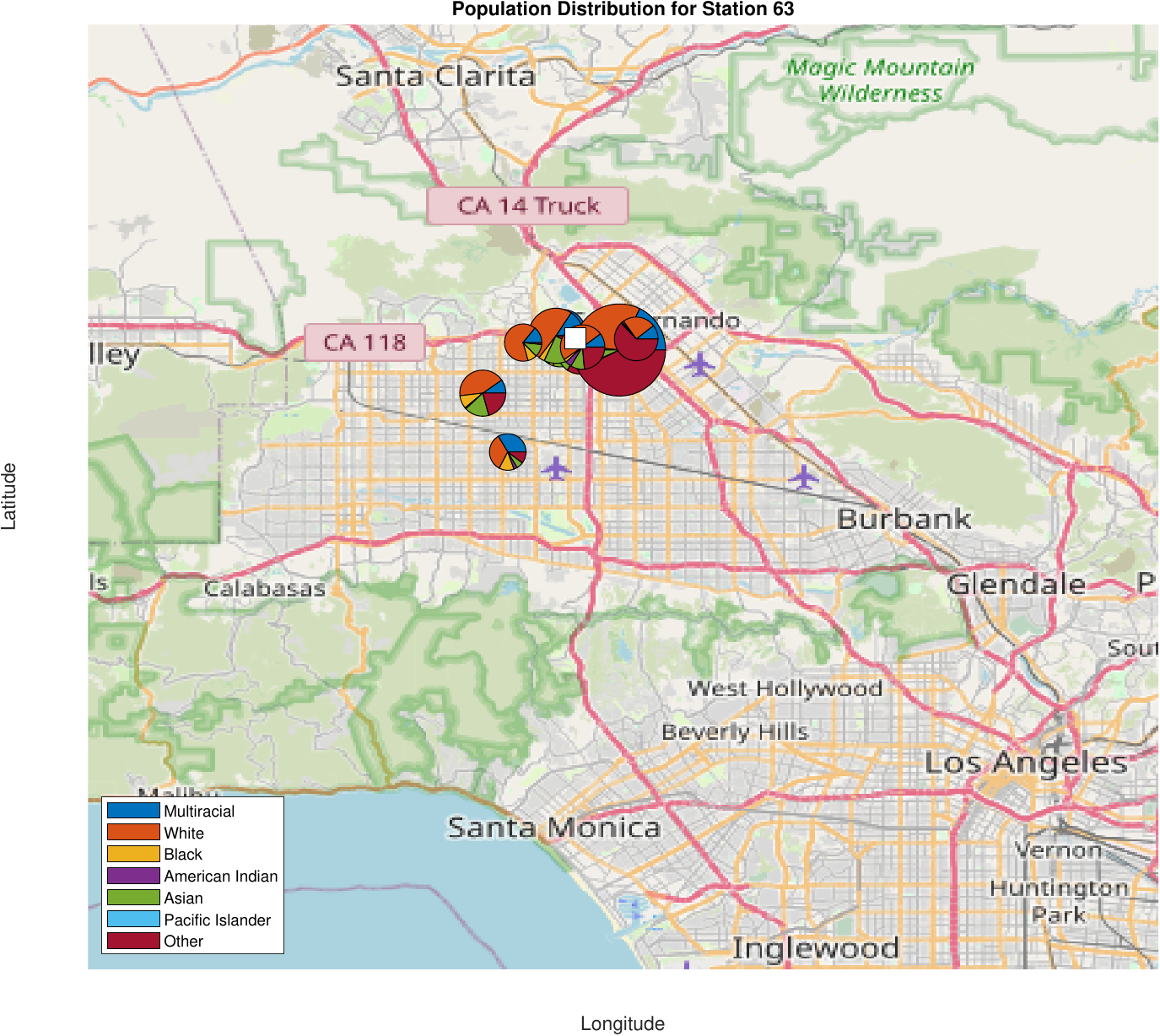}
		\label{fig:27r}
	}
	\subfigure[Facility 63 (squared distance).]{
		\includegraphics[width=0.33\textwidth]{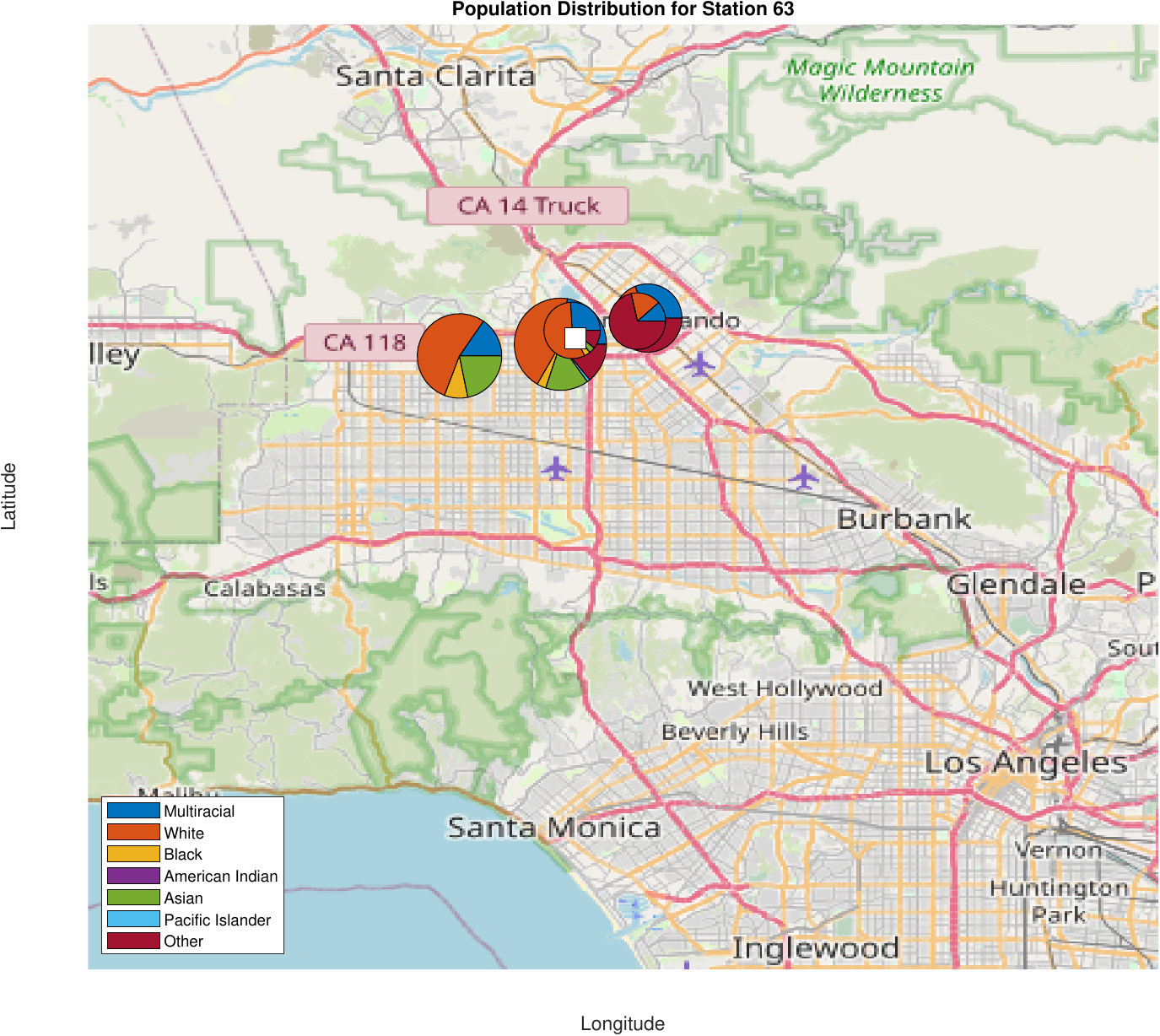}
		\label{fig:27s}
	}
	\subfigure[Facility 76 (travel distance).]{
		\includegraphics[width=0.33\textwidth]{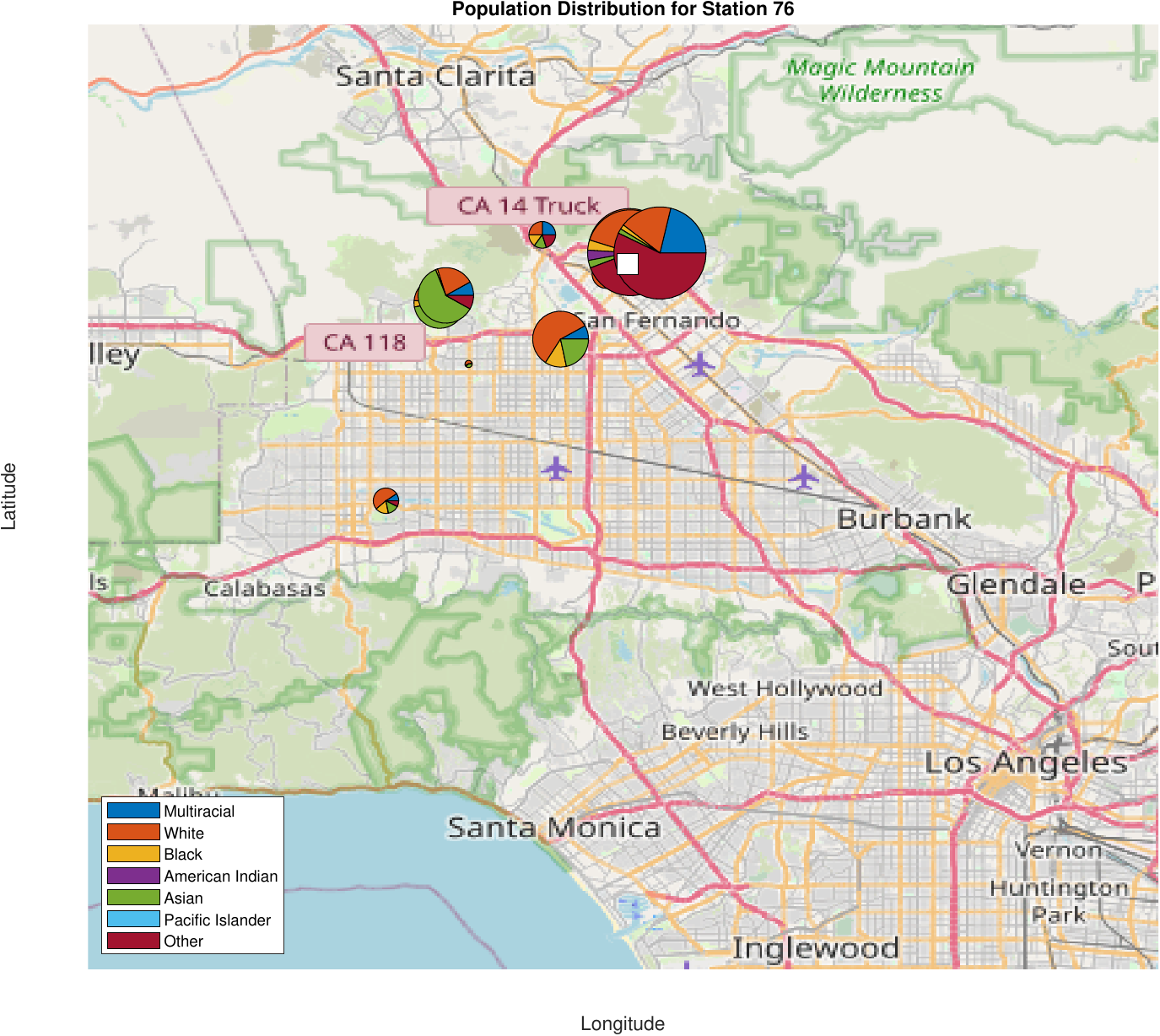}
		\label{fig:32r}
	}
	\subfigure[Facility 76 (squared distance).]{
		\includegraphics[width=0.33\textwidth]{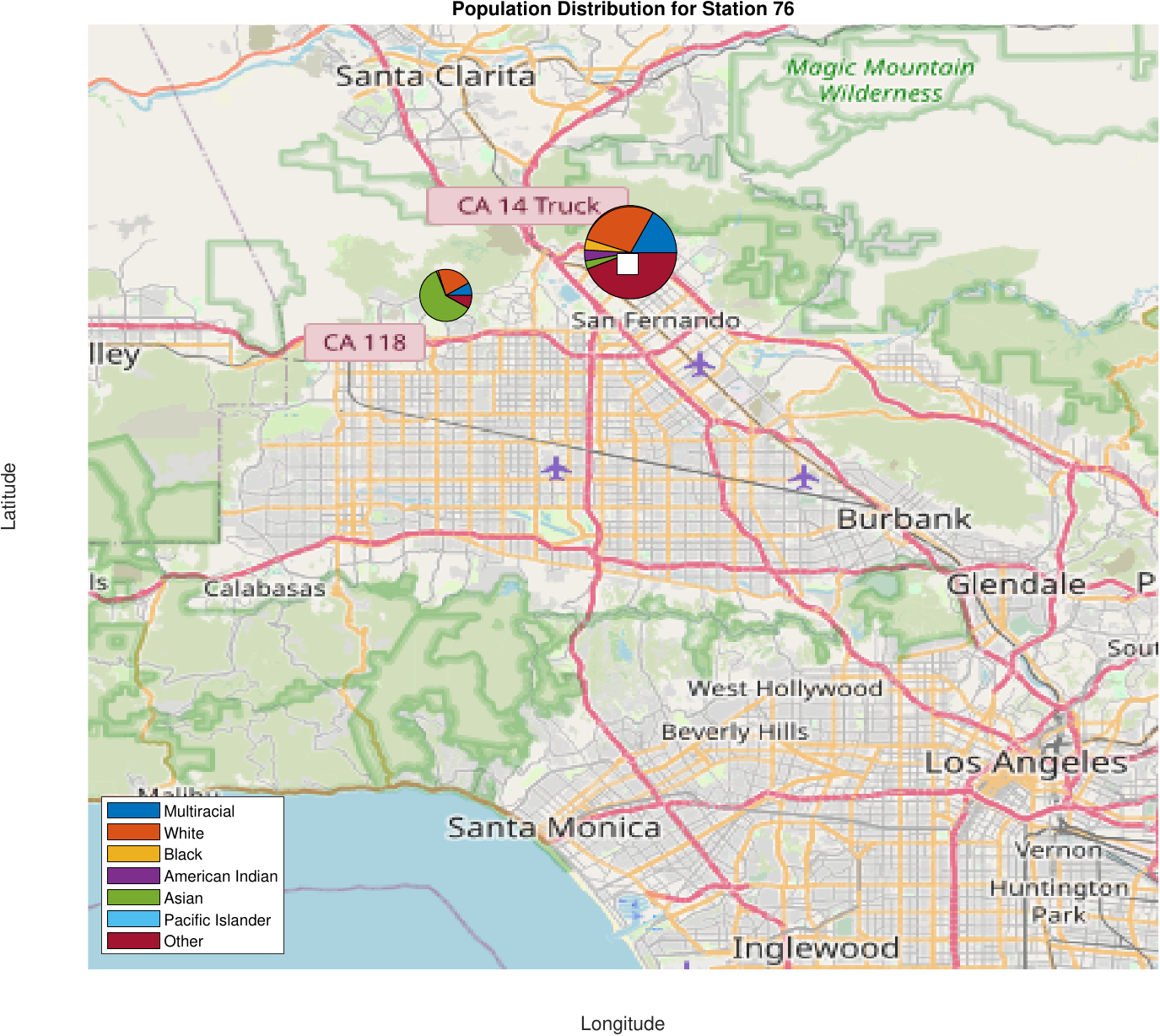}
		\label{fig:32s}
	}
\caption{Four representative districts, drawn using two different cost functions (regular and squared distance).}\label{fig:facilities2}
\end{figure}

Figure \ref{fig:facilities2} revisits the four representative facilities from Figure \ref{fig:facilities} in the three-group context; a complete set of plots can be found in the Appendix. It is especially instructive to compare the district boundaries between the seven-group and three-group settings. For example, in the seven-group case (Figure \ref{fig:facilities}), Facility 63 had three outlying census tracts under regular distance and one under squared distance. In the three-group case (Figure \ref{fig:facilities2}), the most distant outliers have disappeared and the census tracts are now closer to the office. Similarly, in the seven-group case, Facility 76 had several very distant outliers, far southwest and southeast of the district office, under regular distance; in the three-group case, only the southwestern outlier is still present, while the southeastern ones are no longer assigned to this facility. Under squared distance, the census tracts were closer to the facility to begin with, and again the most distant ones have been removed in the three-group case. In other words, running the model with three groups does not guarantee that the districts will become contiguous, but it significantly reduces the fragmentation of the boundaries. It is worth noting that district closures are also reduced: under three groups, $3$ and $5$ offices are closed, respectively, for regular and squared distance (in the seven-group case these numbers were $7$ and $12$). A single office (\#16) is closed in all four versions of the case study.

This discussion shows how our analysis can be used, not only to assess the cost of fairness in a general sense, but to make a more fine-grained determination of which subpopulations could be included in the model. For example, the Pacific Islander population is known to be disadvantaged in education \citep{PaHaPa11}. At the same time, when all seven groups are included in the model, Pacific Islanders suffer disproportionately from increased travel costs, with the median going up by $102\%$ relative to the baseline (under unsquared costs). In the three-group version of the analysis, the increase for that subpopulation is only $29\%$. The precise considerations behind the decision to require fair representation for a particular group are far beyond the scope of our work, but it is conceivable that, in some cases, the cost of representation may be judged to be too prohibitive \textit{for the disadvantaged group itself}. In such cases, reducing the total number of groups can also alleviate the burden on other groups that remain in the model.

\section{Conclusion}

We have introduced a new class of geographical partitioning problems in which the population is heterogeneous, and different subgroups are required to be represented fairly (i.e., proportionally to their incidence in the population as a whole) in each district. Our study is motivated by the phenomenon of socioeconomic segregation in school districts, where some demographic groups are over- or under-represented in certain areas. This problem has been a subject of discussion for many years among social scientists and policymakers, with some authors even recommending a kind of ``reverse gerrymandering'' as a means to achieve integration. Essentially, our work shows how this can be done in a principled and \textit{optimal} way, constructing a fair partition that minimizes expected cost.

We have provided a complete mathematical characterization of this optimal fair partition, showing that it is a novel generalization of a classical geometric structure known as the additively weighted Voronoi diagram. Although the optimization problem we are solving is infinite-dimensional, it has a finite-dimensional dual, whose decision variables are sufficient to fully describe the optimal primal solution. Furthermore, the optimal dual variables can be computed using a stochastic approximation algorithm that is trivial to code, runs efficiently, and does not require a closed form for the cost function. Together, all of these developments answer an open question dating back to \cite{DvWaWo51}, which proved the existence of fair partitions but did not offer any guidance as to how they may be computed.

Our case study sheds light on some of the practical challenges that may arise when considering the implementation of such a model. Fair representation can have unintended consequences, creating non-contiguous districts and significantly increasing travel times for the entire population, particularly for the very subgroups that policymakers aim to help. Nonetheless, our analysis can also be used to explore ways to mitigate these problems, for example by changing the cost function in a manner that reduces very high travel times. There are also difficult political and ethical considerations involved in determining the precise set of subgroups to include in the model. However, we believe that the main value of our paper is in offering a rigorous, evidence-based, politically neutral framework for assessing the feasibility of fair representation in complex regions that are often highly segregated to begin with.

\bibliographystyle{agsm} 
\bibliography{hamsandwich} 

\newpage

\section{Appendix: proofs}\label{sec:proofs}

Below, we provide complete proofs for all results that were stated in the text.

\subsection{Proof of Theorem \ref{thm:duality}}

The proof uses the following result from Sec. 8.6 of \cite{Lu97}.

\begin{thm}\label{thm:luenberger}
Let $\mathfrak{f}$ be a real-valued convex functional defined on a convex subset $\Omega$
of a vector space $\mathfrak{X}$, and let $\mathfrak{G}$ be a convex
mapping of $\mathfrak{X}$ into a normed space $\mathfrak{Z}$. Suppose
there exists an $\mathfrak{x}_{1}$ such that $\mathfrak{G}(\mathfrak{x}_{1})<\theta$,
where $\theta$ is the zero element, and that $\inf\{\mathfrak{f}(x):\mathfrak{x}\in\Omega,\,\mathfrak{G}(\mathfrak{x})\leq\theta\}$
is finite. Then
\begin{eqnarray*}
\inf_{\mathfrak{x}\in\Omega,\,\mathfrak{G}(\mathfrak{x})\leq\theta}\mathfrak{f}(\mathfrak{x}) & = & \max_{\mathfrak{z}^{*}\geq\theta}\inf_{\mathfrak{x}\in\Omega}\mathfrak{f}(\mathfrak{x})+\left\langle \mathfrak{G}(\mathfrak{x}),\mathfrak{z}^{*}\right\rangle
\end{eqnarray*}
and the maximum on the right is achieved by some $\mathfrak{z}_{0}^{*}\geq\theta$.
If the infimum on the left is achieved by some $\mathfrak{x}_{0}\in\Omega$,
then $\left\langle \mathfrak{G}(\mathfrak{x}_{0}),\mathfrak{z}_{0}^{*}\right\rangle =0$,
and $\mathfrak{x}_{0}$ minimizes $\mathfrak{f}(\mathfrak{x})+\left\langle \mathfrak{G}(\mathfrak{x}),\mathfrak{z}_{0}^{*}\right\rangle $
over all $\mathfrak{x}\in\Omega$.
\end{thm}

Consider the problem
\begin{equation}\label{eq:luenberger1}
\max_{u_{kz}}\iint_{\mathcal{X}}\sum_{z}f_{z}(x)\min_{k}\left\{ q_{z}c(x,k)-f_{z}(x)u_{kz}\right\} +\sum_{k,z}p_{k}u_{kz}
\end{equation}
 which has a maximizer $u_{kz}^{*}$ by compactness of $\mathcal{X}$. We may rewrite (\ref{eq:luenberger1}) as an LP over $\mathbb{R}^{KM}\oplus L_{1}(\mathcal{X})$. This problem has the objective
\begin{equation*}
\max_{u_{kz},\sigma(x)}\iint_{\mathcal{X}}\sigma(x)+\sum_{k,z}p_{k}u_{kz}
\end{equation*}
subject to the constraints
\begin{align*}
\sum_{z}f_{z}(x)u_{kz}+\sigma(x) & \leq\left(\sum_{z}q_{z}f_{z}(x)\right)c(x,k)\qquad\forall x,k.
\end{align*}
Equivalently, we may state it as
\begin{equation*}
\min_{u_{kz},\sigma(x)}-\left(\iint_{\mathcal{X}}\sigma(x)+\sum_{k,z}p_{k}u_{kz}\right)
\end{equation*}
subject to
\begin{align*}
\sum_{z}f_{z}(x)u_{kz}+\sigma(x)-\left(\sum_{z}q_{z}f_{z}(x)\right)c(x,k) & \leq0\qquad\forall x,k.
\end{align*}
We thus have
\begin{align*}
\mathfrak{X}=\Omega= & \mathbb{R}^{KZ}\oplus L_{1}(\mathcal{X})\\
\mathfrak{x} & =\left(\begin{array}{c}
u_{kz}\\
\sigma(\cdot)
\end{array}\right)\\
\mathfrak{Z} & =L^{1}(\mathcal{X})^{K}\\
\mathfrak{G}(\mathfrak{x}) & =\sum_{z}f_{z}(x)u_{kz}+\sigma(x)-\left(\sum_{z}q_{z}f_{z}(x)\right)c(x,k)
\end{align*}
According to Theorem \ref{thm:luenberger}, the solution exists and we have $\mathfrak{Z}^{*}=L^{\infty}(\mathcal{X})^{K}$,
so $\mathfrak{z}^{*}=g(x,k)$ and the solution is the same as
\begin{eqnarray*}
\max_{g(\cdot,\cdot)\geq0}\inf_{u_{kz},\sigma(x)}&-&\left(\iint_{\mathcal{X}}\sigma(x)\,dx+\sum_{k,z}p_{k}u_{kz}\right)\\
&+&\sum_{k}\iint_{\mathcal{X}}\left[\left(\sum_{z}f_{z}(x)u_{kz}\right)+\sigma(x)-\left(\sum_{z}q_{z}f_{z}(x)\right)c(x,k)\right]g(x,k)\,dx.
\end{eqnarray*}
Equivalently, we may write
\[
\max_{g(\cdot,\cdot)\geq0}\inf_{u_{kz},\sigma(x)}-\sum_{k,z}p_{k}u_{kz}+\iint_{\mathcal{X}}\sum_{k}\left[\left(\sum_{z}f_{z}(x)u_{kz}\right)+\sigma(x)-\left(\sum_{z}q_{z}f_{z}(x)\right)c(x,k)\right]g(x,k)-\sigma(x)\,dx,
\]
which is the same as
\begin{align}
\max_{g(\cdot,\cdot)\geq0}\inf_{u_{kz},\sigma(x)} & \iint_{\mathcal{X}}\sum_{k}\left[-\left(\sum_{z}q_{z}f_{z}(x)\right)c(x,k)\right]g(x,k)\,dx\nonumber \\
+ & \iint_{\mathcal{X}}\left(\sum_{k}g(x,k)-1\right)\sigma(x)\,dx+\iint_{\mathcal{X}}\sum_{k,z}\left(f_{z}(x)g(x,k)-p_{k}\right)u_{kz}\,dx.\label{eq:balance-constraint}
\end{align}
Because we are taking $\inf_{u_{kz},\sigma(x)}$, (\ref{eq:balance-constraint}) requires $\sum_{k}g(x,k)=1\,\forall x\in\mathcal{X}$,
and $\iint_{\mathcal{X}}g(x,k)f_{z}(x)\,dx=p_{k}\,\forall k,\,\forall z$.
Thus, swapping out the minus sign in the objective function and replacing
it with a minimization, the problem is the same as
\begin{align*}
\min_{g}\sum_{k}\iint_{\mathcal{X}}\sum_{z}q_{z}c(x,k)g(x,k)f_{z}(x)\,dx
\end{align*}
subject to
\begin{align*}
\sum_{k}g(x,k) & =1\qquad\forall x\in\mathcal{X}\\
\iint_{\mathcal{X}}g(x,k)f_{z}(x)\,dx & =p_{k}\qquad\forall k,\,\forall z\\
g & \geq0
\end{align*}
as desired.

\subsection{Proof of Theorem \ref{thm:sa}}

Observing that $v^{n+1}-v^* = v^n - v^* + \alpha_n \zeta^n\left(v^n\right)$, we write
\begin{eqnarray}
\mathbb{E}\left(\|v^{n+1}-v^*\|^2_2\right) &=& \mathbb{E}\left(\|v^n-v^*\|^2_2\right) + \alpha^2_n\mathbb{E}\left(\|\zeta^n\left(v^n\right)\|^2_2\right) + 2\alpha_n \mathbb{E}\left(\left(v^n-v^*\right)^\top \nabla_v h\left(v^n\right)\right)\label{eq:expansion1}\\
&\leq& \mathbb{E}\left(\|v^n-v^*\|^2_2\right) + \alpha^2_n C^2,\label{eq:expansion2}
\end{eqnarray}
where (\ref{eq:expansion1}) is obtained by conditioning the last term on $v^n$, and (\ref{eq:expansion2}) follows by the boundedness of $\zeta^n$ and the concavity of $h$, i.e., $0\leq h\left(v^*\right)-h\left(v^n\right) \leq \nabla_v h\left(v^n\right)^\top\left(v^* - v^n\right)$. Thus, we have
\begin{equation*}
\mathbb{E}\left(\|v^{n}-v^*\|^2_2\right) \leq \|v^0-v^*\|^2_2 + C^2 \sum^{n-1}_{m=0} \alpha^2_m \leq C_1 \log n
\end{equation*}
for some sufficiently large $C_1 > 0$.

Applying the concavity of $h$ twice, we derive
\begin{eqnarray}
h\left(v^*\right) - h\left(\frac{1}{n}\sum^n_{m=1} v^m\right) &\leq & h\left(v^*\right) - \frac{1}{n}\sum^n_{m=1} h\left(v^m\right)\nonumber\\
&=& \frac{1}{n}\sum^n_{m=1} h\left(v^*\right) - h\left(v^m\right)\nonumber\\
&\leq & \frac{1}{n}\sum^n_{m=1} \left(v^*-v^m\right)^\top \nabla_v h\left(v^m\right).\label{eq:useconcavity}
\end{eqnarray}
By (\ref{eq:expansion1}), we have
\begin{equation}\label{eq:expansion3}
\mathbb{E}\left(\left(v^*-v^m\right)^\top \nabla_v h\left(v^m\right)\right) = \frac{\delta^m - \delta^{m+1} + \alpha^2_m \mathbb{E}\left(\|\zeta^m\left(v^m\right)\|^2_2\right)}{2\alpha_m},
\end{equation}
where $\delta^m = \mathbb{E}\left(\|v^m-v^*\|^2_2\right)$. Plugging (\ref{eq:expansion3}) into (\ref{eq:useconcavity}) yields
\begin{eqnarray*}
h\left(v^*\right) - \mathbb{E}\left(h\left(\frac{1}{n}\sum^n_{m=1} v^m\right)\right) &\leq & \frac{1}{n} \sum^n_{m=1} \frac{\delta^m - \delta^{m+1} + \alpha^2_m \mathbb{E}\left(\|\zeta^m\left(v^m\right)\|^2_2\right)}{2\alpha_m}\\
&\leq& \frac{1}{2n} \sum^n_{m=1} \frac{\delta^m - \delta^{m+1}}{\alpha_m} + \frac{C^2}{2n}\sum^n_{m=1}\alpha_m\\
&=& \frac{1}{2n}\left(\sum^n_{m=1} \frac{\delta^m}{\alpha_m} - \sum^n_{m=1} \frac{\delta^{m+1}}{\alpha_m} + \sum^n_{m=1} \delta^{m+1}\left(\frac{1}{\alpha_{m+1}} - \frac{1}{\alpha_m}\right)\right) + \frac{C^2}{2n}\sum^n_{m=1}\alpha_m\\
&=& \frac{1}{2n}\left( \frac{\delta^1}{\alpha_1} - \frac{\delta^{n+1}}{\alpha_{n+1}} + \sum^n_{m=1} \delta^{m+1}\left(\frac{1}{\alpha_{m+1}} - \frac{1}{\alpha_m}\right)\right) + \frac{C^2}{2n}\sum^n_{m=1}\alpha_m\\
&\leq & \frac{1}{2n}\left( \frac{\delta^1}{\alpha_1} - \frac{\delta^{n+1}}{\alpha_{n+1}} + C_1\sum^n_{m=1}\left(\frac{1}{\alpha_{m+1}}-\frac{1}{\alpha_m}\right)\log m\right) + \frac{C^2}{2n}\sum^n_{m=1}\alpha_m\\
&\leq & \frac{1}{2n}\left( \frac{\delta^1}{\alpha_1} - \frac{\delta^{n+1}}{\alpha_{n+1}} + C_1\log n\sum^n_{m=1}\frac{1}{\alpha_{m+1}}-\frac{1}{\alpha_m}\right) + \frac{C^2}{2n}\sum^n_{m=1}\alpha_m\\
&=& \frac{1}{2n}\left( \frac{\delta^1}{\alpha_1} - \frac{\delta^{n+1}}{\alpha_{n+1}} + C_1\log n\left(\frac{1}{\alpha_{n+1}}-\frac{1}{\alpha_1}\right)\right) + \frac{C^2}{2n}\sum^n_{m=1}\alpha_m\\
&\leq & \frac{C_1}{2n}\cdot\frac{1}{\alpha_{n+1}} \cdot \log n + \frac{C^2}{2n}\sum^n_{m=1}\alpha_m.
\end{eqnarray*}
The desired result follows after plugging in the definition of $\left\{\alpha_n\right\}$.

\section{Appendix: complete visuals for case study}

In this section, we present a complete set of plots showing the district boundaries for all $78$ facilities in Sections \ref{sec:laseven}-\ref{sec:lathree}. The plots were generated in the same way as in Figures \ref{fig:facilities}-\ref{fig:facilities2}. A blank plot indicates that the district office in question was closed, i.e., assigned $p^*_k = 0$ by the relevant optimization model. 

There are $312$ plots in all, presented in groups of $20$. We first provide plots for the seven-group setting from Section \ref{sec:laseven}. The first set of $78$ is for regular (unsquared) travel distance as the cost function, and the second set is for squared distance. We then provide two more sets for the three-group case (Section \ref{sec:lathree}).

\begin{figure}[h!]
	\centering
	\includegraphics[width=0.99\textwidth]{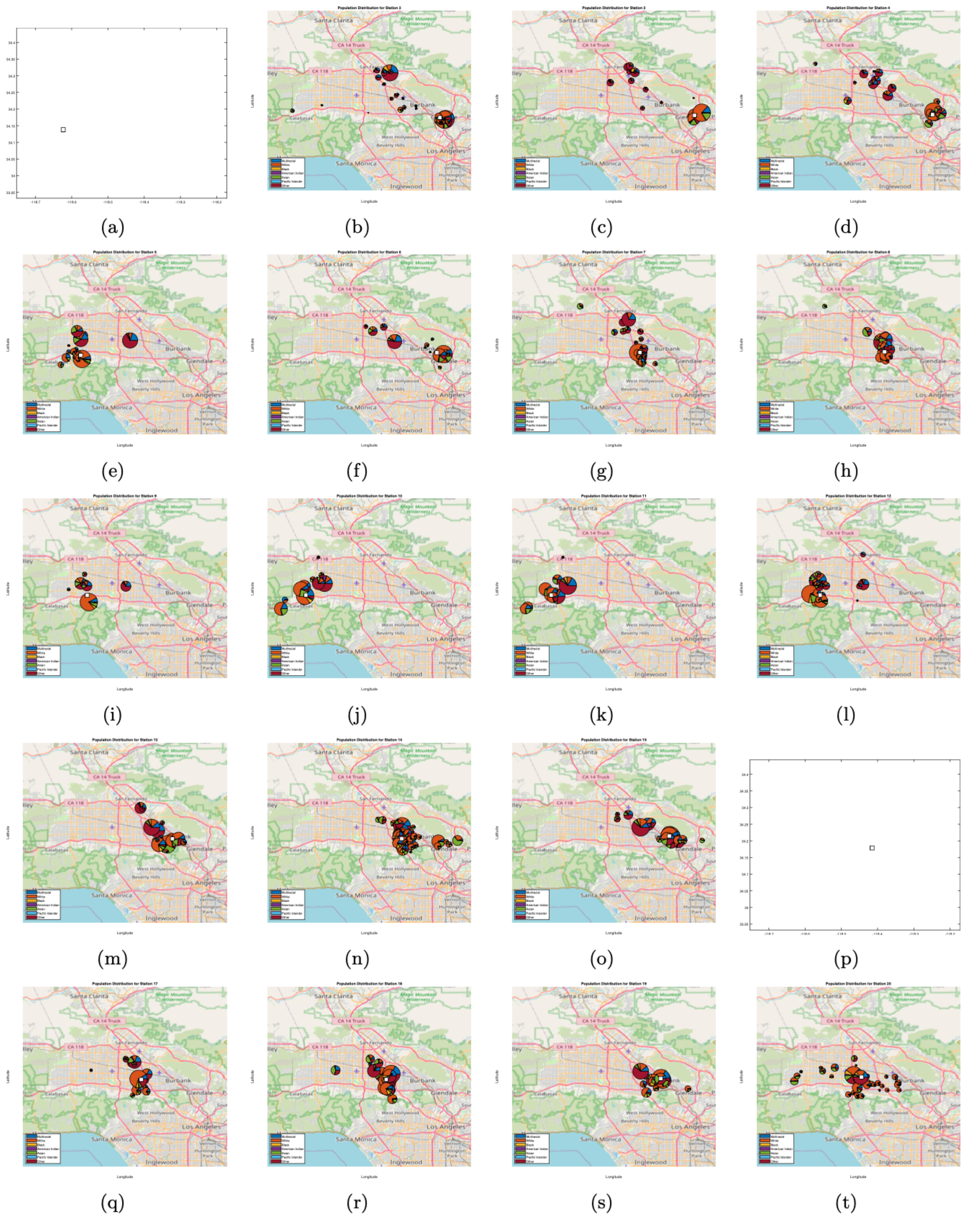}
\caption{District boundaries for district offices 1-20 under regular distance (seven groups).}
\end{figure}

\begin{figure}[h!]
	\centering
\includegraphics[width=0.99\textwidth]{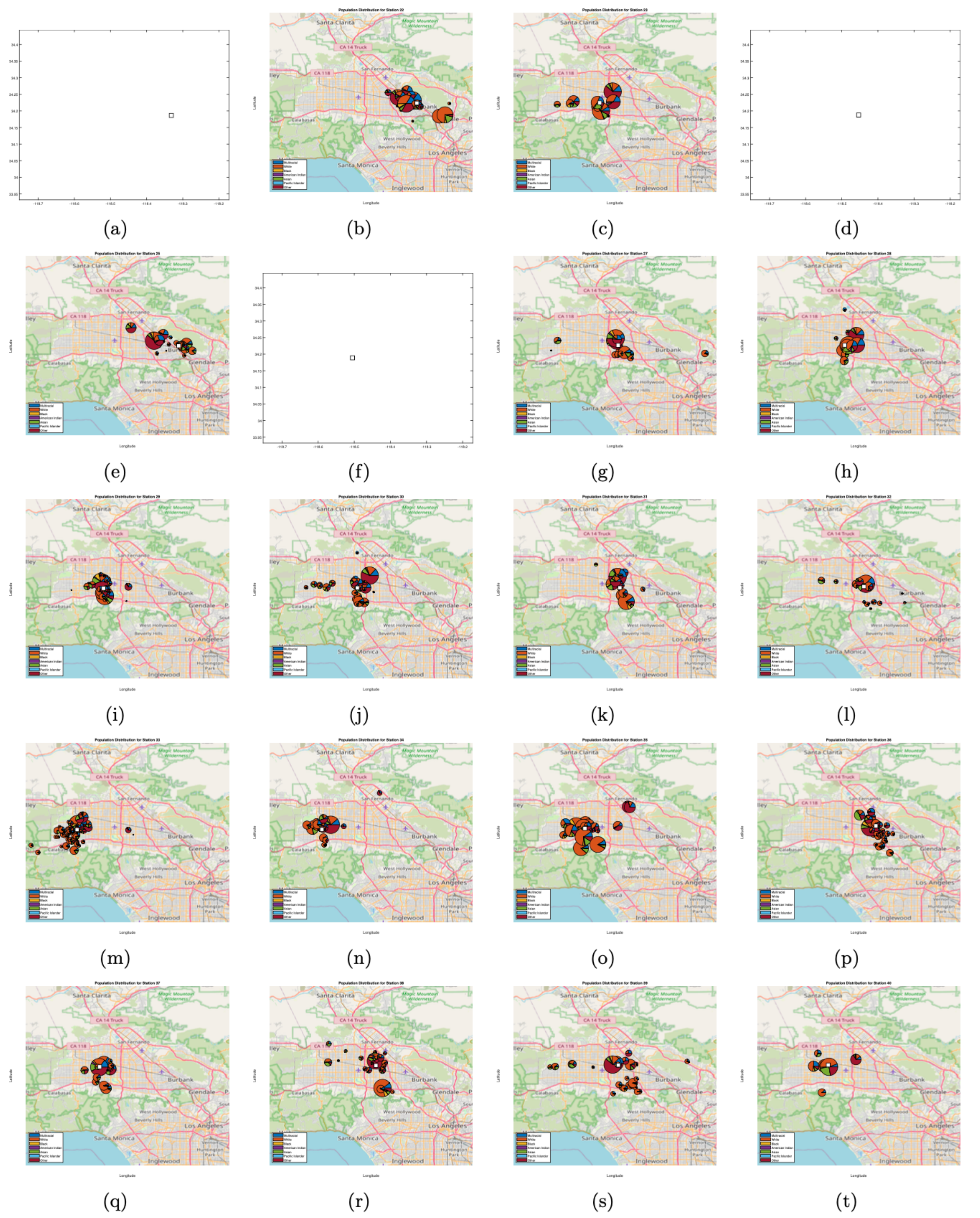}
\caption{District boundaries for district offices 21-40 under regular distance (seven groups).}
\end{figure}

\begin{figure}[h!]
	\centering
\includegraphics[width=0.99\textwidth]{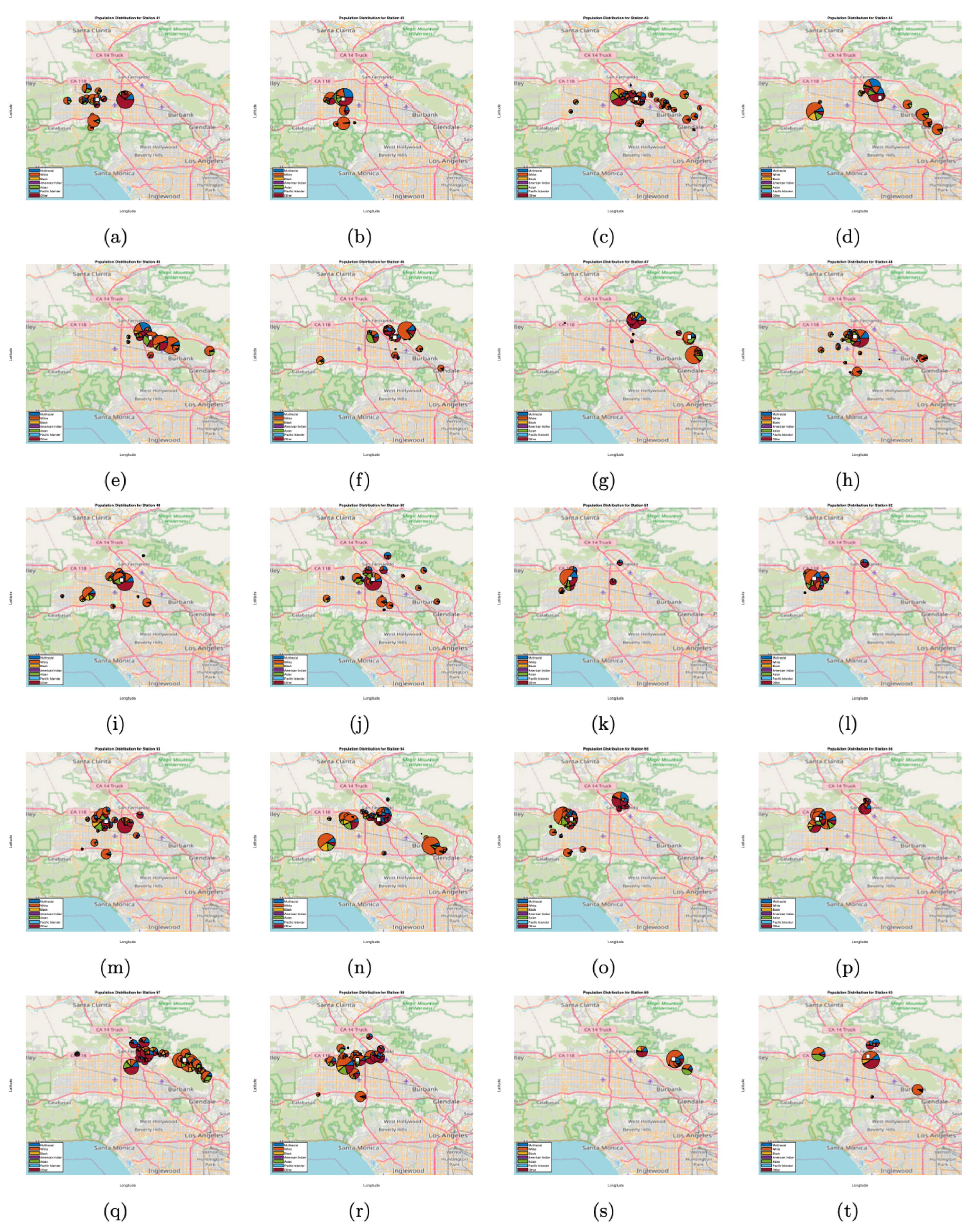}
\caption{District boundaries for district offices 41-60 under regular distance (seven groups).}
\end{figure}

\begin{figure}[h!]
	\centering
\includegraphics[width=0.99\textwidth]{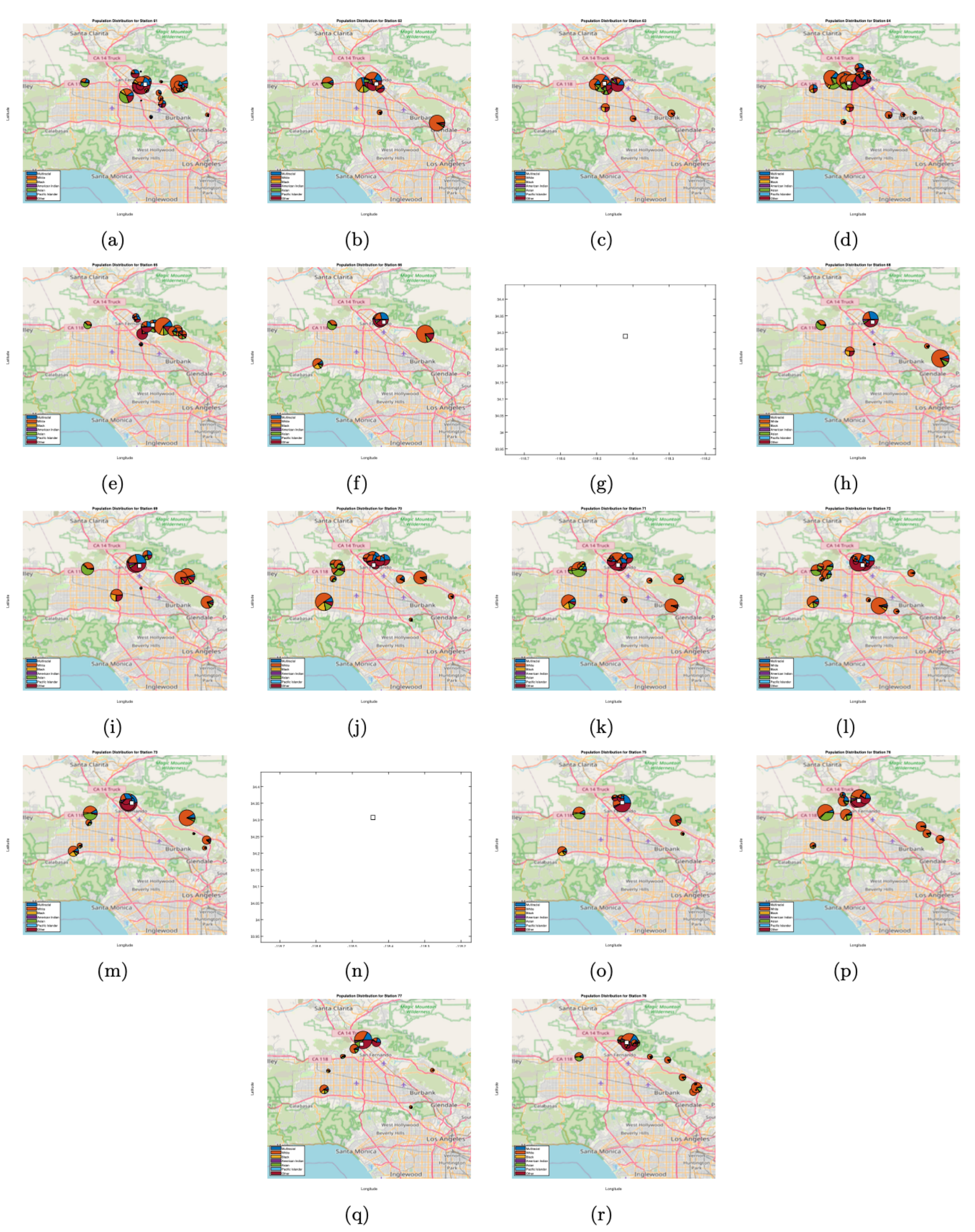}
\caption{District boundaries for district offices 61-78 under regular distance (seven groups).}
\end{figure}

\begin{figure}[h!]
	\centering
\includegraphics[width=0.99\textwidth]{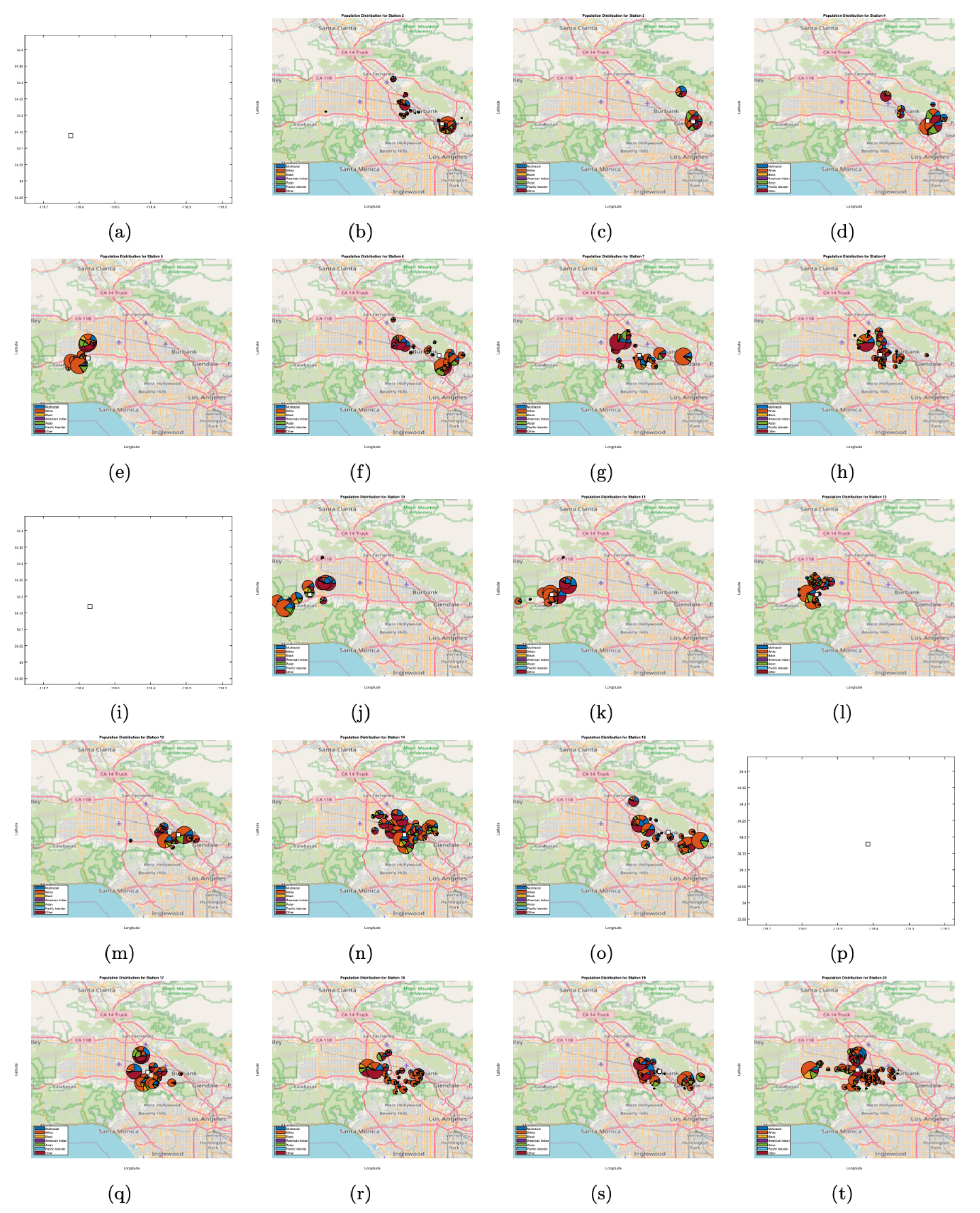}
\caption{District boundaries for district offices 1-20 under squared distance (seven groups).}
\end{figure}

\begin{figure}[h!]
	\centering
\includegraphics[width=0.99\textwidth]{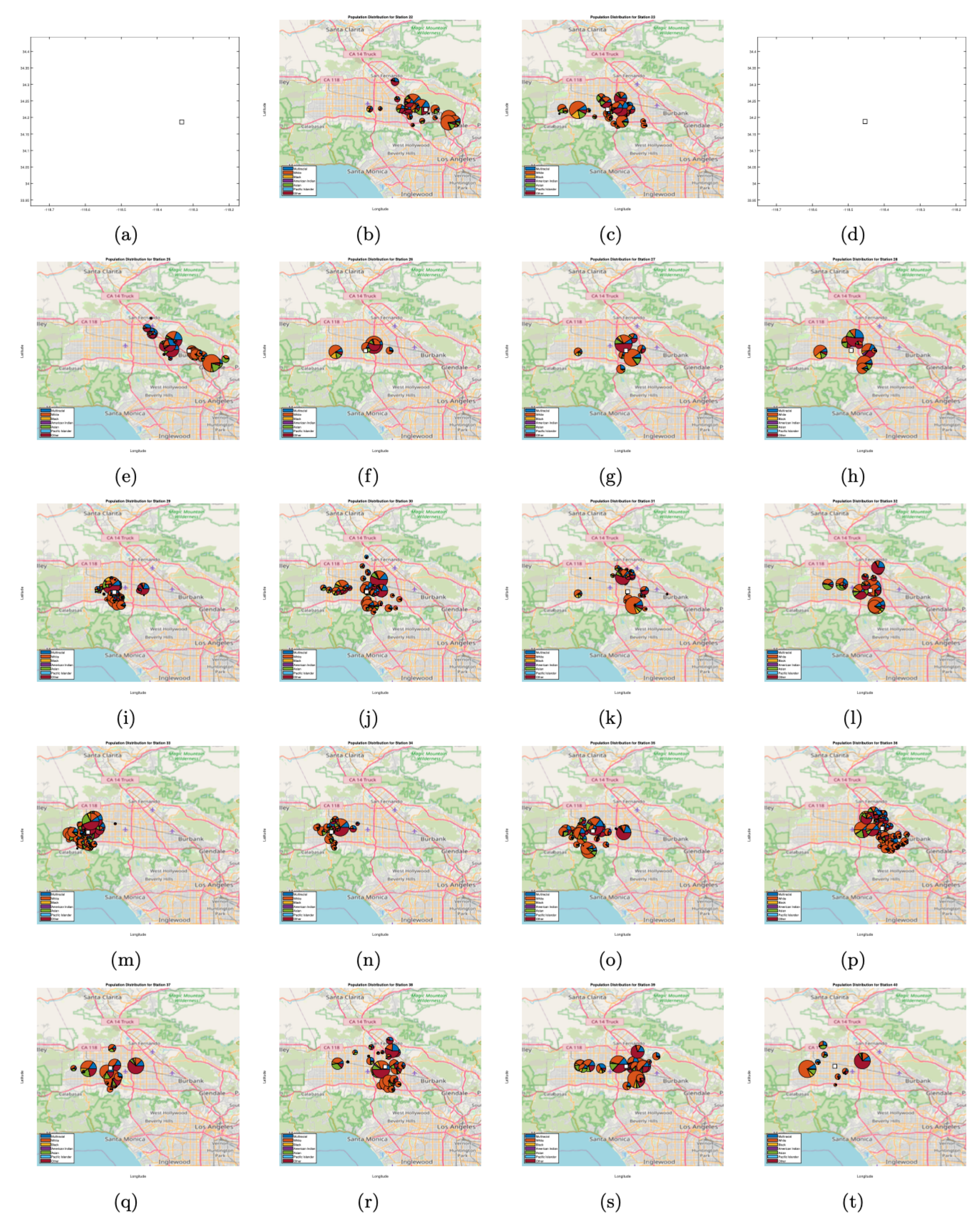}
\caption{District boundaries for district offices 21-40 under squared distance (seven groups).}
\end{figure}

\begin{figure}[h!]
	\centering
\includegraphics[width=0.99\textwidth]{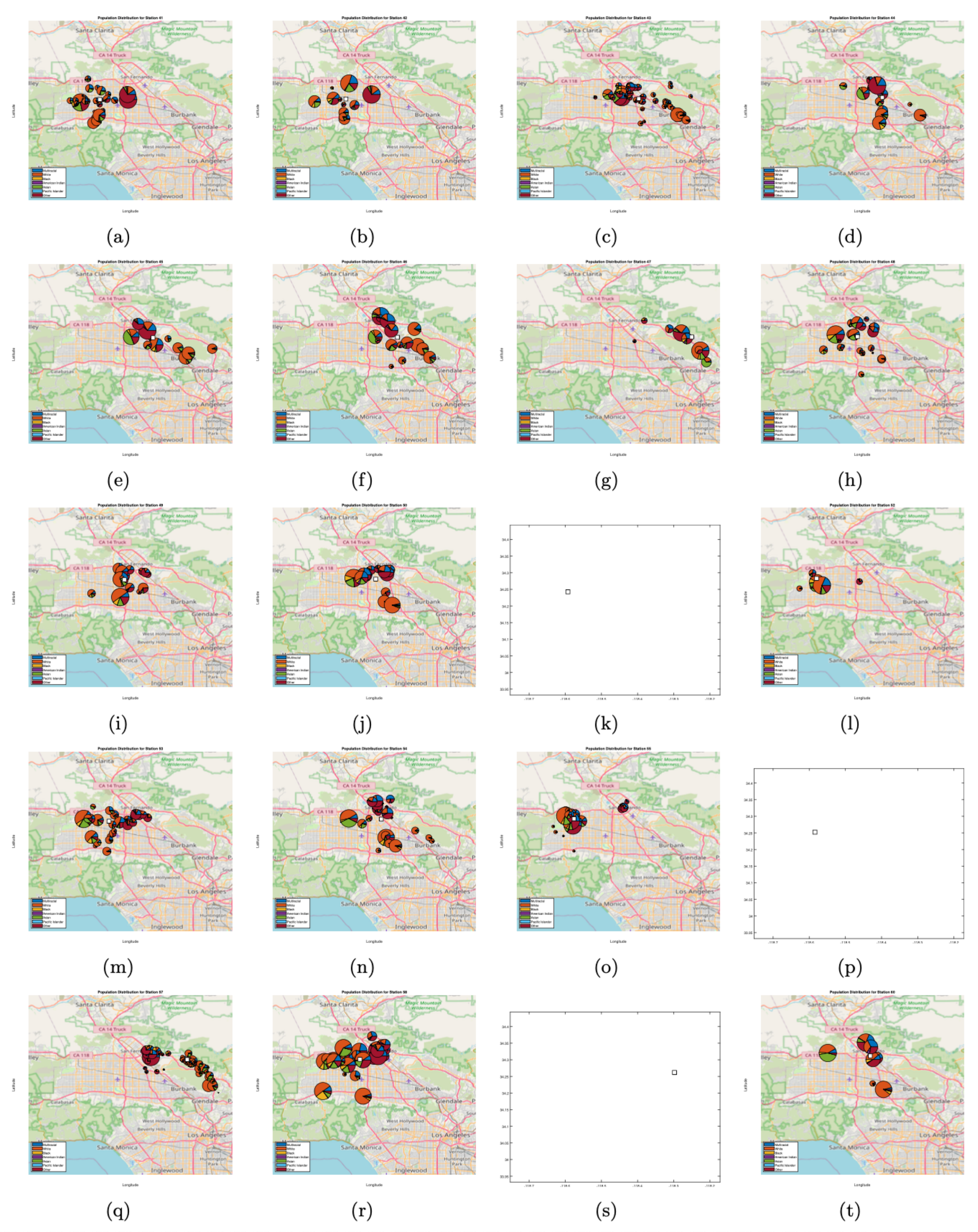}
\caption{District boundaries for district offices 41-60 under squared distance (seven groups).}
\end{figure}

\begin{figure}[h!]
	\centering
\includegraphics[width=0.99\textwidth]{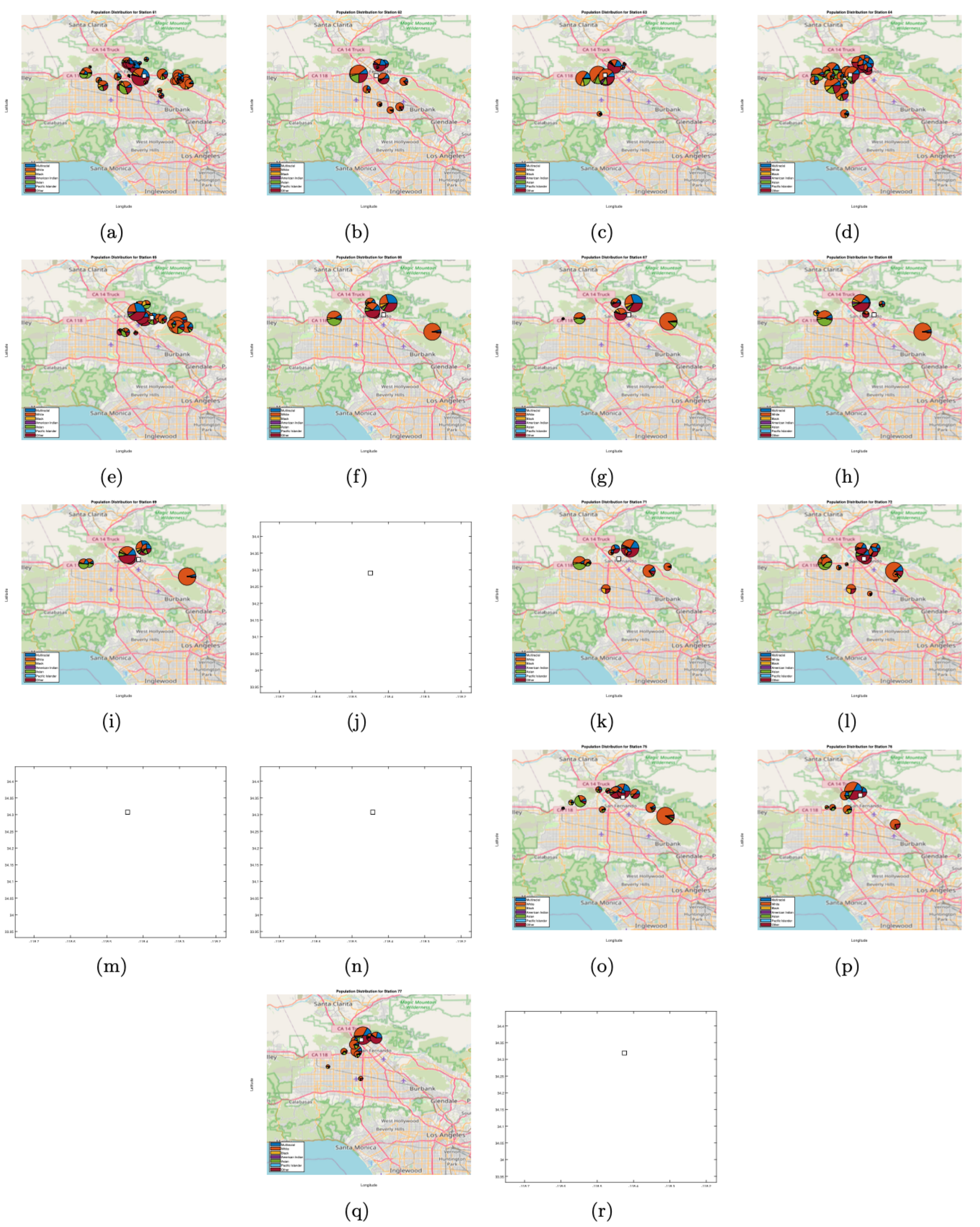}
\caption{District boundaries for district offices 61-78 under squared distance (seven groups).}
\end{figure}


\begin{figure}[h!]
	\centering
\includegraphics[width=0.99\textwidth]{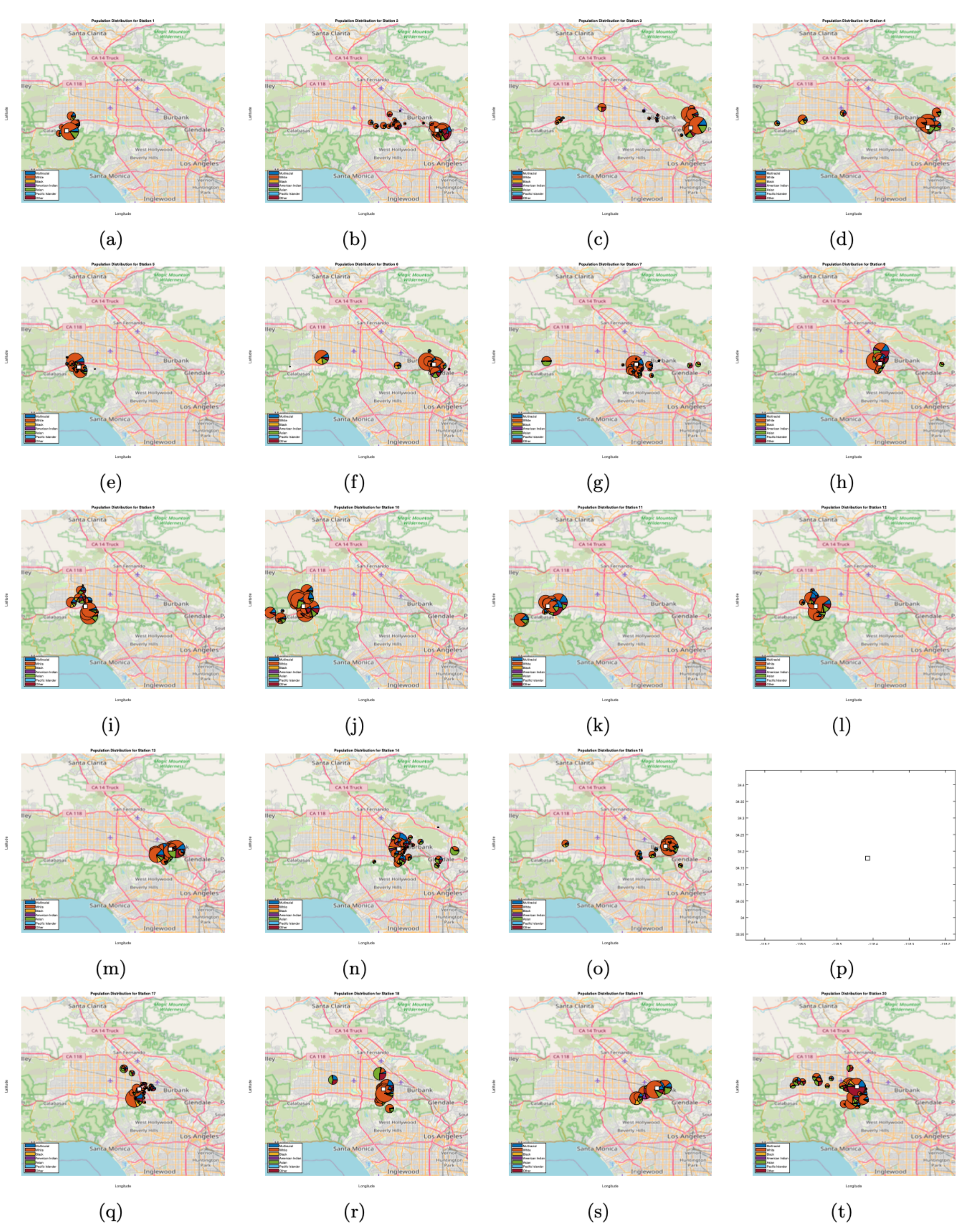}
\caption{District boundaries for district offices 1-20 under regular distance (three groups).}
\end{figure}

\begin{figure}[h!]
	\centering
\includegraphics[width=0.99\textwidth]{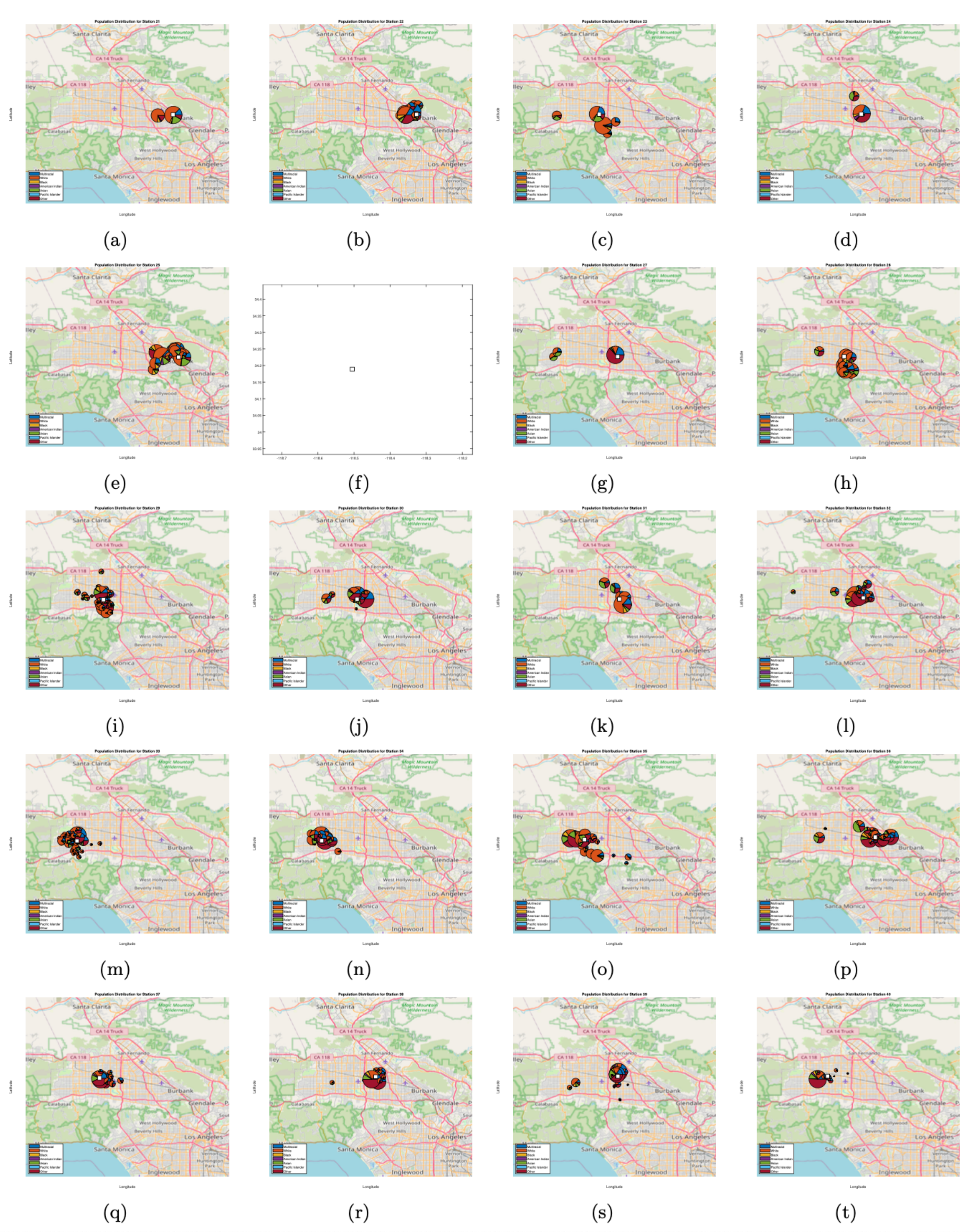}
\caption{District boundaries for district offices 21-40 under regular distance (three groups).}
\end{figure}

\begin{figure}[h!]
	\centering
\includegraphics[width=0.99\textwidth]{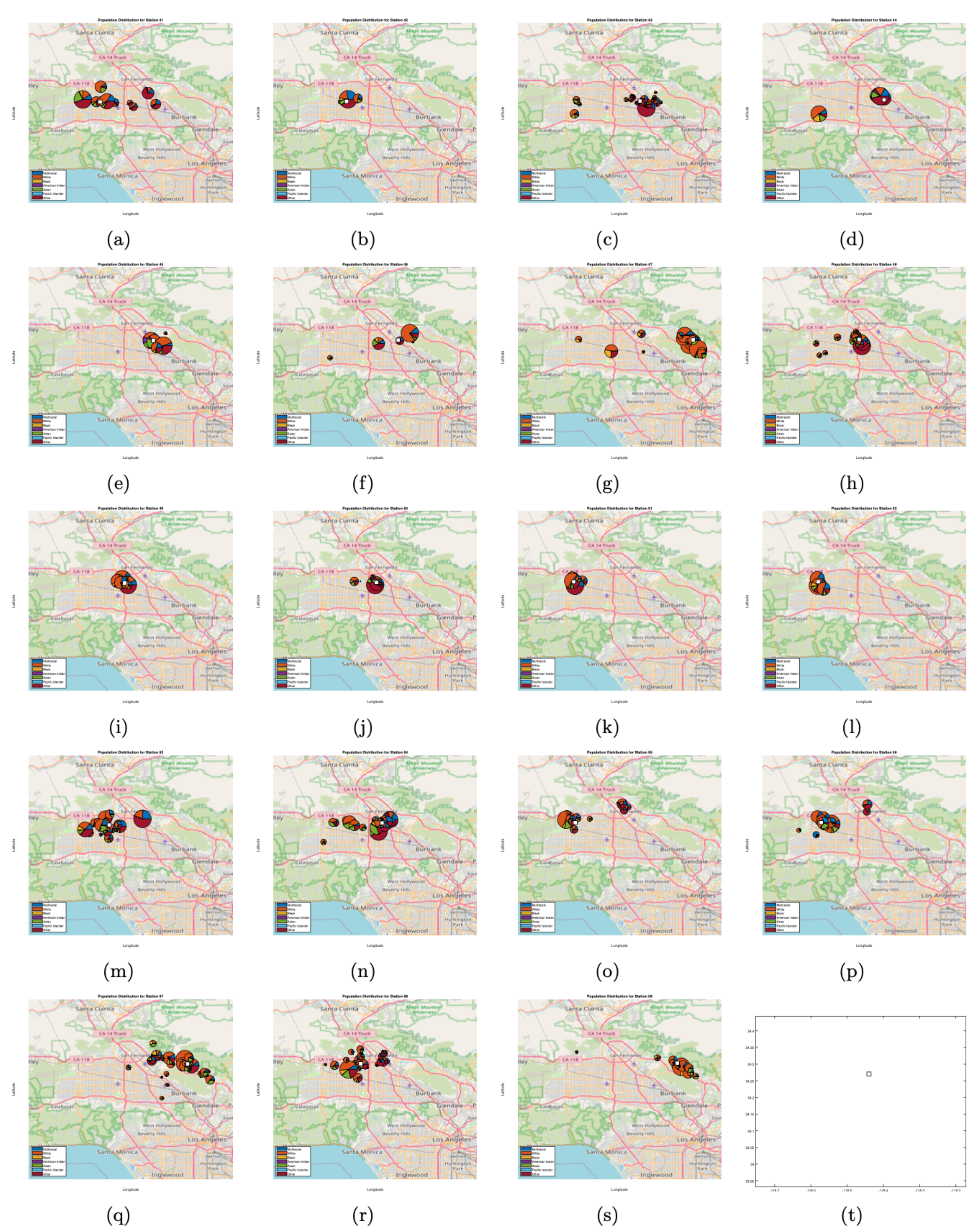}
\caption{District boundaries for district offices 41-60 under regular distance (three groups).}
\end{figure}

\begin{figure}[h!]
	\centering
\includegraphics[width=0.99\textwidth]{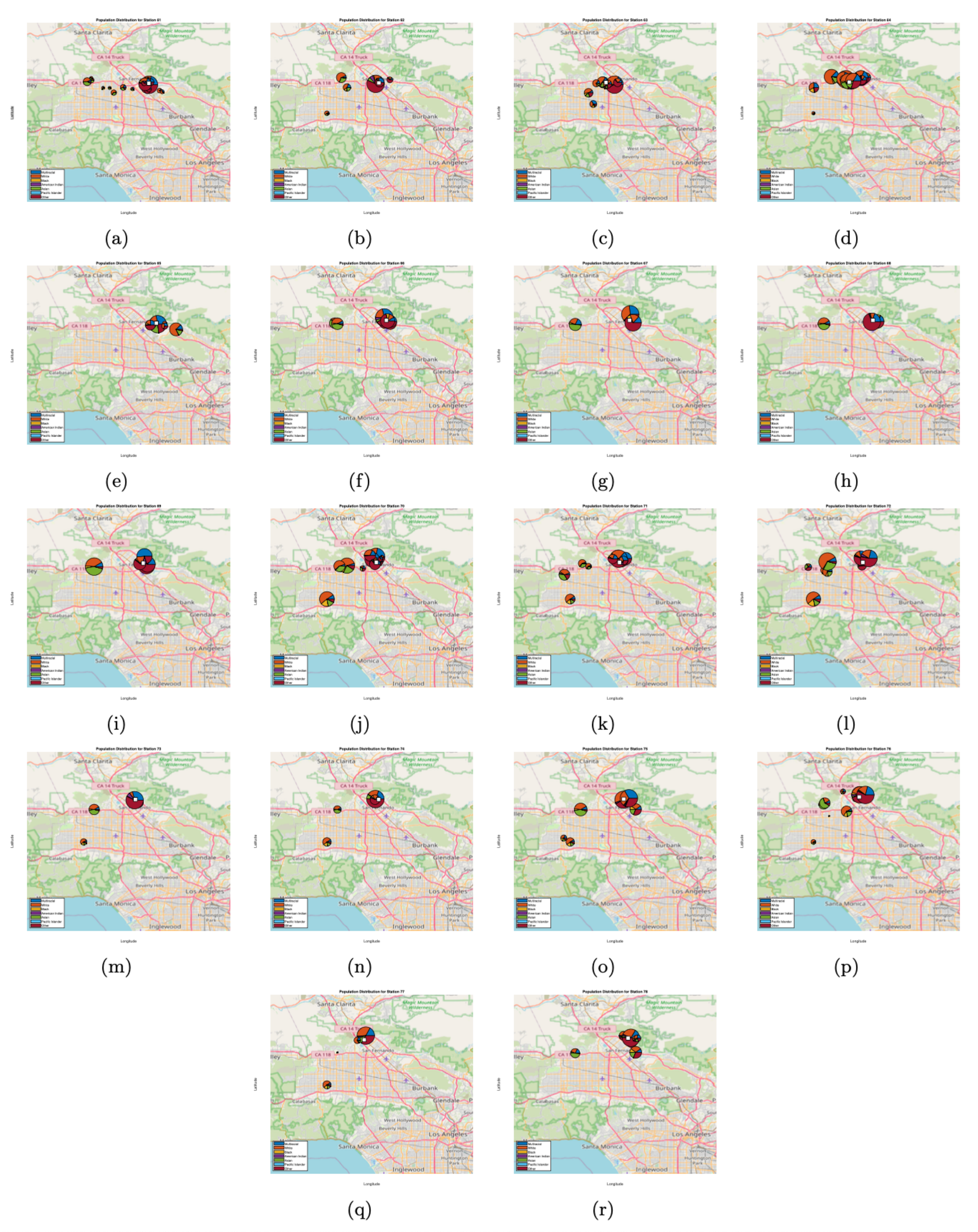}
\caption{District boundaries for district offices 61-78 under regular distance (three groups).}
\end{figure}

\begin{figure}[h!]
	\centering
\includegraphics[width=0.99\textwidth]{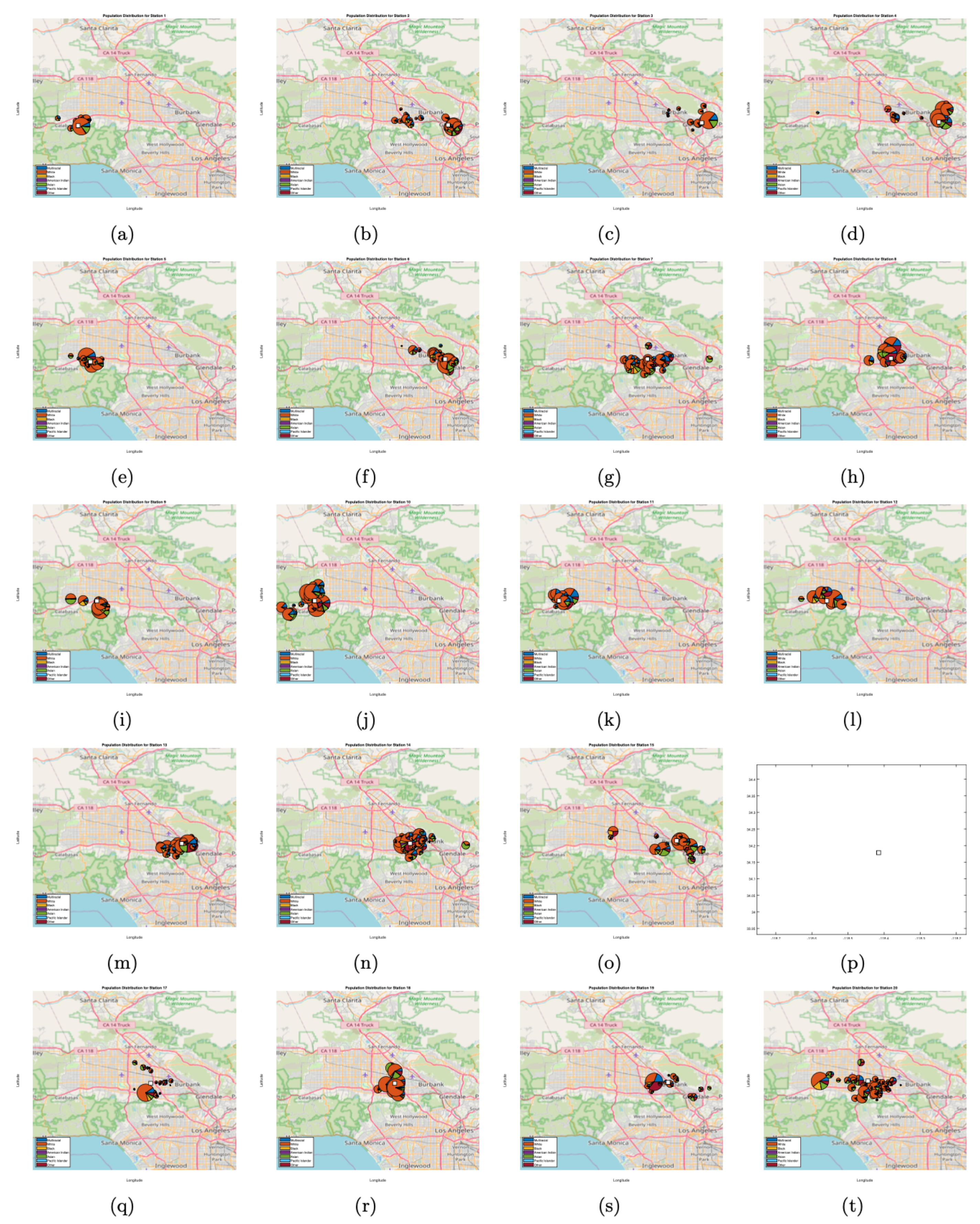}
\caption{District boundaries for district offices 1-20 under squared distance (three groups).}
\end{figure}

\begin{figure}[h!]
	\centering
\includegraphics[width=0.99\textwidth]{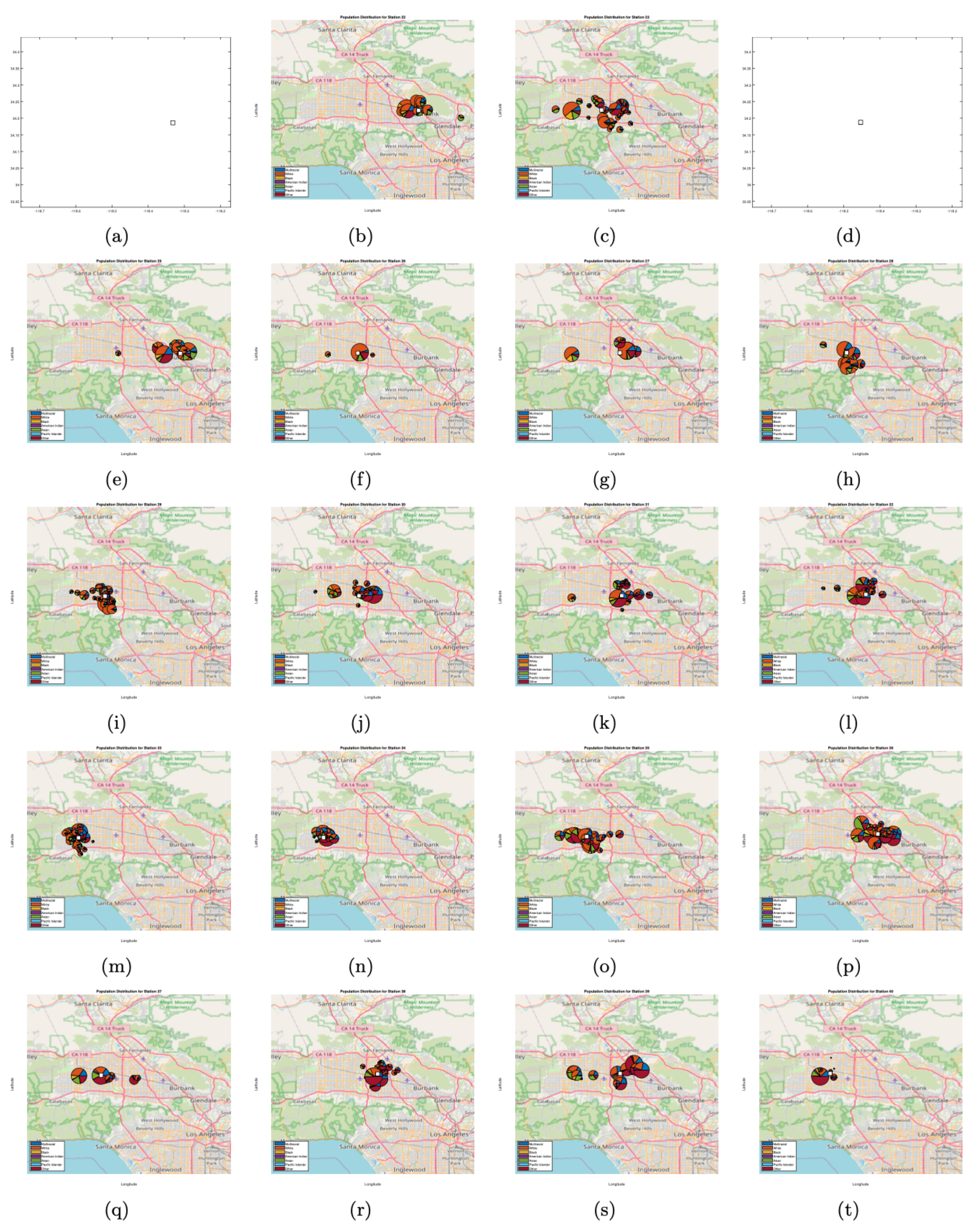}
\caption{District boundaries for district offices 21-40 under squared distance (three groups).}
\end{figure}

\begin{figure}[h!]
	\centering
\includegraphics[width=0.99\textwidth]{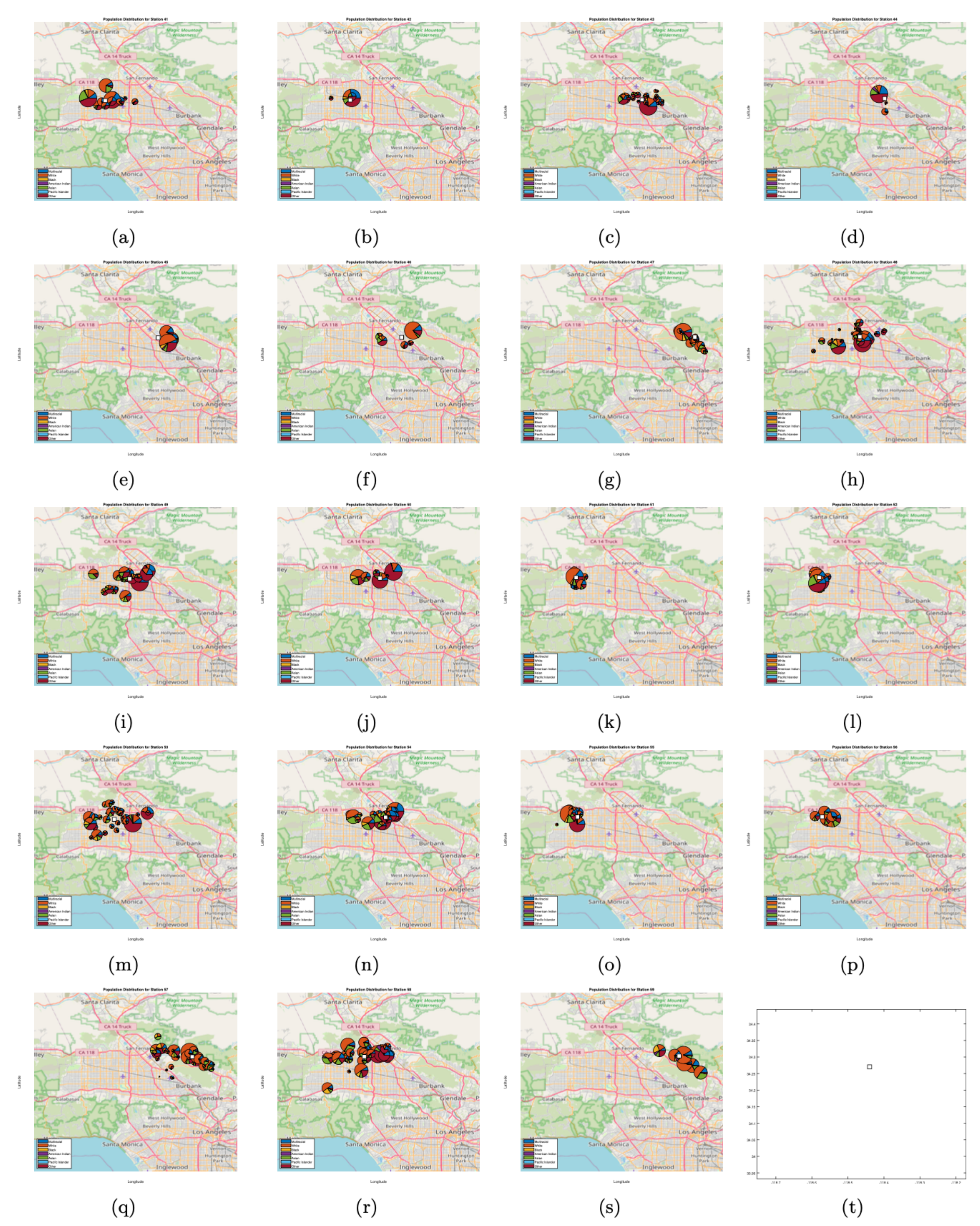}
\caption{District boundaries for district offices 41-60 under squared distance (three groups).}
\end{figure}

\begin{figure}[h!]
	\centering
\includegraphics[width=0.99\textwidth]{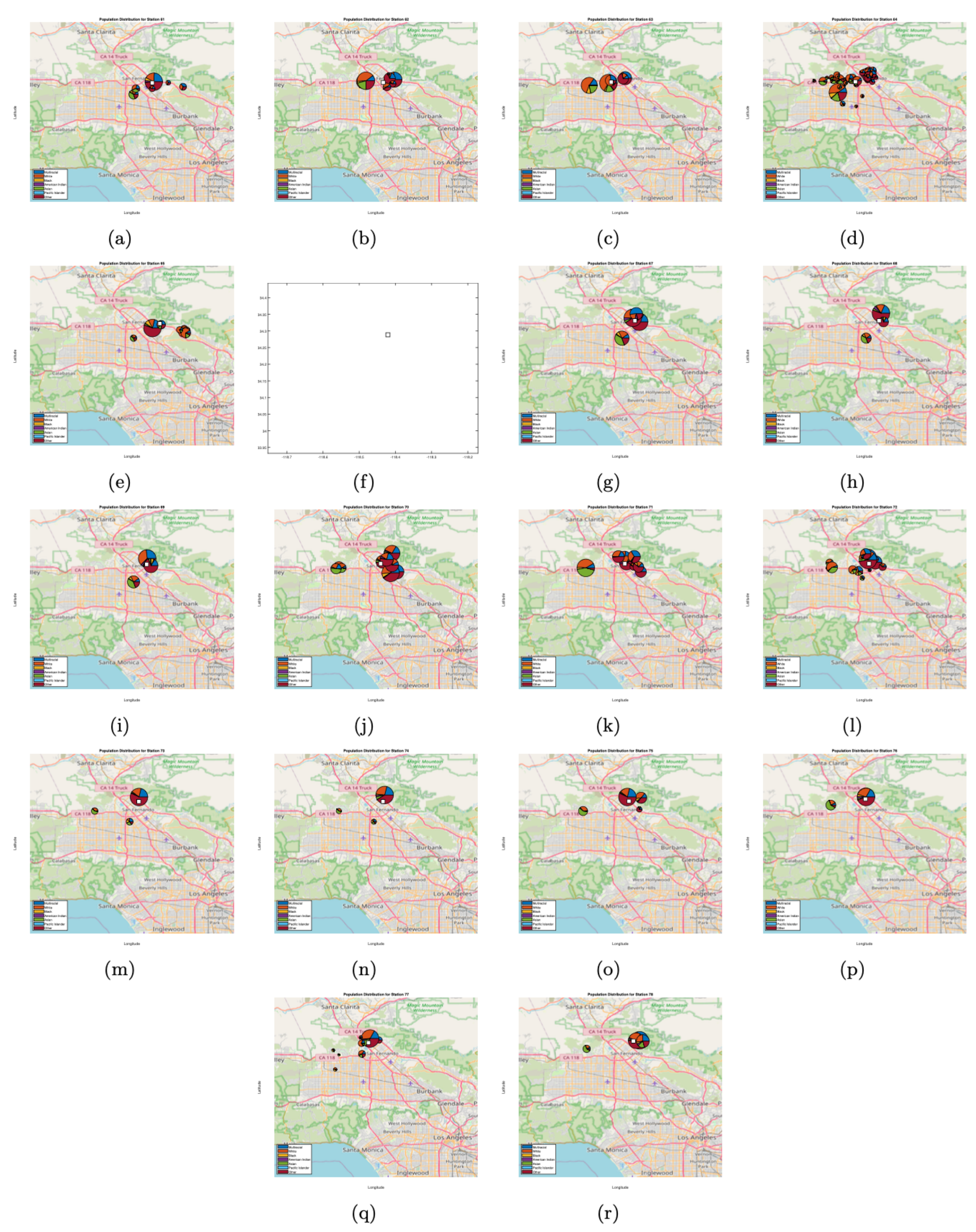}
\caption{District boundaries for district offices 61-78 under squared distance (three groups).}
\end{figure}

\end{document}